\documentclass[aps,prd,preprint,superscriptaddress,amssymb,nofootinbib]{revtex4}

\usepackage{color}
\usepackage{epsfig}
\usepackage{graphicx}
\usepackage{epstopdf}

\def\fmax{F_{MAX}}
\def\falam{F_{\alam}}
\def\fakap{F_{\akap}}

\def\hi{h_1}
\def\hii{h_2}
\def\ai{a_1}
\def\mhi{m_{h_1}}
\def\mhii{m_{h_2}}
\def\mai{m_{a_1}}
\def\mueff{\mu_{\rm eff}}
\def\mtau{m_\tau}
\def\noi{\noindent}

\def\ptl{\partial}
\def\lam{\lambda}
\def\kap{\kappa}
\def\alam{A_\lambda}
\def\akap{A_\kappa}

\def\ptl{\partial}
\def\lam{\lambda}

\def\calo{{\cal O}}

\def\mhusq{m_{H_u}^2}
\def\mhdsq{m_{H_d}^2}
\def\mssq{m_S^2}
\def\mqsq{m_Q^2}
\def\musq{m_U^2}
\def\mdsq{m_D^2}
\def\mlsq{m_L^2}
\def\mesq{m_E^2}

\def\h{h}

\def\what{\widehat}

\def\hbar{\overline h}
\def\lam{\lambda}

\def\mx{M_X}

\def\mz{m_Z}

\def\hi{h_i^0}
\def\mhi{m_{\hi}}

\def\h{h}

\def\lam{\lambda}

\def\nn{\nonumber}

\def\what{\widehat}
\def\tauptaum{\tau^+\tau^-}

\def\lsim{\mathrel{\raise.3ex\hbox{$<$\kern-.75em\lower1ex\hbox{$\sim$}}}}
\def\gsim{\mathrel{\raise.3ex\hbox{$>$\kern-.75em\lower1ex\hbox{$\sim$}}}}
\def\ifmath#1{\relax\ifmmode #1\else $#1$\fi}
\def\half{\ifmath{{\textstyle{1 \over 2}}}}

\def\vev#1{\langle #1 \rangle}
\def\lam{\lambda}

\def\mhi{m_{h_1}}

\def\calo{{\cal O}}
\def\eg{{\it e.g.}}

\def\susy{{\rm SUSY}}

\def\calo{{\cal O}}
\def\eg{{\it e.g.}}

\def\hh{H}
\def\ha{A}

\def\tanb{\tan\beta}

\def\mb{m_b}
\def\mz{m_Z}
\def\mw{m_W}
\def\mgut{M_U}
\def\mx{M_X}

\def\MPL #1 #2 #3 {{\sl Mod.~Phys.~Lett.}~{\bf#1} (#3) #2}
\def\NPB #1 #2 #3 {{\sl Nucl.~Phys.}~{\bf #1} (#3) #2}
\def\PLB #1 #2 #3 {{\sl Phys.~Lett.}~{\bf #1} (#3) #2}
\def\PR #1 #2 #3 {{\sl Phys.~Rep.}~{\bf#1} (#3) #2}
\def\PRD #1 #2 #3 {{\sl Phys.~Rev.}~{\bf #1} (#3) #2}
\def\PRL #1 #2 #3 {{\sl Phys.~Rev.~Lett.}~{\bf#1} (#3) #2}
\def\RMP #1 #2 #3 {{\sl Rev.~Mod.~Phys.}~{\bf#1} (#3) #2}
\def\ZPC #1 #2 #3 {{\sl Z.~Phys.}~{\bf #1} (#3) #2}
\def\IJMP #1 #2 #3 {{\sl Int.~J.~Mod.~Phys.}~{\bf#1} (#3) #2}
\def\NIM #1 #2 #3 {{\sl Nucl.~Inst.~and~Meth.}~{\bf#1} {#3} #2}

\def\lam{\lambda}
\def\br{{\rm Br}}
\def\tauptaum{\tau^+\tau^-}

\def\gam{\gamma}

\def\anti{\overline}
\def\epem{e^+e^-}

\def\ie{{\it i.e.}}
\def\eg{{\it e.g.}}

\def\anti{\overline}

\def\ai{a_1}
\def\aii{a_2}
\def\mai{m_{\ai}}
\def\maii{m_{\aii}}

\def\gev{~{\rm GeV}}

\def\mb{m_b}

\def\hi{\h_1}
\def\hii{\h_2}
\def\hiii{\h_3}
\def\mhi{m_{\hi}}
\def\mhii{m_{\hii}}
\def\mhiii{m_{\hiii}}

\newcommand{\nc}{\newcommand}
\nc{\beq}{\begin{equation}}   \nc{\eeq}{\end{equation}}
\nc{\bea}{\begin{eqnarray}}   \nc{\eea}{\end{eqnarray}}
\nc{\baa}{\begin{array}}      \nc{\eaa}{\end{array}}
\nc{\bit}{\begin{itemize}}    \nc{\eit}{\end{itemize}}
\nc{\ben}{\begin{enumerate}}  \nc{\een}{\end{enumerate}}
\nc{\bce}{\begin{center}}     \nc{\ece}{\end{center}}
\def\beqa{\begin{eqnarray}}
\def\eeqa{\end{eqnarray}}
\def\bed{\begin{description}}
\def\eed{\end{description}}

\def\mhi{m_{h_1}}

\def\calo{{\cal O}}
\def\eg{{\it e.g.}}

\def\half{\frac{1}{2}\,}

\def\tanb{\tan\beta}

\def\simle{
    \mathrel{\rlap{\raise 0.511ex
        \hbox{$<$}}{\lower 0.511ex \hbox{$\sim$}}}}

\def\slashchar#1{\setbox0=\hbox{$#1$}           
   \dimen0=\wd0                                 
   \setbox1=\hbox{/} \dimen1=\wd1               
   \ifdim\dimen0>\dimen1                        
      \rlap{\hbox to \dimen0{\hfil/\hfil}}      
      #1                                        
   \else                                        
      \rlap{\hbox to \dimen1{\hfil$#1$\hfil}}   
      /                                         
   \fi}

\def\lam{\lambda}

\def\ls#1{\ifmath{_{\lower1.5pt\hbox{$\scriptstyle #1$}}}}
\def\lss#1{\ifmath{^{\,\lower2.5pt\hbox{$\scriptstyle #1$}}}}

\begin{document}

\preprint{UCD-2006-16}
\title{\boldmath The Next-to-Minimal Supersymmetric Model 
Close to the R-symmetry Limit and Naturalness in $h
  \to aa$ Decays for  $m_a<2\mb$.}

\author{Radovan Derm\' \i\v sek}
\email[]{dermisek@ias.edu}

\affiliation{School of Natural Sciences, Institute for Advanced Study, Princeton,
NJ 08540}

\author{John F. Gunion}
\email[]{gunion@physics.ucdavis.edu}

\affiliation{Department of Physics, University of California at Davis, Davis, CA 95616}

\date{April 18, 2007}

\begin{abstract}

Dominant decay of a SM-like Higgs boson into particles beyond those
contained in the minimal supersymmetric standard model has been
identified as a natural scenario to avoid fine tuning in
electroweak symmetry breaking while satisfying all LEP limits.
 In the simplest such extension,
the next-to-minimal supersymmetric model, the lightest CP-even
Higgs boson can decay into two pseudoscalars. In the
scenario with least fine tuning the lightest CP-even Higgs boson
has mass of order $100\gev$. In order to escape LEP limits it must
decay to a pair of the lightest CP-odd Higgs bosons with Br$(h\to
aa)>.7$ and $m_a<2m_b$ (so that $a \to \tau^+ \tau^-$ or light quarks and
gluons). The mass of the lightest CP-odd
Higgs boson is controlled by the soft-trilinear couplings, $A_\lambda(\mz)$
and $A_\kappa(\mz)$. We identify the region of parameter space where
this situation occurs and discuss how natural this scenario is. It
turns out that in order to achieve $m_a < 2 m_b$ with $A_\lambda(\mz)$,
$A_\kappa(\mz)$ of order the typical radiative corrections, the required tuning of
trilinear couplings need not be larger than 5-10 \%. Further, the
necessity for this tuning can be eliminated in specific SUSY
breaking scenarios. Quite interestingly, $\br(h \to aa) $ is
typically above 70 \% in this region of parameter space and thus
an appropriately large value requires no additional tuning.

\end{abstract}

\pacs{}
\keywords{}

\maketitle






\section{Introduction}

The Minimal Supersymmetric Standard Model (MSSM) is a widely studied
possibility for physics beyond the standard model (SM). Its
content in the matter and gauge sectors is fixed by all known
particles and assumed superpartners. However, the choice of
a two-doublet Higgs sector is made purely on the basis of
minimality arguments. It is exactly the Higgs sector, namely the
non-observation of the Higgs boson, that casts a shadow on the
whole MSSM. It is becoming obvious that if the MSSM is the correct
description of nature then the supersymmetric (SUSY) spectrum has to be
quite unusual: heavy enough that sparticles escape direct detection, but not so
heavy that electroweak symmetry
breaking becomes unnatural. And, if the sparticles are not
particularly heavy, the MSSM parameters must be special enough
that the Higgs boson mass is pushed (through radiative corrections) above the
experimental bound.~\footnote{Scenarios that lead to such a special
SUSY spectrum were recently found, see for example mixed
anomaly-modulus mediation~\cite{modulus_anomaly} or gauge
mediation with gauge messengers~\cite{gauge_messengers}. Both
scenarios generate large mixing in the stop sector which maximizes
the Higgs mass, allowing all experimental limits to be satisfied with
a fairly light SUSY spectrum.} Another possibility, which does not
require any special assumptions about the SUSY spectrum and
parameters, is to abandon
the minimal Higgs sector, for which there was never any
deep reason anyway and which gives rise to the famous $\mu$-problem.
In this  case, the expectations for Higgs
phenomenology are modified and the tension from not having observed the
Higgs boson at LEP can be eliminated~\cite{Dermisek:2005ar}.

A particularly appealing extension of the SM or  MSSM is the
introduction of a completely new sector of particles which are
singlets under the SM gauge symmetry. As such, this extra (E)
sector would not spoil any of the virtues of the MSSM, including
the possibility of gauge coupling unification and matter particles
fitting into complete GUT multiplets. In addition, E-sector
particles would have easily escaped direct detection. Of course, if
this E-sector is completely decoupled from the SM then it plays no
role in particle physics phenomenology at accelerators. 
However, it is possible that this
sector couples to the MSSM through the Higgs fields. For example,
the superpotential can contain a term in which the two Higgs
doublets are combined in a SM-singlet form. Without additional
Higgs fields, the coefficient of this form must have dimensions of
mass, and the resulting superpotential component is the so-called
$\mu$-term. In contrast, the E-sector can couple to this
SM-singlet form in many ways, including at the renormalizable
(dimensionless coupling) level. Such couplings would have a
negligible effect on the phenomenology involving SM matter fields,
whereas they can dramatically alter  Higgs physics. For example,
they would allow the lightest CP-even Higgs boson $h$ to decay
into two of these E-fields if the E-fields are light enough. In
that case, the usual Higgs decay mode, $h \to b \bar b$, might no
longer be dominant and, since the $h b\bar b$ coupling is small,
$\br(h\to b\bar b)$ can be suppressed by a large factor. The
strategy for Higgs discovery would then depend on the way the
E-fields appearing in the decays of the $h$ themselves decay. They
might decay predominantly into other stable E-fields, in which
case the MSSM-like $h$ decays mainly invisibly. If such decays are
kinematically impossible or suppressed, then, given that 
couplings between the MSSM and E-sector Higgs fields are generically
present and imply that the mass eigenstates are mixed, 
the mostly E-field light Higgses will decay
into $b \bar b$, $\tau^+ \tau^-$ or other quarks or leptons
depending on the model. Although E-particles would have small
direct production cross sections and it would be difficult to
detect them directly, their presence would be manifest through the
dominant Higgs decay modes being $h \to 4f$, where $4f$
symbolically means four SM fields, \eg\ $b \bar b b \bar b$,
$b\bar b\tau^+\tau^-$, $\tau^+ \tau^- \tau^+ \tau^-$, $4 \gamma$
and so on~\footnote{The situation can be even more complicated if
the E-field decays into other E-fields before the latter finally
decay to SM fields. In such a case, the SM-like Higgs would
effectively decay into $8f$. Even more complicated variations can
also be constructed.}. This would imply a very complicated Higgs
phenomenology. Nevertheless, such a scenario is a simple
consequence of having an extra sector which couples to the SM or
MSSM through mixing of the Higgs sectors with one of the extra
Higgs mass eigenstates  being light enough that the SM-like Higgs
can decay into a pair them.

The situation described in the previous paragraph already occurs
in the simplest extension of the MSSM, the next-to-minimal
supersymmetric model (NMSSM) 
which adds only one singlet chiral
superfield, $\widehat{S}$. The very attractive nature
of the NMSSM extension of the MSSM on general grounds has been
discussed for many years \cite{allg}; in particular, it avoids
the need for the $\mu$ parameter of the MSSM
superpotential term $\mu \what H_u\what H_d$. The NMSSM particle content differs from the
MSSM by the addition of one CP-even and one CP-odd state in the
neutral Higgs sector (assuming CP conservation), and one
additional neutralino.  We will follow the conventions of
\cite{Ellwanger:2004xm}.  Apart from the usual quark and lepton
Yukawa couplings, the scale invariant superpotential is
\vspace*{-.07in} 
\beq \label{1.1} 
\lambda \ \widehat{S}
\widehat{H}_u \widehat{H}_d + \frac{\kappa}{3} \ \widehat{S}^3
\vspace*{-.11in} 
\eeq 
\noi depending on two dimensionless
couplings $\lambda$, $\kappa$ beyond the MSSM.  [Hatted (unhatted)
capital letters denote superfields (scalar superfield
components).]  An effective $\mu$ term arises from the first term of
Eq.~(\ref{1.1}) when the scalar component of $\what S$ acquires a
vacuum expectation value, $s\equiv\vev{\what S}$, yielding 
\beq
\mueff=\lam s\,.
\eeq
The trilinear soft terms associated with the superpotential terms in
Eq.~(\ref{1.1}) are
\vspace*{-.1in}
\beq \label{1.2}
\lambda A_{\lambda} S H_u H_d + \frac{\kappa} {3} A_\kappa S^3 \,.
\vspace*{-.1in}
\eeq
The final input parameter is
\vspace*{-.1in}
\beq \label{1.3} \tan \beta = h_u/h_d\,, 
\vspace*{-.07in}
\eeq
where $h_u\equiv
\vev {H_u}$, $h_d\equiv \vev{H_d}$.
The vevs $h_u$, $h_d$ and $s$, along with $\mz$, can be viewed as
determining the three \susy\ breaking masses squared for $H_u$, $H_d$
and $S$ (denoted $\mhusq$, $\mhdsq$ and $\mssq$)
through the three minimization equations of the scalar potential.
Thus, as compared to the three
independent parameters needed in the
MSSM context (often chosen as $\mu$, $\tan \beta$ and $M_A$), the
Higgs sector of the NMSSM is described by the six parameters
\vspace*{-.1in}
\beq \label{6param}
\lambda\ , \ \kappa\ , \ A_{\lambda} \ , \ A_{\kappa}, \ \tan \beta\ ,
\ \mu_\mathrm{eff}\ .
\vspace*{-.1in}
  \eeq
(We employ a convention in which all parameters are evaluated at scale $\mz$ unless otherwise stated.)
We will choose sign conventions for the fields
such that $\lambda$ and $\tan\beta$ are positive, while $\kappa$,
$A_\lambda$, $A_{\kappa}$ and $\mu_{\mathrm{eff}}$ should be allowed
to have either sign.
In addition, values must be input for the gaugino masses ($M_{1,2,3}$)
and for the soft terms related to the (third generation)
squarks and sleptons ($\mqsq$, $\musq$, $\mdsq$, $\mlsq$,  $\mesq$,
$A_t$, $A_b$ and $A_\tau$)
that contribute to the
radiative corrections in the Higgs sector and to the Higgs decay
widths.  For moderate $\tanb$, the soft parameters which play
the most prominent role are $\mqsq$, $\musq$, $\mdsq$ and $A_t$.

Of all the  possible new phenomena, the additional Higgses in the NMSSM can
lead to, perhaps the most intriguing one is the possibility of the
lightest CP-even Higgs decaying into a pair of the two
lightest CP-odd Higgses, $\hi\to\ai\ai$,  where the latter are
mostly
singlets~\cite{Gunion:1996fb,Dermisek:2005ar,Dermisek:2005gg,newewsb,Dobrescu:2000jt}.
Not only would $\hi\to\ai\ai$ decays complicate
Higgs searches, but also it is found that precisely this scenario
can essentially eliminate the fine tuning of EWSB in
the NMSSM for $\mhi\sim 100\gev$~\cite{Dermisek:2005ar,Dermisek:2005gg,newewsb}.
If $\br(\hi\to\ai\ai)>0.7$ and $\mai<2\mb$, the usual
(LEP) limit on the Higgs boson mass does not apply and
the SUSY spectrum can be arbitrarily
light, perhaps just above the experimental bounds, \ie\ certainly
light enough for natural EWSB~\cite{Dermisek:2005ar,Dermisek:2005gg,newewsb}.

In this paper, we identify the region of parameter space where this
situation occurs and discuss how natural this scenario is. The
mass of the lightest CP-odd Higgs boson is controlled by
the soft-trilinear couplings $A_\lambda$ and $A_\kappa$ and vanishes
in the R-symmetry limit, $A_\lambda, \, A_\kappa \to 0$. However,
both $A_\lambda$ and $A_\kappa$ receive radiative corrections from
gaugino masses and so arbitrarily small values of trilinear
couplings would require cancellation between the bare values and
the radiative corrections. It turns out that in order to achieve $\mai <
2 m_b$ with $A_\lambda$, $A_\kappa$ of order the typical radiative
corrections, a tuning of trilinear couplings at the level of
$5\%-10\%$ might be required, but that even less fine-tuning is
possible in the context of various types of models. 
Quite interestingly, $\br(\hi\to\ai\ai)>70~\%$
is automatic in this region of parameter space, implying
no need for additional tuning in order to achieve 
$\br(\hi\to b\anti b)$ small enough to escape LEP limits~\footnote{Some aspects of EWSB and the possibility of having a light CP-odd
Higgs boson in the NMSSM in the 
R-symmetry limit were also recently discussed in Ref.~\cite{Schuster:2005py}. Results and conclusions presented 
in~Ref.~\cite{Schuster:2005py} are based on studying a region of parameter space very near the R-symmetry 
limit with soft-trilinear couplings much below the typical size of radiative corrections.
The conclusions of this reference do not necessarily apply to the
scenario discussed above that assumes soft-trilinear couplings
$A_\lambda$ and $A_\kappa$ of order the typical radiative corrections
(they could, however, originate from an exact symmetry limit at some high scale). We refer to our scenario as {\it NMSSM close to the R-symmetry limit}.}.

It is important to note that to a large extent the tuning in $\alam$
and $\akap$ required for $\br(\hi\to\ai\ai)> 0.7$ and $\mai<2\mb$
can be separated from fine-tuning/naturalness for proper EWSB.  
The value of $\mz$, as obtained from
the renormalization group (RG) equations after evolving from the
GUT-scale, $\mgut$,
down to $\mz$, is primarily sensitive to $M_3(\mgut)$, $\mhusq(\mgut)$,
$\mhdsq(\mgut)$, $\mssq(\mgut)$, $\mqsq(\mgut)$, $\musq(\mgut)$, $\mdsq(\mgut)$, and $A_t(\mgut)$ (and $A_b(\mgut)$ if
$\tanb$ is large). In some cases, $M_2(\mgut)$ can contribute
significantly to the standard measure of EWSB fine-tuning with respect
to the GUT-scale parameters.
In contrast, $M_1(\mgut)$, $\alam(\mgut)$, and $\akap(\mgut)$ have to take
very large values in order to contribute significantly to fine-tuning.
The scenarios that have small fine-tuning that we focus on are ones in
which these parameters are small enough that they do not affect the
measure of the fine-tuning associated with EWSB.
In a companion paper~\cite{newewsb}, we discuss
EWSB fine-tuning in detail, expanding upon the earlier discussions
in Refs.~\cite{Dermisek:2005ar,Dermisek:2005gg}. 

Recently various scenarios, similar in spirit, which improve on the
naturalness of EWSB by modifying Higgs decays or which, in general, suggest the
possibility of an extra sector near the EW scale have been discussed; see
\eg\ \cite{Chang:2005ht,Barger:2006dh,Strassler:2006im,Carpenter:2006hs}.
Phenomenological consequences and possible collider signatures for the
NMSSM and, more specifically, 
for some of these scenarios have also been discussed in
Refs.~\cite{Graham:2006tr,Strassler:2006ri,Arhrib:2006sx,Strassler:2006qa}.

The paper is organized as follows. In the next section, 
we discuss the NMSSM close to the
R-symmetry limit. Numerical results are presented in
Sec.~\ref{sec:results} and we conclude in Sec.~\ref{conclusions}.

\section{NMSSM close to the R-symmetry limit}

In the NMSSM, one of the CP-odd Higgses is massless in the
Peccei-Quinn symmetry limit, $\kappa \to 0$, or in the R-symmetry
limit, $A_\kappa, A_\lambda \to 0$~\cite{Dobrescu:2000jt}. We
focus here on the second case because it was identified as the
easiest way to achieve EWSB without large fine tuning in soft SUSY
breaking terms~\cite{Dermisek:2005ar}. This scenario requires
that the lightest CP-even Higgs boson, $\hi$,
decays to a pair of the lightest CP-odd Higgs bosons
 with $\br(\hi\to \ai\ai)$ large enough that
the $\hi\to b \bar b$ signal is highly reduced  (compared to the SM signal).
Furthermore, it was found that this scenario with reduced $\br(\hi
\to b \bar b)$ is consistent with the observed excess of events at
$m_h \simeq 100$ GeV in $Zh$ production at LEP~\cite{Dermisek:2005gg}.

The masslessness of $a_1$ in the limit $A_\kappa, A_\lambda \to 0$
can be understood as a consequence of a global $U(1)_R$ symmetry
of the superpotential under which the charge of $S$ is half of the
charge of $H_u H_d$. In the limit $A_\kappa, A_\lambda \to 0$ it
is also a symmetry of the scalar potential. This symmetry is
spontaneously broken by the vevs of $H_u$, $H_d$ and $S$,
resulting in a Nambu-Goldstone boson in the spectrum. Soft
trilinear couplings explicitly break $U(1)_R$ and thus lift the
mass of the $a_1$. For small trilinear couplings, the mass of the
lightest CP-odd Higgs boson is approximately given as:
\begin{equation}
m_{a_1}^2  \simeq 3s \left( \frac{3 \lambda A_\lambda \cos^2
\theta_A }{2 \sin 2 \beta}
 - \kappa  A_\kappa \sin^2 \theta_A \right), \label{eq:ma1_first}
\end{equation}
where $\cos \theta_A$ measures the doublet component of the
lightest CP-odd Higgs mass eigenstate,
\begin{equation}
a_1 =  \cos \theta_A \, A_{MSSM} + \sin \theta_A \, A_S.
\label{eq:a1_composition}
\end{equation}
In the limit of large $\tan \beta$ or $|s| \gg v$, $\cos\theta_A$ can be
approximated by
\begin{equation}
\cos \theta_A  \simeq \frac{v\sin2\beta}{s }. \label{eq:cos_th_a}
\end{equation}
In this limit, the $\ai$ mass eigenstate is mostly singlet and
\beq
\mai^2\sim 3s\left({3\lam\alam v^2\sin 2\beta\over 2 s^2 }-\kap\akap\right)\,.
\label{eq:ma1}
\eeq

Naively, an arbitrarily small mass for the $\ai$ is achievable provided
small values of $A_\kappa$ and $A_\lambda$ are generated by a SUSY
breaking scenario. Indeed, there are SUSY breaking scenarios, \eg\
gauge mediation or gaugino mediation, which have zero soft
trilinear couplings in the leading order. However, even if zero
values of $A_\kappa$ and $A_\lambda$ are generated at the SUSY
breaking scale, the corresponding electroweak (EW) scale values
will be shifted due to radiative corrections from gaugino masses.
The typical EW scale values can be estimated from the one-loop renormalization
group equations for $A_\lambda$ and $A_\kappa$,
\begin{eqnarray}
{d A_\lambda \over d t} &=& {1\over 16\pi^2}\left[
 6 A_t \lambda_t^2 + 8 \lambda^2 A_\lambda + 4 \kappa^2  A_\kappa + 6 g_2^2 M_2 +
              \left({6\over 5}\right) g_1^2  M_1 \right], \label{RG_Al}\\
{d A_\kappa \over d t} &=& {12\over 16\pi^2} \left( \lambda^2 A_\lambda + \kappa^2
A_\kappa  \right),\label{RG_Ak}
\end{eqnarray}
where $t=\log(Q/\mz)$, $A_t$ is the top soft trilinear coupling
and $M_2$ and $M_1$ are the masses of the SU(2) and U(1) gauginos,
and we have neglected terms proportional to $A_b$ and $A_\tau$.
Starting with $A_\lambda=0$ at the GUT scale, we find that
$A_\lambda (\mz) \sim M_2$ because the $\log (\mgut/\mz)$ coming
from the integration approximately cancels the $(6g_2^2)/(16\pi^2)$ loop
factor. On the other hand, $A_\kappa$ receives contributions from gaugino
masses only at the two-loop level implying that $A_\kappa (\mz)$
is expected to be much smaller than $A_\lambda (\mz)$. Assuming
gaugino masses of order 100 GeV, we should naturally expect
$A_\lambda (\mz) \simeq 100\gev$ and $A_\kappa(\mz) \sim few\gev$.
Much smaller values would require cancellations between the values
of $A_\lambda$, $A_\kappa$ coming from a particular SUSY breaking
scenario and the contributions from the radiative corrections.

For sizable $A_\lambda (\mz)$, Eq.~(\ref{eq:ma1}) is no longer
a good approximation for the mass of the lightest CP-odd Higgs.
In order to understand the numerical results presented later,
we have developed a more accurate
formula. The $2 \times 2$ mass matrix squared for the CP-odd Higgs
bosons, in the basis $(A_{MSSM}, A_S)$, has the following matrix
elements~\cite{Ellwanger:2004xm}:
\begin{eqnarray}
M_{11}^2 &=& \frac{ 2 \lambda s}{\sin 2 \beta} \left(A_\lambda + \kappa s \right), \label{eq:M11sq}\\
M_{12}^2 &=& \lambda v \left( A_\lambda - 2 \kappa s \right), \label{eq:M12sq} \\
M_{22}^2 &=& 2 \lambda \kappa v^2 \sin 2 \beta + \lambda A_\lambda
\frac{v^2 \sin 2 \beta}{2s} - 3 \kappa A_\kappa s.
\label{eq:M22sq}
\end{eqnarray}
The eigenstate masses are
\beq
m_{_{ \ai \atop \aii}}^2=\half\left[M_{11}^2+M_{22}^2\mp \sqrt{(M_{11}^2-M_{22}^2)^2+4(M_{12}^2)^2}\right]\,.
\eeq
The mixing angle for the diagonalization process is obtained from
\bea
\sin 2\theta_A&=&-{2M_{12}^2\over
  \sqrt{(M_{11}^2-M_{22}^2)^2+4(M_{12}^2)^2}}\\
\cos 2\theta_A&=&-{M_{11}^2-M_{22}^2\over
  \sqrt{(M_{11}^2-M_{22}^2)^2+4(M_{12}^2)^2}}\,,
\eea
where we are using the convention defined in
Eq.~(\ref{eq:a1_composition}). Obviously, the value of $\theta_A$
is only determined ${\rm mod}(\pi)$.
In our numerical work, we employ the program
NMHDECAY\cite{Ellwanger:2004xm} which adopts the convention 
$0\leq \theta_A\leq \pi$.
If $M_{11}^2$ is much larger in magnitude than the other entries, then
it must be positive and 
the mass squared of the lightest CP-odd Higgs boson is then given by
\begin{equation}
m_{a_1}^2 \simeq \frac{M_{11}^2 M_{22}^2 - (M_{12}^2)^2}{M_{11}^2
+ M_{22}^2}.\label{ma1start}
\end{equation}
For typical values of trilinear couplings, $|A_\kappa| \ll |A_\lambda|
\sim v \ll |s|$, and $\tan \beta \gsim few$  we find:
\begin{equation}
m_{a_1}^2 \simeq 3 s \left( \frac{3 \kappa \lambda^2 A_\lambda
v^2}{{2\lambda s\over \sin 2\beta} (A_\lambda + \kappa s)   - 3 \kappa
A_\kappa s } - \kappa A_\kappa \right), \label{eq:ma1_better}
\end{equation}
which reduces to Eq.~(\ref{eq:ma1}) if we neglect $A_\lambda$
and $A_\kappa$ compared to $s$. Similarly, the mixing angle is
determined by
\bea
\cos2\theta_A&=&2\cos^2\theta_A-1\simeq \frac{2 (M_{12}^2)^2}{(M_{11}^2 -
  M_{22}^2)^2}-1,\\
\sin2\theta_A&=&2\sin\theta_A\cos\theta_A\simeq -{2M_{12}^2\over M_{11}^2-M_{22}^2}\,,
\eea
from which we obtain (using the conventions defined earlier)
$\sin\theta_A\sim 1$ and the doublet component of $\ai$ is given by
\beq
\cos\theta_A\simeq -{M_{12}^2\over M_{11}^2-M_{22}^2}\simeq  -
\frac{\lambda v (A_\lambda - 2 \kappa s) \sin 2 \beta}{2 \lambda s
(A_\lambda + \kappa s) + 3 \kappa A_\kappa s \sin 2 \beta}\,.
\label{ctaform}
\eeq
This reduces to Eq.~(\ref{eq:cos_th_a}) for very small
$A_\lambda$ and $A_\kappa$. 

Both Eq.~(\ref{eq:ma1}) and Eq.~(\ref{eq:ma1_better}) indicate that
sizable $A_\lambda$ could give a large contribution to
the mass of the lightest CP-odd Higgs. However this term is highly
suppressed if the lightest CP-odd Higgs is mostly singlet. In this
case, both terms in Eq.~(\ref{eq:ma1}) or Eq.~(\ref{eq:ma1_better})
are comparable and then it depends on their relative sign as to whether
they contribute constructively or destructively. One measure of the
tuning in $A_\lambda(\mz)$ and $A_\kappa(\mz)$ necessary to achieve $\mai <2 m_b$ is
\beq
\fmax = \max \left\{|\falam|,|\fakap|\right\}\,,\label{eq:F_max}
\eeq
where
\beq
\falam\equiv \frac{\alam(\mz)}{\mai^2} \frac{d
\mai^2}{d \alam(\mz)}  \,, \ \fakap\equiv \frac{\akap(\mz)}{\mai^2} \frac{d
\mai^2}{d \akap(\mz)} \label{eq:F_max2}
\eeq
are evaluated for given choices of $\alam(\mz)$ and
$\akap(\mz)$ that yield a given $\mai$.  This definition, which
reflects the fact that $\mai$ is determined by these
parameters for fixed $\lam,\kap,\mueff,\tanb$, will prove useful below
for discussing the sensitivity of $\mai$ to GUT-scale parameters.  We argue below
that $\fmax$ is typically an upper bound on the magnitude of fine
tuning with respect to GUT scale parameters.  Thus, $\fmax$ provides
a useful first measure for determining a ``preferred'' region of parameter
space where a small mass for the lightest CP-odd Higgs boson is
achieved with the least tension. However, we will also find that in
large classes of models $\fmax$ greatly over-estimates the fine-tuning.

The fine-tuning measure for $\mai^2$ relative to GUT scale parameters
that is completely analogous to that employed for EWSB is 
\beq
F_{\mai}\equiv {\rm max}_p f_p\,,\quad \mbox{with}\quad f_p\equiv {d\log \mai^2\over d\log p}\,,
\eeq
where $p$ is any GUT-scale parameter,
\beq
p= M_i\  (i=1,2,3),\  \mhusq,\  \mhdsq,\  \mssq,\ 
A_t,\  \akap,\  \alam, \ \mqsq,\  \musq,\  \mdsq 
\eeq
(to name the most important GUT-scale parameters). 
However, for our purposes we can simplify this general computation
because we are only interested in $\mai^2$ fine tuning for cases in
which the EWSB fine tuning is already known to be small, that is cases
in which 
\beq
\mz^2=2\left[-\lam^2(\mz) s^2(\mz)+{\mhdsq(\mz)-\tan^2\beta(\mz)\mhusq(\mz)\over \tan^2\beta(\mz)-1}\right]
\eeq
is insensitive to the parameters $p$. 
 Small fine tuning for $\mz^2$ means
that
$v(\mz)$ (which sets $\mz$), $s(\mz)$, $\tanb(\mz)$, $\mhusq(\mz)$ and $\mhdsq(\mz)$ are not
fine-tuned with respect to the various $p$ listed above. 
The only additional parameters upon
which $\mai^2$ depends that could still be sensitive to the GUT-scale
parameters $p$ when $\mz$ is not are $\alam(\mz)$ and $\akap(\mz)$.  
For any of the $p$, we can then approximate 
\beq
f_p\sim {p\over \alam(\mz)} \falam {d\alam(\mz)\over d p}+{p\over
  \akap(\mz)} \fakap {d\akap(\mz)\over dp}\,.
\eeq
To understand the implications of this formula, we need to solve the RG
equations and express $\alam(\mz)$ and $\akap(\mz)$
in terms of the GUT-scale values of all the soft-SUSY-breaking parameters. 
 (Of course, we are taking the GUT scale as an example;
a similar exercise can be done for any scale.)
 The solution depends on $\lam$, $\kap$ and $\tanb$; we
give only a representative example. For $\lambda = 0.2$, $\kappa =
\pm 0.2$ and $\tan \beta = 10$ we find:
\begin{eqnarray}
A_\lambda (\mz) &=& -0.03 A_\kappa + 0.93 A_\lambda - 0.35 A_t -
    0.03 M_1 - 0.37 M_2 + 0.66 M_3, \label{alrun}\\
A_\kappa (\mz) &=& 0.90 A_\kappa - 0.11 A_\lambda + 0.02 A_t +
0.003 M_1 + 0.025 M_2 - 0.017 M_3\label{akrun}\,,
\end{eqnarray}
where the parameters on the right-hand side of these equations are
the GUT-scale values.

Before discussing the fine-tuning implications of Eqs.~(\ref{alrun}) and
(\ref{akrun}),
it is useful to understand a few of their features.
First, we note that while the coefficients in the
$A_\lambda(\mz)$ expansion do not change much when changing $\lambda$
and $\kappa$, the coefficients in the $A_\kappa(\mz)$ expansion are
quite sensitive to changes of the $\lam,\kap$ Yukawa couplings. The
reason is that gaugino masses enter $A_\kappa$ through the $A_\lambda$
term in the RG equation, see Eqs.~(\ref{RG_Al})--(\ref{RG_Ak}), the
strength of which is controlled by $\lambda$. The opposite sign of the
coefficient in front of $M_3$ in the expansion of $A_\lambda$ as
compared to the coefficients in front of $M_1$ and $M_2$ is due to the
fact that $M_3$ enters only through $A_t$ in the RG evolution, while
both $M_1$ and $M_2$ enter directly. Similarly, the opposite sign in
front of the gaugino masses in the $A_\lambda$ expansion as compared
to the $A_\kappa$ expansion is due to the fact that gaugino masses
enter $A_\kappa$ through $A_\lambda$ in the evolution.  The above
expansions provide several ways to achieve $A_\kappa \ll A_\lambda
\sim \mz$. The easiest way is to assume that a SUSY breaking scenario
generates negligible trilinear couplings at the GUT scale. 

We now return to a consideration of the fine-tuning implications of Eqs.~(\ref{alrun}) and
(\ref{akrun}). 
If there is a $p_\lam$   that dominates $\alam(\mz)$
and a $p_\kap$ that dominantes $\akap(\mz)$, and these are different, then
\beq
{p_\lam\over \alam(\mz)}{d \alam(\mz)\over dp_\lam}\sim
  \calo\left(1\right)\,,\quad\mbox{and}\quad {p_\kap\over \akap(\mz)}{d \akap(\mz)\over dp_\kap}\sim
  \calo\left(1\right)\,,
\eeq
and roughly $F_{\mai}\sim \fmax$.  If the same $p$ dominates both
$\alam(\mz)$ and $\akap(\mz)$ then
$F_{\mai}\sim f_p\sim \falam+\fakap$. This result also holds  if
the GUT-scale parameters are correlated.  For example,
consider the case of universal
gaugino masses and zero trilinear couplings at the GUT scale, for which
$A_\lambda (\mz)=0.26 M_{1/2}$ and $A_\kappa (\mz) = 0.01
M_{1/2}$.  Then, 
\beq
{M_{1/2}\over \alam(\mz)}{d\alam(\mz)\over dM_{1/2}}={M_{1/2}\over \akap(\mz)}{d\akap(\mz)\over dM_{1/2}}=1
\eeq
and it is quite precisely the case that
\beq
F_{\mai}=f_{M_{1/2}}\sim \falam+\fakap
\label{fpresult}
\eeq
As we shall see in the numerical section of this paper, 
$\falam$ and $\fakap$ are typically opposite in
sign and of similar magnitude. This can already be seen from the
approximate formula of Eq.~(\ref{eq:ma1}) where the linearity of
$\mai^2$ in $\alam(\mz)$ and $\akap(\mz)$ would give
(neglecting the mild dependence of $s(\mz),\ldots$ on $M_{1/2}$ for
the points of interest)
\beq
\falam+\fakap \sim 1\,,
\eeq 
and fine tuning with respect to $M_{1/2}$ would be small.  The more
precise (but still approximate) result of Eq.~(\ref{eq:ma1_better})
gives somewhat larger results for $\falam+\fakap$ for some parameter
choices, but a subset of choices still gives $\falam+\fakap\sim
\calo\left(1\right)$. Later, we will present numerical results for
$\falam$ and $\fakap$ that confirm that they largely cancel against
one another for a significant fraction of parameter choices.  The
result of Eq.~(\ref{fpresult}) clearly applies whenever the gaugino
masses are correlated (in any way) and trilinears (at the GUT scale)
are small or correlated with the gaugino masses.  As already noted,
this same result also applies whenever a single term dominates in
Eqs.~(\ref{alrun}) and (\ref{akrun}). Many models fall into one or the
other of these categories.

Let us reemphasize that tuning in $\alam(\mz)$ and $\akap(\mz)$ is
completely unnecessary to achieve a light CP-odd Higgs boson in models
with specific relations among GUT-scale parameters.  Any SUSY breaking
scenario that determines all soft trilinear couplings 
 and gaugino masses from a SUSY breaking scale will
automatically give $A_\lambda (\mz)= c_\lambda M_{SUSY}$ and $A_\kappa
(\mz) = c_\kappa M_{SUSY}$, where $c_\lambda$ and $c_\kappa$ depend
(given the known values of $g_1$ and $g_2$) only on the couplings
$\lam$, $\kap$ and $\lam_t$ (equivalently, $\tanb$, given the known
value of $\mw$). The mass of the lightest CP odd Higgs boson will be
given as $m_{a_1}^2 = f(\lambda, \kappa, \tan \beta) M_{SUSY}^2$ and
will either be small or not.  This means that, in any SUSY breaking
scenario that is determined by a SUSY breaking scale only, there is no
tuning of $m_{a_1}$ with respect to the SUSY breaking scale.
(Algebraically, ${d \log \mai^2\over d\log M_{SUSY}^2}=1$.)  This
result holds even if there are large cancellations among the RG
contributions to $\alam(\mz)$ and $\akap(\mz)$.  Whether or not a
light CP odd Higgs boson is possible simply depends on the above
couplings.

Let us now turn our attention to $\br(\hi\to\ai\ai)$, a completely
general expression for which is the following (neglecting
phase space suppression):
\bea
\Gamma(\hi\to\ai\ai)&\sim& {\mw^2\over 32 \pi g_2^2
  \mhi}\Biggl[{g_1^2+g_2^2\over 2}\cos
2\beta\sin(\beta+\alpha)\cos^2\theta_A\cos\theta_S\nn\\
&&+2\lam^2
\biggl(\sin(\beta-\alpha)\sin^2\theta_A\cos\theta_S+{s\over
  v}\cos^2\theta_A\sin\theta_S 
\nn\\
&&\hspace*{.5in}
+(\cos\alpha\sin^3\beta-\sin\alpha\cos^3\beta)
\cos^2\theta_A\cos\theta_S
\biggr)\nonumber\\
&&+2\lam\kap\Biggl(\cos(\beta+\alpha)\sin^2
  \theta_A\cos\theta_S-\sin2\theta_A\sin\theta_S
\nonumber\\
&&\hspace*{.5in}-{ s\over
    v}\sin(\beta-\alpha) \sin 2\theta_A \cos\theta_S
+{ s\over v}\sin2\beta\cos^2\theta_A\sin\theta_S\Biggr)\nonumber\\
&&+{\lam\alam\over
    v}\left(\sin(\beta-\alpha)\sin2\theta_A\cos\theta_S+\sin2\beta\cos^2\theta_A\sin\theta_S\right)\nonumber\\
&&+\left({4\kap^2s\over
      v}-{2\kap\akap\over
      v}\right)\sin^2\theta_A\sin\theta_S\Biggr]^2\label{aawid}
\eea
This is to be contrasted with the dominant SM decay channel, $\hi\to
b\bar b$, the width for which is (neglecting phase space suppression):
\bea
\Gamma(\hi\to b\bar b)&\sim & {3g_2^2\over 32 \pi \mw^2}\left({\cos\alpha\over\cos\beta}\right)^2\mhi\mb^2\cos^2\theta_S\,,\label{bbwid}
\eea
Our conventions are those 
of NMHDECAY in which the $WW$ coupling
relative to SM strength is given by $\sin(\beta-\alpha)\cos\theta_S$ and
$v^2=2\mw^2/g_2^2$; $\mb$ is to be evaulated at scale $\mhi$. 
In  Eq.~(\ref{eq:a1_composition}),
$\cos\theta_A$ gives the MSSM doublet component of the $\ai$
and $\sin\theta_S$ is the coefficient of the singlet component of the
$\hi$; both are small numbers in our scenario. 
Detailed numerical results and discussion for the
$\hi\to\ai\ai$ width will be given in the
next section.  However, it is important to understand the limit in
which $\alam,\akap\to 0$. This is nicely illustrated in
the case of $v/s\ll 1$. In this limit we have:  $\alpha\to\beta-\pi/2$ 
(more precisely, $\sin(\beta-\alpha)\to \sin 2\beta
+\calo\left({v^2/s^2}\right))$; 
$\cos\theta_A\to (v/s)\sin2\beta+\calo\left({v^3/s^3}\right)$ (see
Eq.~(\ref{eq:cos_th_a})); $\sin\theta_A\to
1-\calo\left({v^2/s^2}\right)$; $\sin\theta_S\to
-(v/s)(\lam^2/(2\kap^2))(1-(\kap/\lam)\sin2\beta)+\calo\left({v^3/s^3}\right)$;
$\cos\theta_S\to 1-\calo\left({v^2/s^2}\right)$. With these inputs, 
we find 
\bea
\Gamma(\hi\to\ai\ai)&\sim& {\mw^2\over 32 \pi g_2^2
  \mhi}\left[2\lam^2+2\lam\kap\sin2\beta-{4\lam\kap s\over
    v}\cos\theta_A+{4\kap^2 s\over v}\sin\theta_S+\calo\left({v^2/s^2}\right)\right]^2\nonumber\\
&\sim&{\mw^2\over 32 \pi g_2^2
  \mhi}\left[\calo\left({v^2/s^2}\right)\right]^2\,.
\eea
In particular, the $\left[\ldots\right]$ does not actually vanish 
for $\alam=\akap=0$ (there is no
exact symmetry argument).
One finds that $\br(\hi\to\ai\ai)$ can approach the
$[0.1-0.2]$ range for the smallest $\kap$ values allowed by
general theoretical consistency (good EWSB vacuum, \ldots) for a given $\lam$ value. 
This is, of course, insufficient for escaping LEP constraints when
$\mhi\sim 100\gev$. Thus, having an adequate 
size for $\br(\hi\to\ai\ai)$ for escaping the LEP constraints depends critically
upon having non-zero values for $\akap$ and, in particular, $\alam$.
From the numerical results presented in the next section, it will be
clear that magnitudes for $\alam$ and $\akap$ of order those developed
from RGE running beginning with $\alam(\mgut)=\akap(\mgut)=0$ are
sufficient to give large $\br(\hi\to\ai\ai)$.

\section{Results \label{sec:results}}

As discussed in the introduction, the Higgs sector in the NMSSM is
determined by 6 basic parameters, given in Eq.~(\ref{1.3}),
as well as subsidiary parameters entering through loop corrections.
(In this section, all parameters are defined at scale $\mz$.)
Consequently, a
complete survey of the parameter space is difficult. To present
results in a manageable way, we fix $\mu$ and $\tan \beta$ together
with all soft SUSY breaking masses and scan over trilinear  and
soft-trilinear couplings. We will plot results in various 2-parameter
planes. The parameters are scanned over the following regions with
fixed steps: $\lambda \in (0,0.5)$ using 30 steps of size $0.01666$; $\kappa \in
(-0.5, 0.5)$ using 70 steps of size $0.014286$; $A_\lambda \in (-300 \, {\rm GeV},
300 \, {\rm GeV} )$ using 100 steps of size $6\gev$; and finally $A_\kappa \in
(-20 \, {\rm GeV}, 20 \, {\rm GeV} )$ using 100 steps of size
$0.4\gev$. Varying the fixed
soft SUSY breaking masses does not have any significant effect
on the results while different choices of $\mu$ and $\tan \beta$ lead
to important changes. Thus, we will present results for several
choices of $\mu$ and $\tan \beta$ keeping all SUSY-breaking masses
fixed at $M_{SUSY}=300\gev$. Let us recall that we are interested in looking
for parameters yielding $\mhi>90\gev$ and $\mai<2\mb$. The latter
is imposed so that the LEP limits on $\hi\to\ai\ai\to 4b$ do not apply
(since if the $\ai$ decayed primarily to $b\bar b$
they would require $\mhi\gsim 110\gev$, \ie\ above our
preferred $\mhi\sim 100\gev$ value).  The $\mhi>90\gev$ restriction
implies that we are above the maximum value for which
LEP limits on $\hi\to\ai\ai\to 4\tau$ are available.
In our plots, the small dark blue diamonds are all points that
satisfy the above constraints, while the large light blue crosses are
those which satisfy all experimental limits, the main
experimental constraint  being that $\br(\hi\to b\bar b)$ must be
suppressed sufficiently by a large $\br(\hi\to\ai\ai)$ that LEP
limits on the $Z\hi\to Zb\bar b$ channel are satisfied. Roughly,
this requires $\br(\hi\to\ai\ai)\gsim 0.7$.

\begin{figure}
\includegraphics[width=2.4in,angle=90]{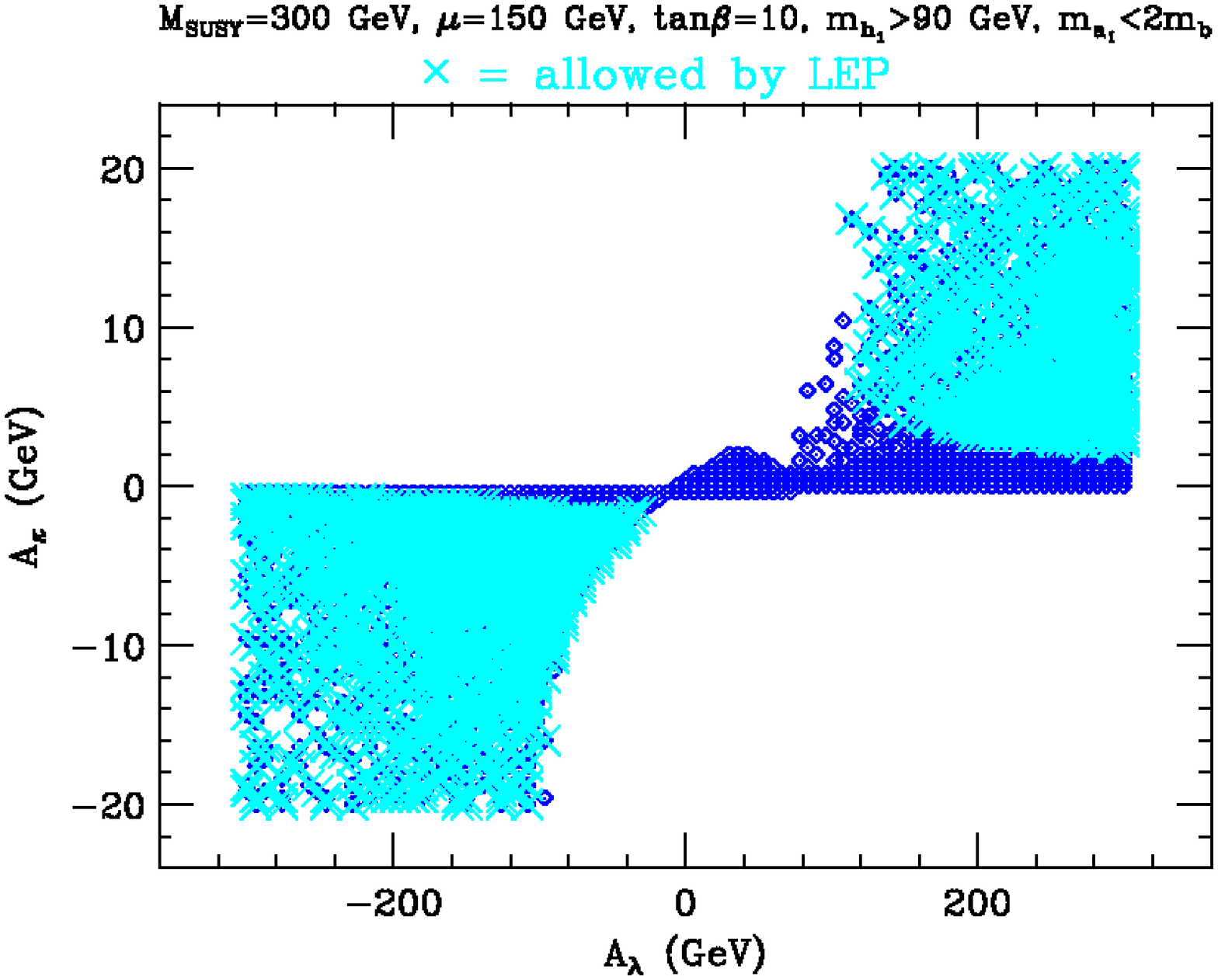}
\hspace{0.2cm}
\includegraphics[width=2.4in,angle=90]{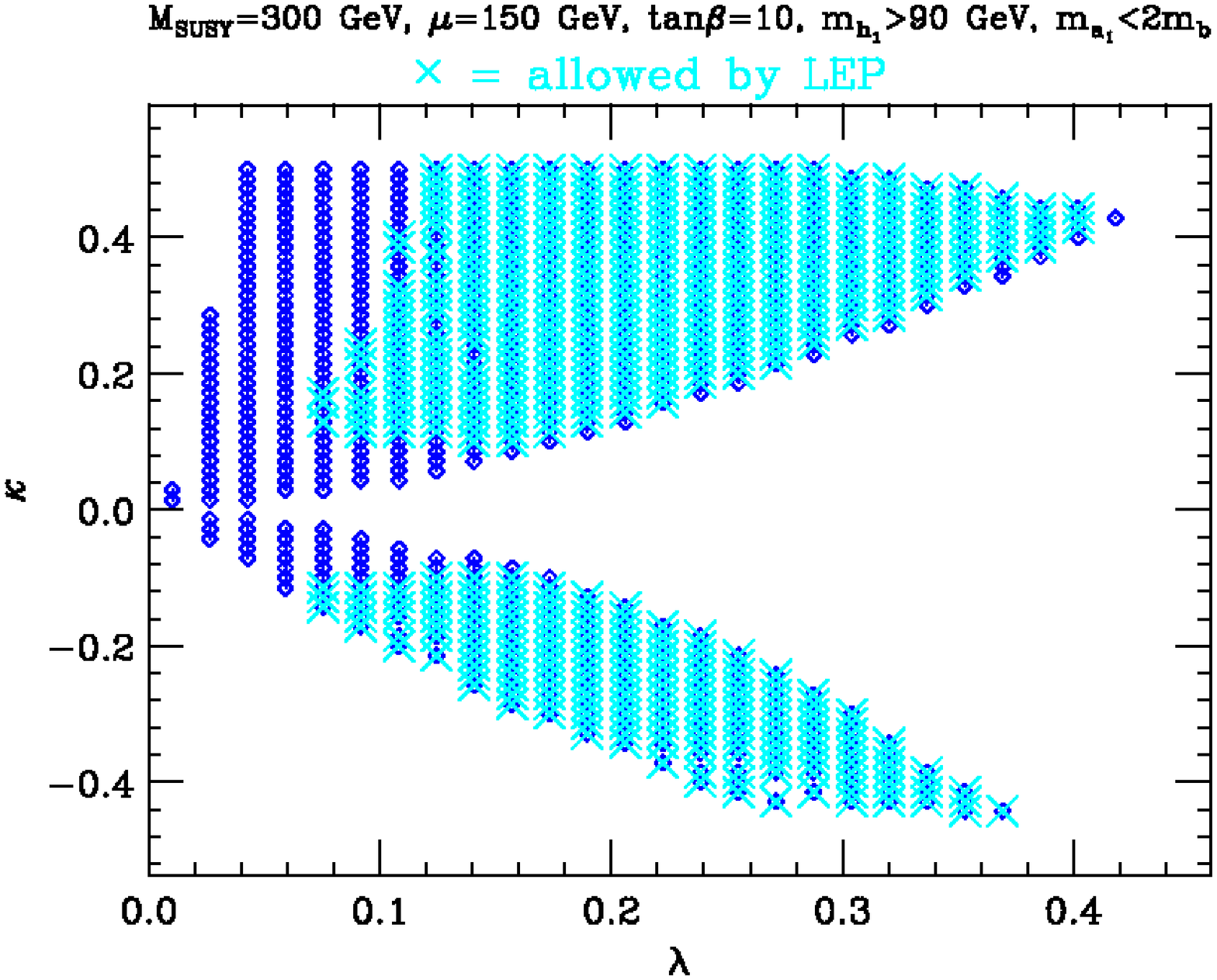}
\caption{Allowed parameter space in the $A_\kappa - A_\lambda$ and
  $\kappa - \lambda$ planes. Light grey (cyan) large crosses are
  points that satisfy all
experimental limits. The dark (blue) diamonds are those points that
do not have large enough $\br(\hi\to\ai\ai)$ so as to suppress $\br(\hi\to
b\bar b)$ sufficiently to escape LEP limits.}
\label{fig:AkAl_and_kl}
\end{figure}

\begin{figure}
\includegraphics[width=2.4in,angle=90]{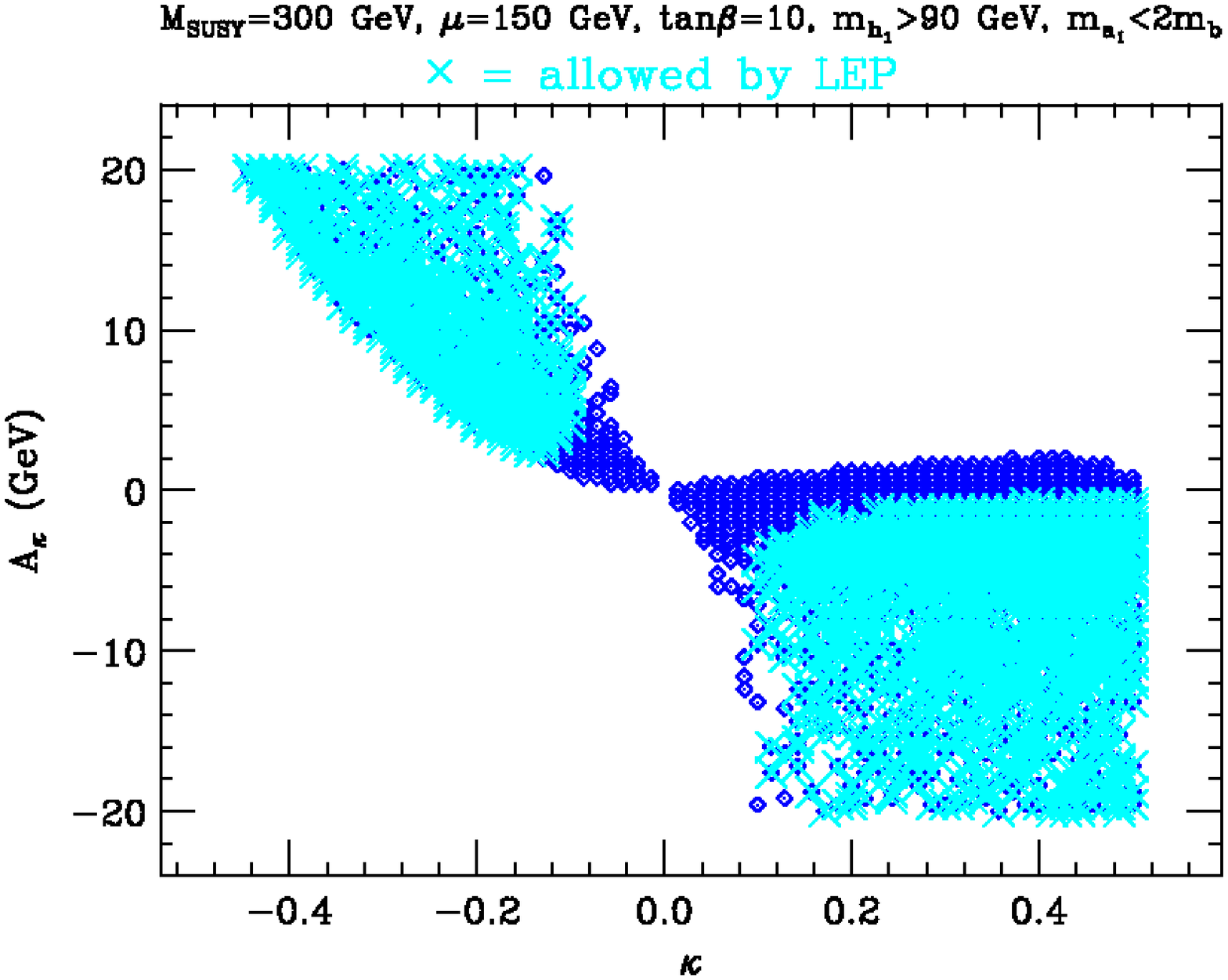}
\hspace{0.2cm}
\includegraphics[width=2.4in,angle=90]{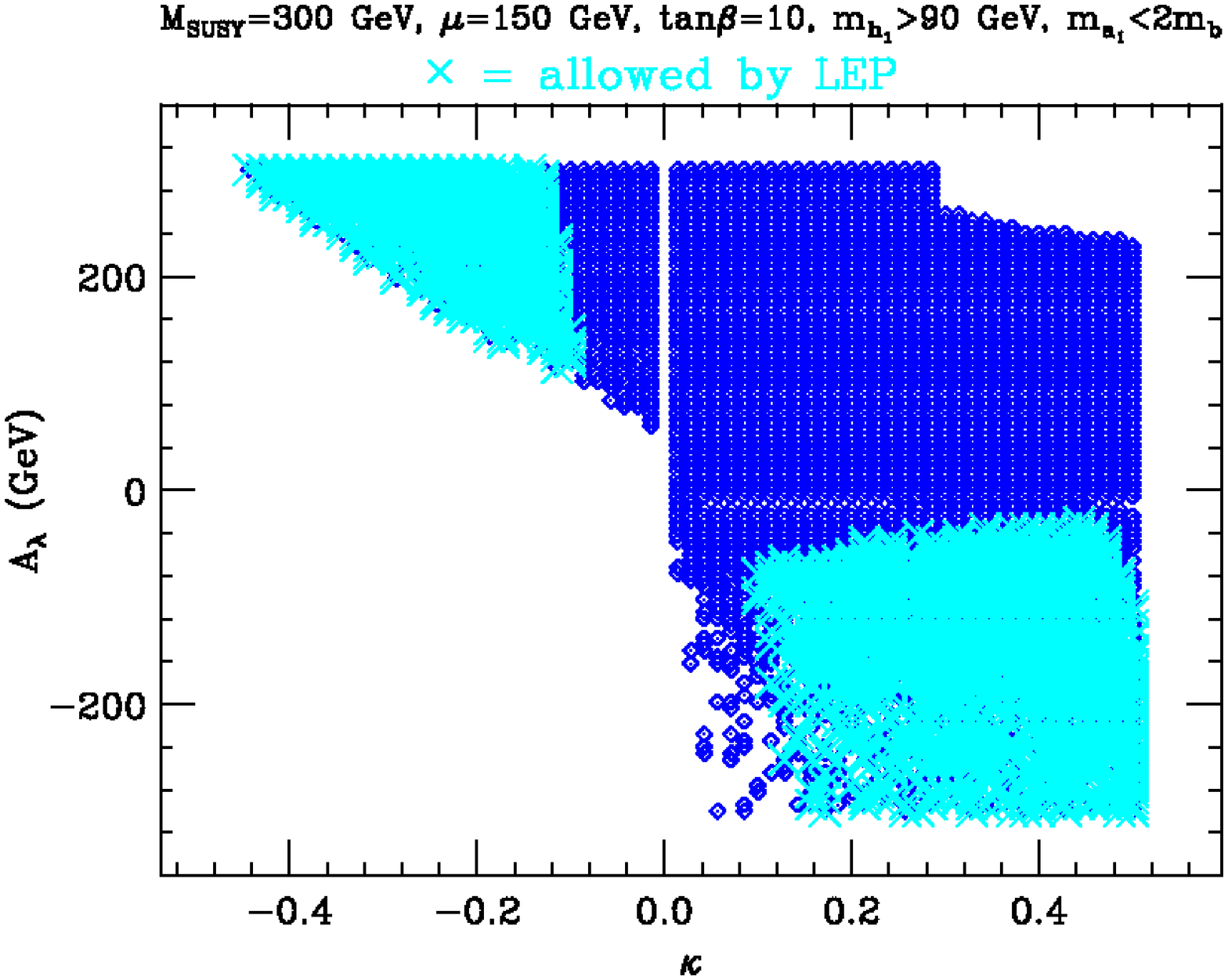}
\caption{Allowed parameter space in the $A_\kappa - \kappa$ and
  $A_\lambda - \kappa$ planes. 
Point conventions as in Fig.~\ref{fig:AkAl_and_kl}.} 
\label{fig:Akk_and_Alk}
\end{figure}

\begin{figure}
\includegraphics[width=2.4in,angle=90]{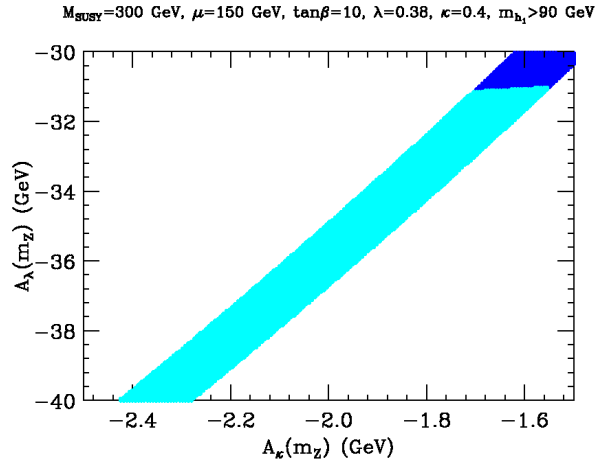}
\caption{A selected region of the allowed parameter space in the
$A_\kappa - A_\lambda$ plane for fixed values of $\lambda=0.38$
and  $\kappa=0.4$. Point conventions as in Fig.~\ref{fig:AkAl_and_kl}.
} \label{fig:AkAl_fixed_lk}
\end{figure}

We first present results for $M_{SUSY} = 300$ GeV, $\mu = 150$ GeV
and $\tan \beta = 10$. In Fig.~\ref{fig:AkAl_and_kl} we plot the
allowed region
 of parameter space in the $A_\kappa -
A_\lambda$ and $\kappa - \lambda$ planes. Similarly, in
Fig.~\ref{fig:Akk_and_Alk} we plot the allowed region in the
$A_\kappa - \kappa$ and $A_\lambda - \kappa$ planes. In
Fig.~\ref{fig:AkAl_fixed_lk}, we plot a selection of the range of
$\akap$ and $\alam$ values that have $\mai<2\mb$ for fixed values
of $\lambda$ and $\kappa$. From Fig.~\ref{fig:AkAl_fixed_lk}, we
see that for a given value of $\alam(\mz)$ keeping $\mai < 2 m_b$
requires that $\akap(\mz)$ be adjusted to a level of order 10\%;
at fixed $\akap(\mz)$, $\alam(\mz)$ must lie within about a 5\%
range. As quantified later, this is a rough measure of the tuning
required to achieve consistency of the envisioned scenario with
LEP constraints when $\mhi\sim 100\gev$. Note that $\akap,\alam$
tuning could be made arbitrarily mild if very small values of
$\akap(\mz)$ were allowed.  However, as illustrated by the dark
blue points, too small a value for $\akap(\mz)$ leads to a value
for $\br(\hi\to\ai\ai)$ that is small and therefore a value of
$\br(\hi\to b\bar b)$) that is too large for consistency with LEP
constraints on the $Zh\to Zb\bar b$ channel. In any case, as we
have discussed, very small values of $\akap(\mz)$ and $\alam(\mz)$
would be purely accidental from the RGE point of view.

The correlations between $\akap(\mz)$ and $\alam(\mz)$ required to
achieve $\mai<2\mb$ can be understood from
Eqs.~(\ref{eq:M11sq})--(\ref{eq:ma1_better}). For the following
discussion, we consider the
case of $\mu>0$, implying $s>0$. Clearly, in order for both
eigenvalues of the CP-odd Higgs mass matrix squared to be
positive, both diagonal elements have to be positive. In the limit
$|A_\kappa| \ll |A_\lambda| \sim v \ll s$, and $\tan \beta \gsim few$,
$M_{22}^2$ is dominated by the last term (unless $|A_\kappa|$ is
too small), and so $M_{22}^2
> 0$ leads to the condition $\kappa A_\kappa < 0$. This means that
the contribution of the $\kappa A_\kappa$ term to the mass of the
lightest CP-odd Higgs is positive, see Eqs.~(\ref{eq:ma1}) and
(\ref{eq:ma1_better}).  Next, we note that
$M_{11}^2 > 0$ requires $A_\lambda + \kappa s > 0$.
Given that $\kap\akap<0$, this implies that the denominator of
the first term in the $\mai^2$ expression 
given in Eq.~(\ref{eq:ma1_better}) is positive, and that this
first term will therefore have the
same sign as $\kap\alam$. Thus, in order for the terms proportional to
$\kap\alam$ and $-\kap\akap$ in Eq.~(\ref{eq:ma1_better})
to cancel so as to give small $\mai^2$, $\kap\alam$ must have
the same sign as $\kap\akap$.  Given that $\kap\akap<0$, we must therefore
also have $\kap\alam < 0$.
Altogether, we have two disconnected regions of allowed parameter
space. The first one is for $\kappa
>0$, $A_\lambda <0$ and $A_\kappa <0$. It is the largest region
because the condition $A_\lambda + \kappa s
> 0$ is easily satisfied, especially with smaller values of
$A_\lambda$. The second region is for $\kappa <0$, $A_\lambda
>0$ and $A_\kappa >0$ which is further constrained by $A_\lambda + \kappa s >
0$ and thus requires larger values of $A_\lambda$ and consequently
larger values of $A_\kappa$.  

The above discussion is not valid when $\akap$ is so small that the
$\kap\akap$ term does not dominate $M_{22}^2$. For such parameters,
$\mai<2\mb$ can still be achieved, but $\br(\hi\to\ai\ai)>0.7$ is not
possible. This region is seen in Fig.~\ref{fig:Akk_and_Alk} as a large
dark (blue) region with a large range of $\alam>0$ 
with $\kap>0$ in the right-hand
plot and a narrow band of very small $\akap$ values and 
$\kap>0$ in the left-hand plot. For this region, $\lam<0.1$ is
typical. This region can be understood by noting that when $\kap\akap$
is negligible, then  Eq.~(\ref{eq:ma1_better}) reduces to
\beq
\mai^2 \buildrel{\akap\to0}\over {\to}{9\alam\kap\lam v^2\sin 2\beta\over 
 2 (\alam+\kap s)}\,.
\eeq
Since we require both $\mai^2>0$ and $M_{11}^2\propto (\alam+\kap s)>0$, see Eq.~(\ref{eq:M11sq}), we must have
$\alam>0$ and $\kap>0$.
Small $\mai^2$ is easily achieved in a variety of ways, for example if
 $\alam$ is small compared to
$\kap s$ or, more generally, for small $\lam\alam$ and large $\tanb$.
Radiative corrections play a significant role also, reducing the
tree-level prediction given above by a substantial amount.

\begin{figure}
\includegraphics[width=2.4in,angle=90]{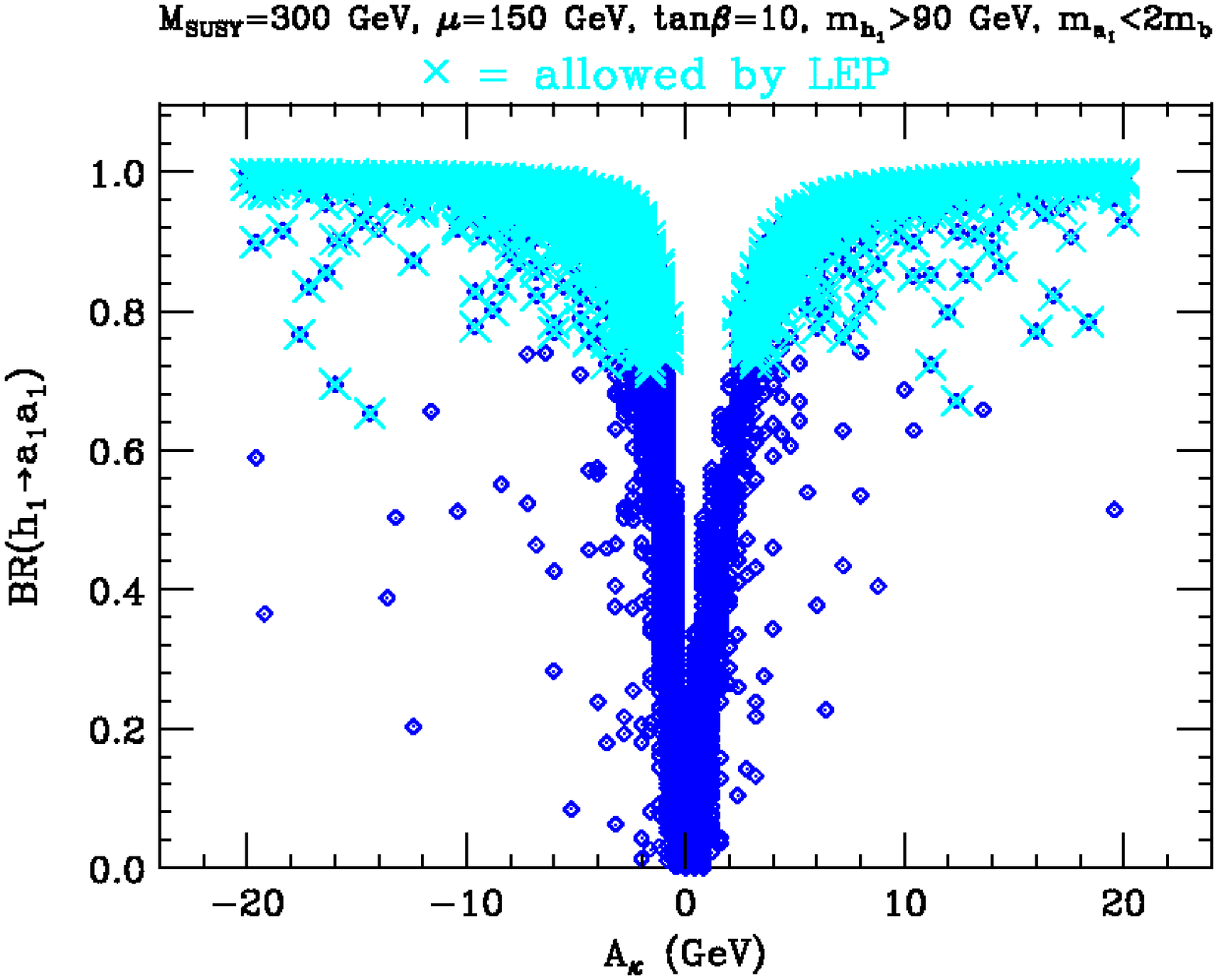}
\hspace{0.2cm}
\includegraphics[width=2.4in,angle=90]{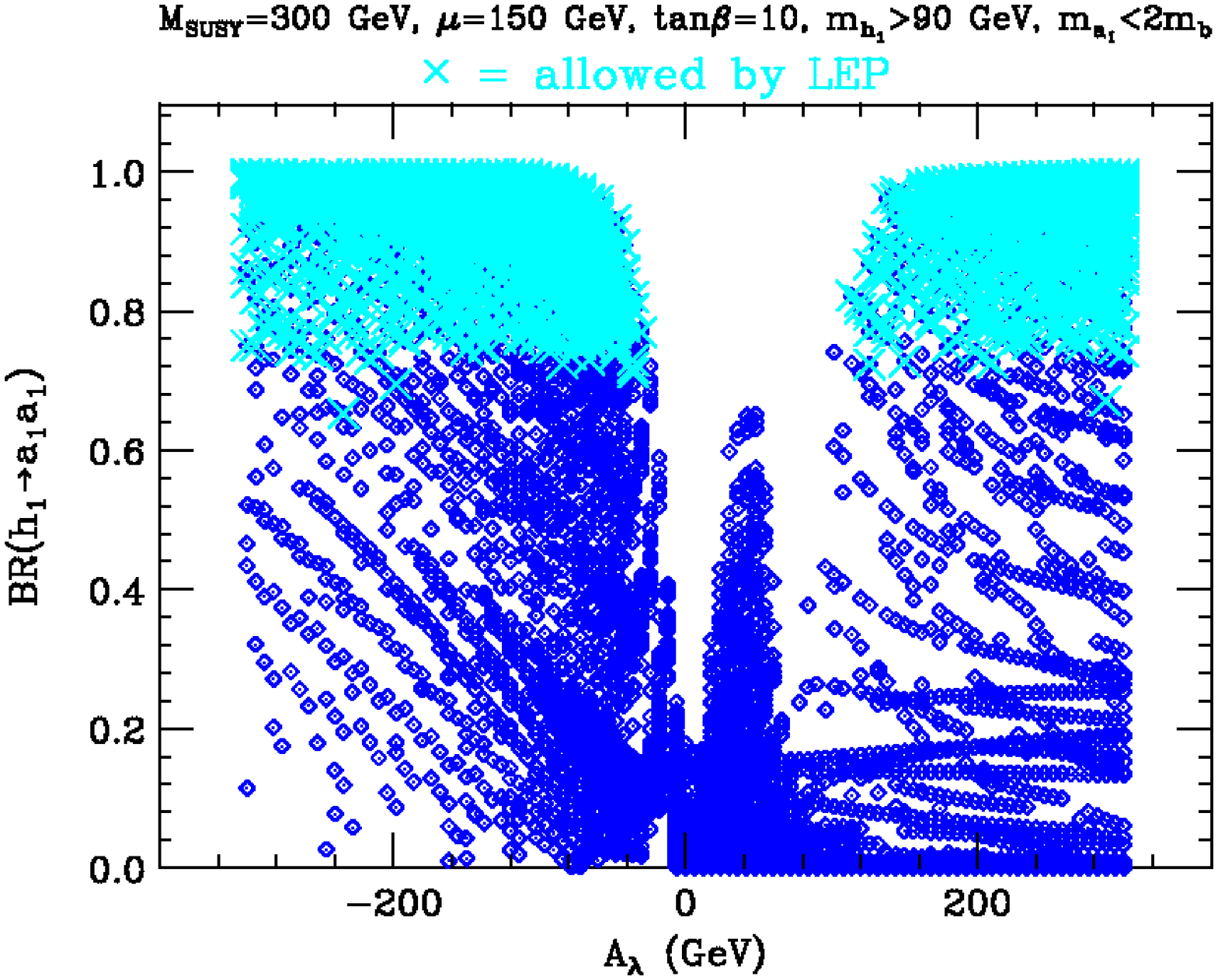}
\caption{Br($\hi\to\ai\ai$) vs. $A_\kappa$ and $A_\lambda$
for $\mu=150\gev$ and $\tanb=10$. Point conventions as in
Fig.~\ref{fig:AkAl_and_kl}.} 
\label{fig:Br_vs_Ak_and_Al}
\end{figure}

The dependence of the branching ratio for $\hi \to \ai\ai$ on $A_\kappa$ and
$A_\lambda$ is given in Fig.~\ref{fig:Br_vs_Ak_and_Al}. As we have stressed, 
a large value for $\br(\hi\to\ai\ai)$ is crucial for 
an $\mhi\sim 100\gev$ light SM-like Higgs to have escaped LEP constraints.
(It is the constraint $\mai<2\mb$ which guarantees
that when $\akap\sim 0$ then $\lam\alam$ must be very small as well,
and vice versa, thereby implying small $\br(\hi\to\ai\ai)$.)
It is interesting to see that if $|A_\kappa| \gsim 2$ GeV and
$|\alam|\gsim 30\gev$  (as typical for the sizes of
the RGE-induced contributions) then $\br(\hi\to\ai\ai)$ is almost always large
enough for the $\hi$ to have escaped detection
through the $\hi \to b \bar b$ channel. These plots also show clearly
that $\br(\hi\to\ai\ai)$ approaches a small (most typically, extremely
small) value when $\akap,\alam\to 0$. This suppression was discussed
analytically in the small $v/s$ limit in the previous section.

\begin{figure}
\includegraphics[width=2.4in,angle=90]{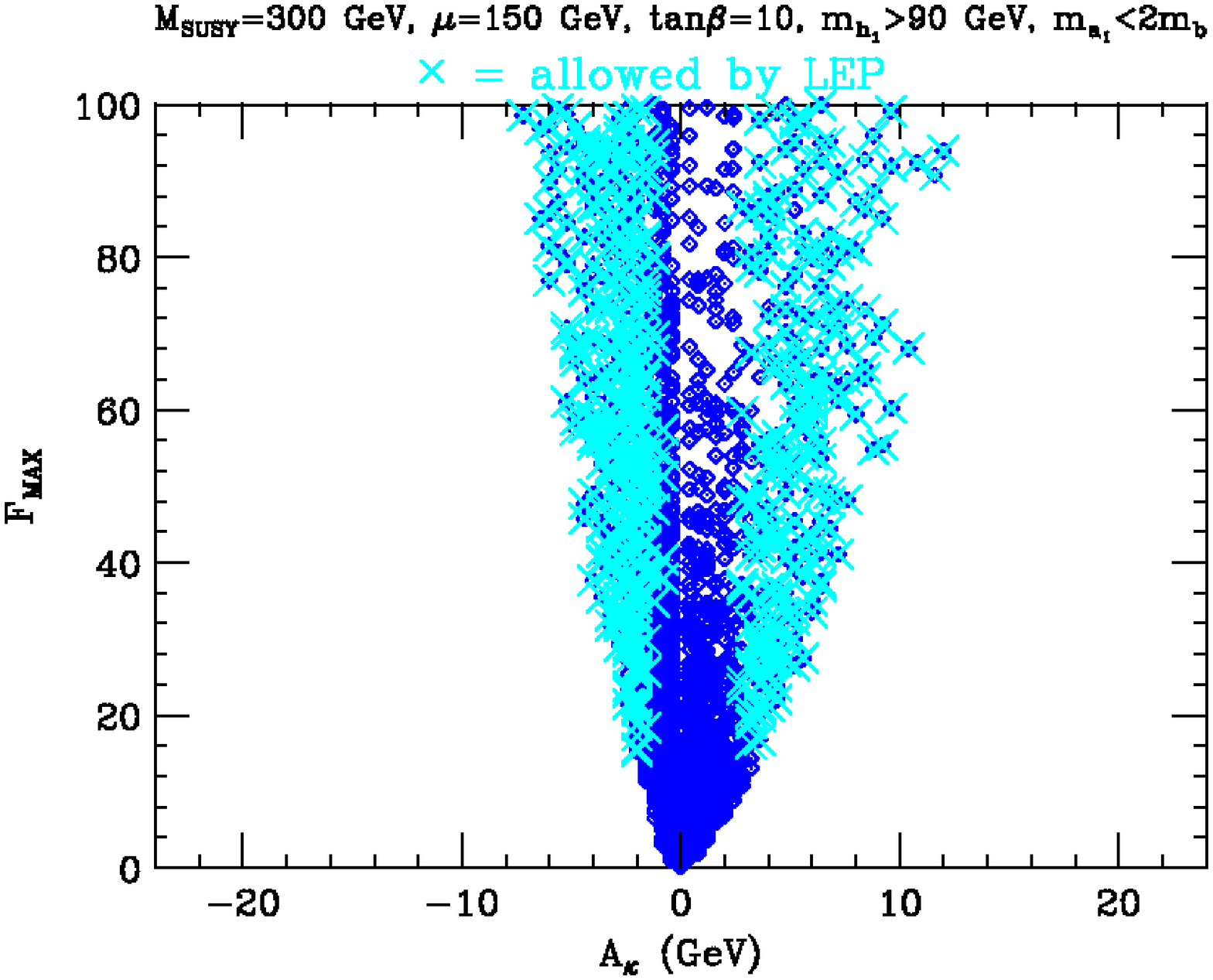}
\hspace{0.2cm}
\includegraphics[width=2.4in,angle=90]{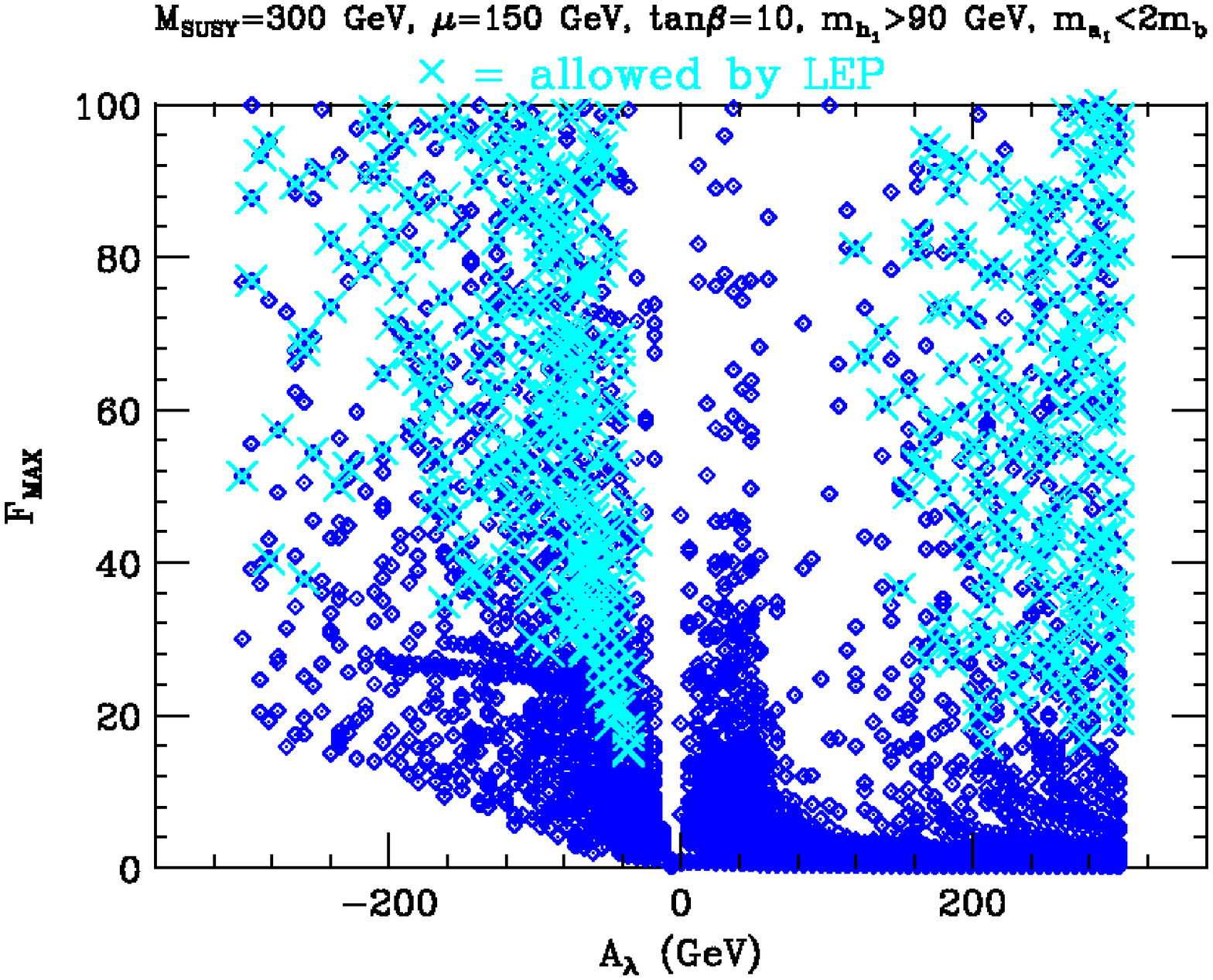}
\caption{Tuning in $m_{a_1}^2$ vs. $A_\kappa$ and $A_\lambda$.
Point conventions as in Fig.~\ref{fig:AkAl_and_kl}.}
\label{fig:fmax_vs_Ak_and_Al}
\end{figure}

\begin{figure}
\includegraphics[width=2.4in,angle=90]{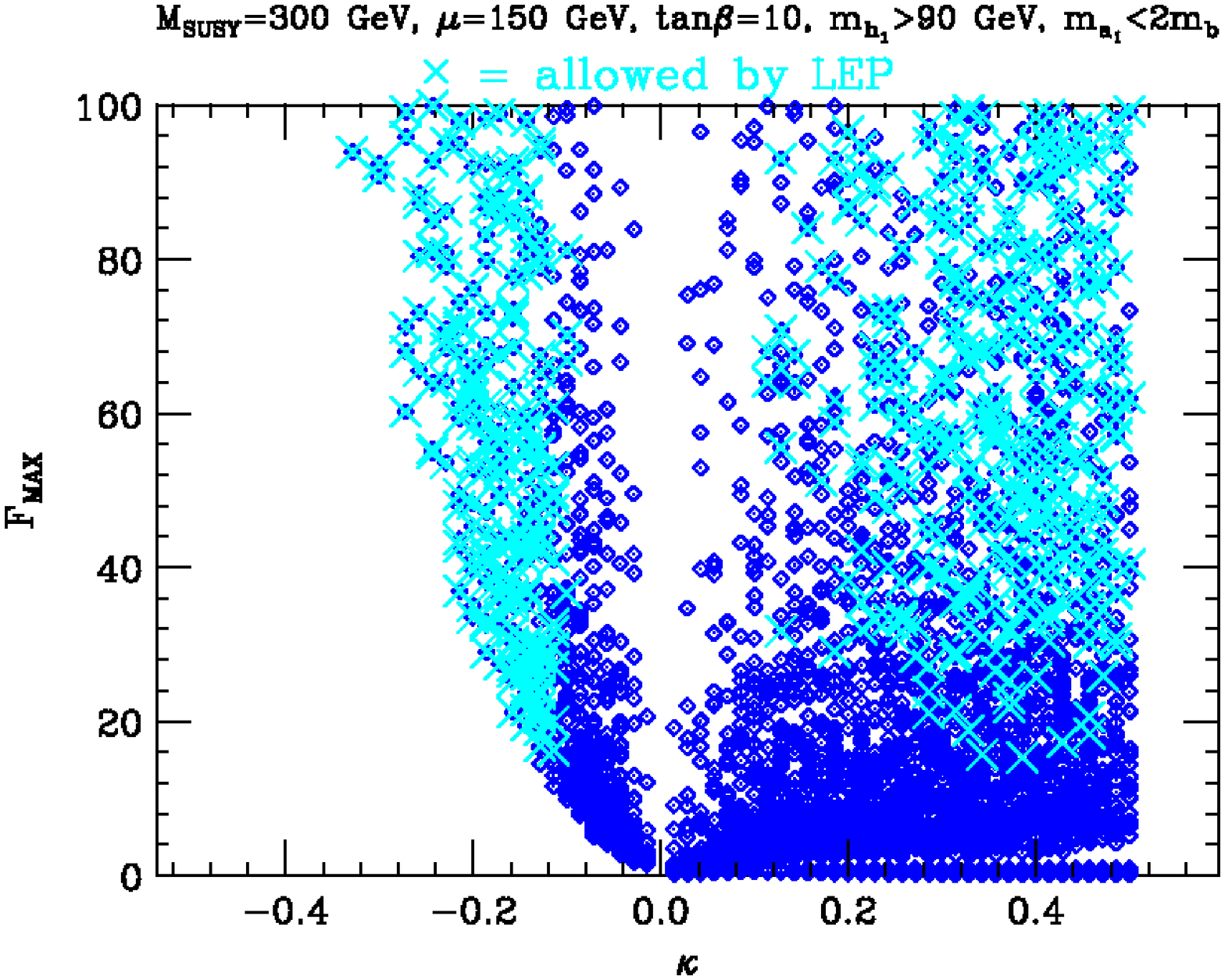}
\hspace{0.2cm}
\includegraphics[width=2.4in,angle=90]{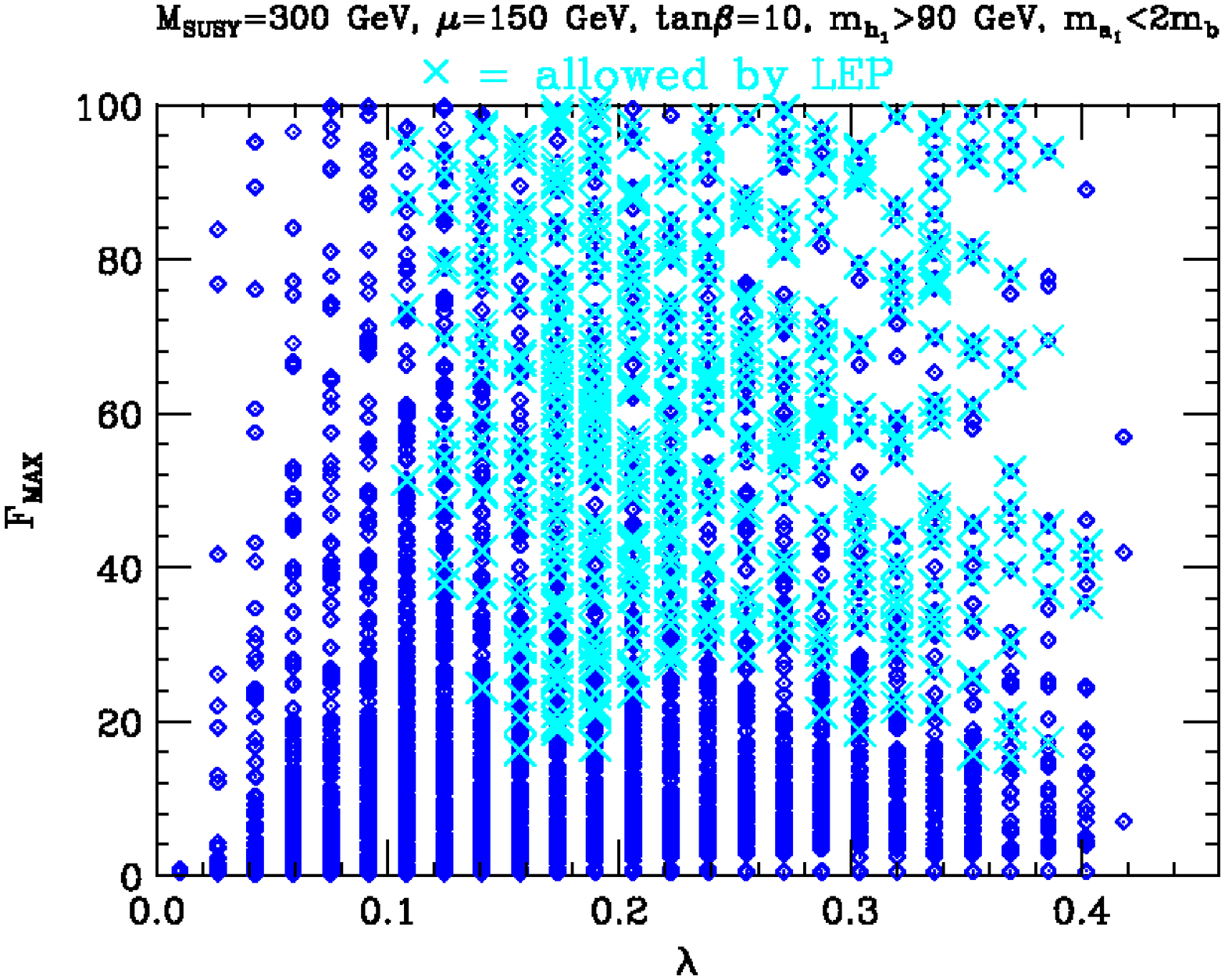}
\caption{Tuning in $m_{a_1}^2$ vs. $\kappa$ and $\lambda$.
Point conventions as in Fig.~\ref{fig:AkAl_and_kl}.}
\label{fig:fmax_vs_k_and_l}
\end{figure}

Another way of characterizing the limited allowed region in
$A_\kappa$ and $A_\lambda$ for which $\mai$ stays below $2 m_b$
(shown in Fig.~\ref{fig:AkAl_fixed_lk}), is to calculate the
tuning [as defined in Eq.~(\ref{eq:F_max})] of $A_\lambda$ and
$A_\kappa$ necessary to achieve this situation. The dependence of
$\fmax$ on $A_\kappa$ and $A_\lambda$ is given in
Fig.~\ref{fig:fmax_vs_Ak_and_Al}. Similarly, the dependence of
$\fmax$ on $\kappa$ and $\lambda$ is given in
Fig.~\ref{fig:fmax_vs_k_and_l}. As expected, the smallest fine
tuning or sensitivity is achieved for as small $A_\kappa$ and
$A_\lambda$ as possible. Of course, as we discussed earlier, very
small values of $A_\kappa$ and $A_\lambda$ would require
cancellations between the bare values and the RGE-induced
radiative corrections and this kind of cancellation would not be
visible from the definition of $\fmax$ given in
Eq.~(\ref{eq:F_max}). However, this is not a particularly
worrisome point given that, as discussed with regard to
Fig.~\ref{fig:Br_vs_Ak_and_Al}, very small values of $A_\kappa$
and $A_\lambda$ do not in any case lead to large enough
$\br(\hi\to\ai\ai)$ that $\br(\hi\to b\anti b)$ is adequately
suppressed. Thus, in the region of parameter space where soft
trilinear couplings are at least of order the typical RGE-induced
contributions, which is also the region where $\br(\hi\to\ai\ai)$
is large, the $\fmax$ measure of the
tuning of the soft trilinear couplings can be as small
as $\calo(5\% - 10\%)$.

\begin{figure}
\includegraphics[width=2.4in,angle=90]{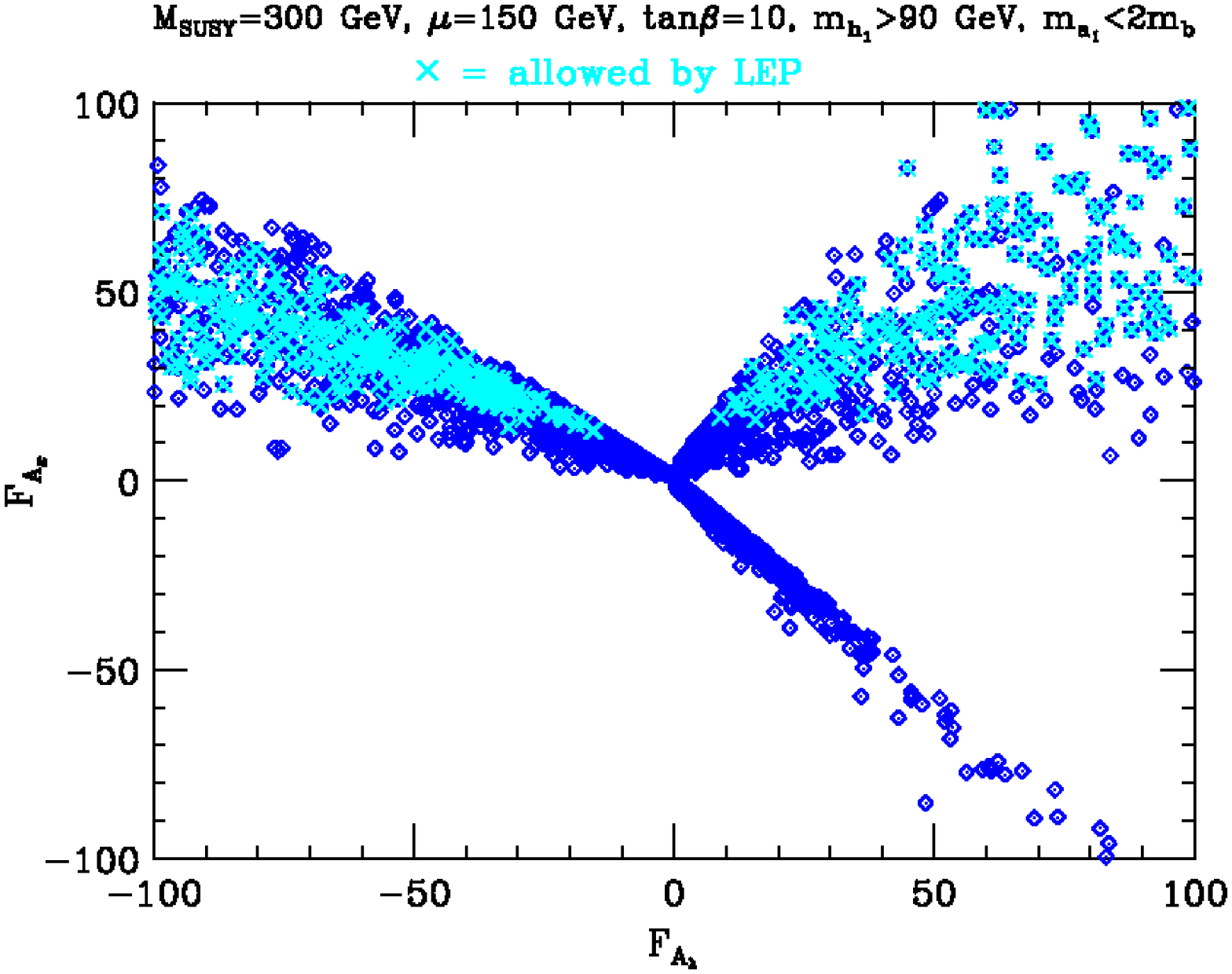}
\hspace{0.2cm}
\includegraphics[width=2.4in,angle=90]{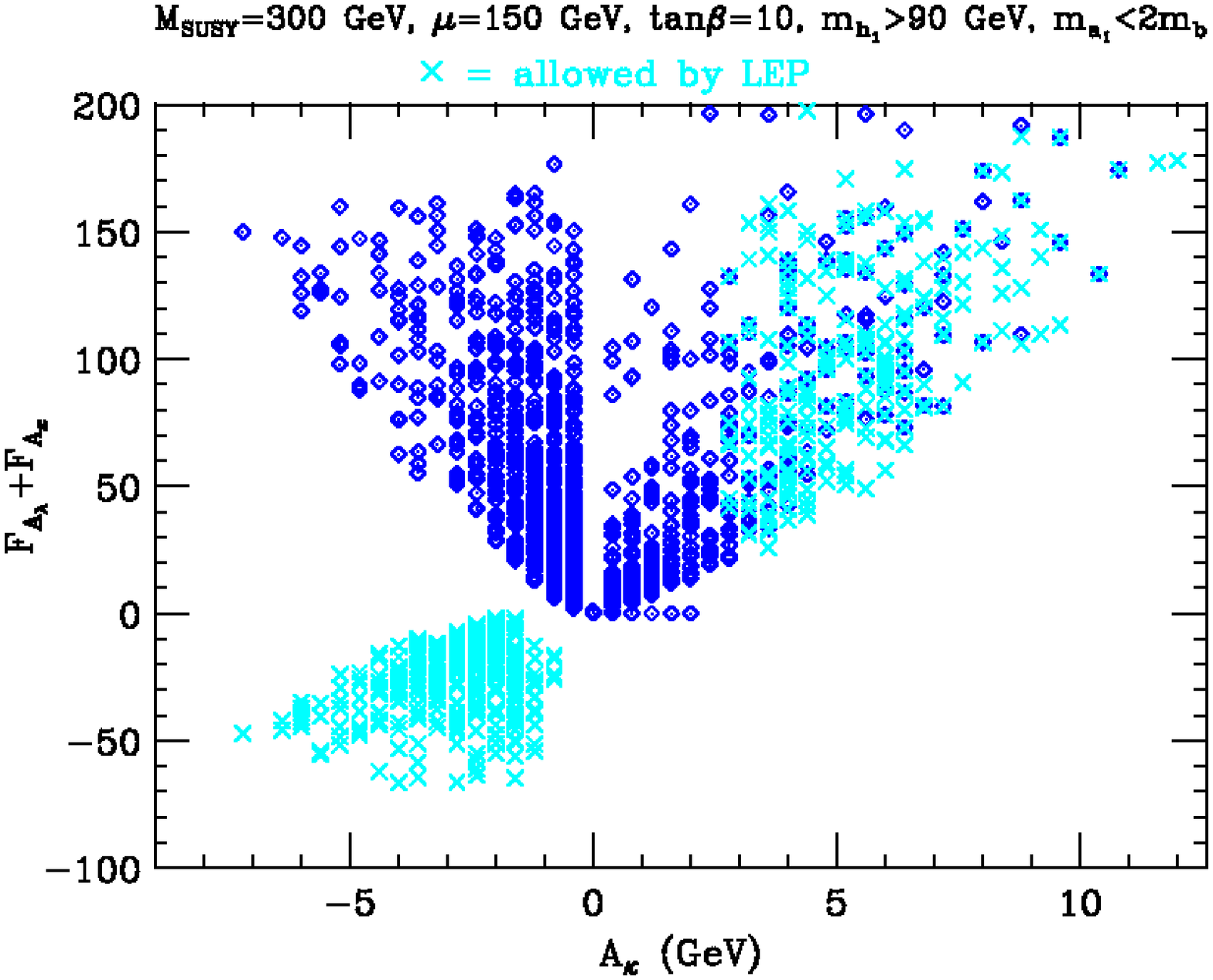}
\caption{In the left-hand frame, we plot $\fakap$ vs. $\falam$
  for the points with $\fmax<100$.  In the right-hand frame, we plot
  $\falam+\fakap$ vs. $\akap$ for these same points.}
\label{fakfal}
\end{figure}

\begin{figure}
\includegraphics[width=2.4in,angle=90]{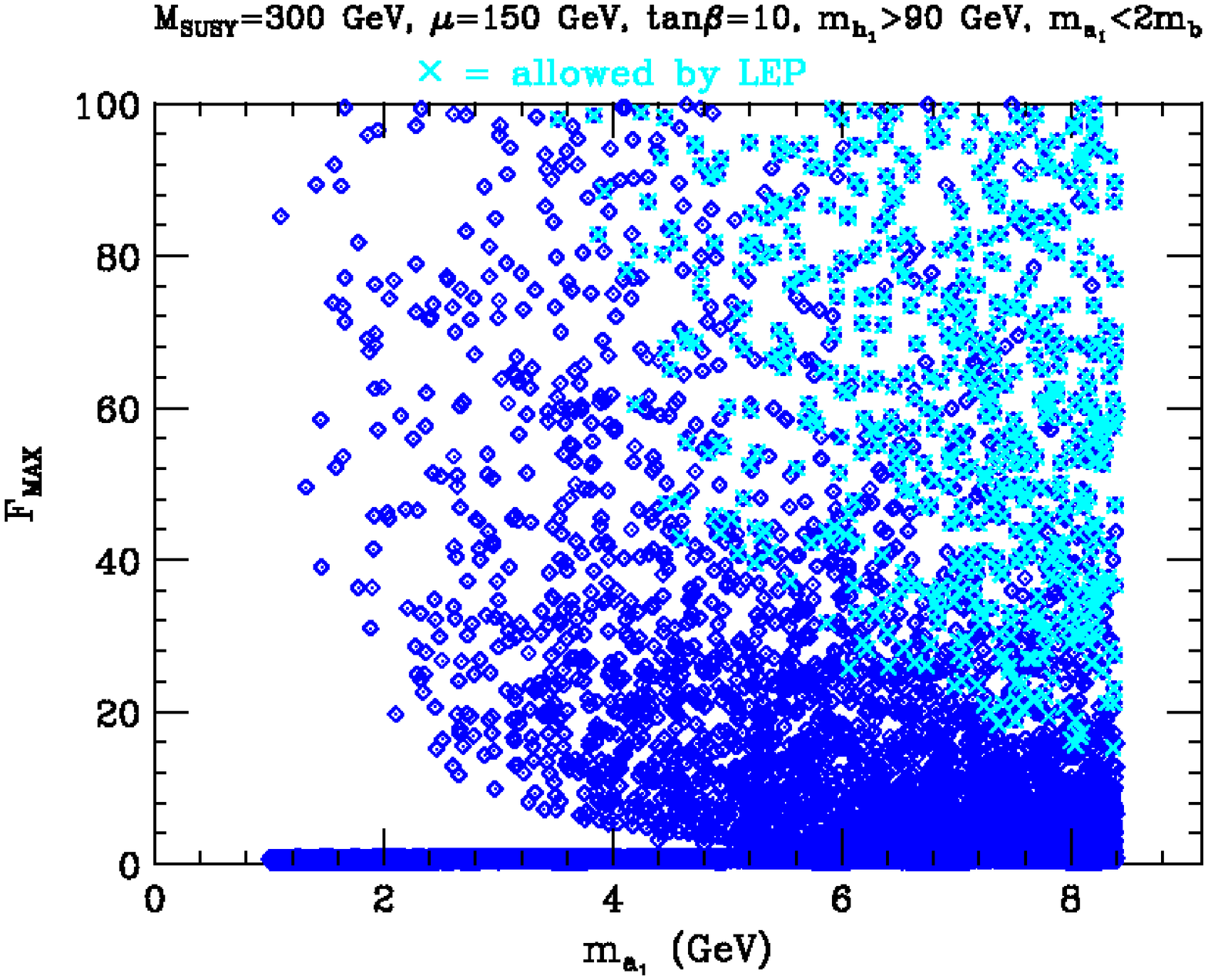}
\hspace{0.2cm}
\includegraphics[width=2.4in,angle=90]{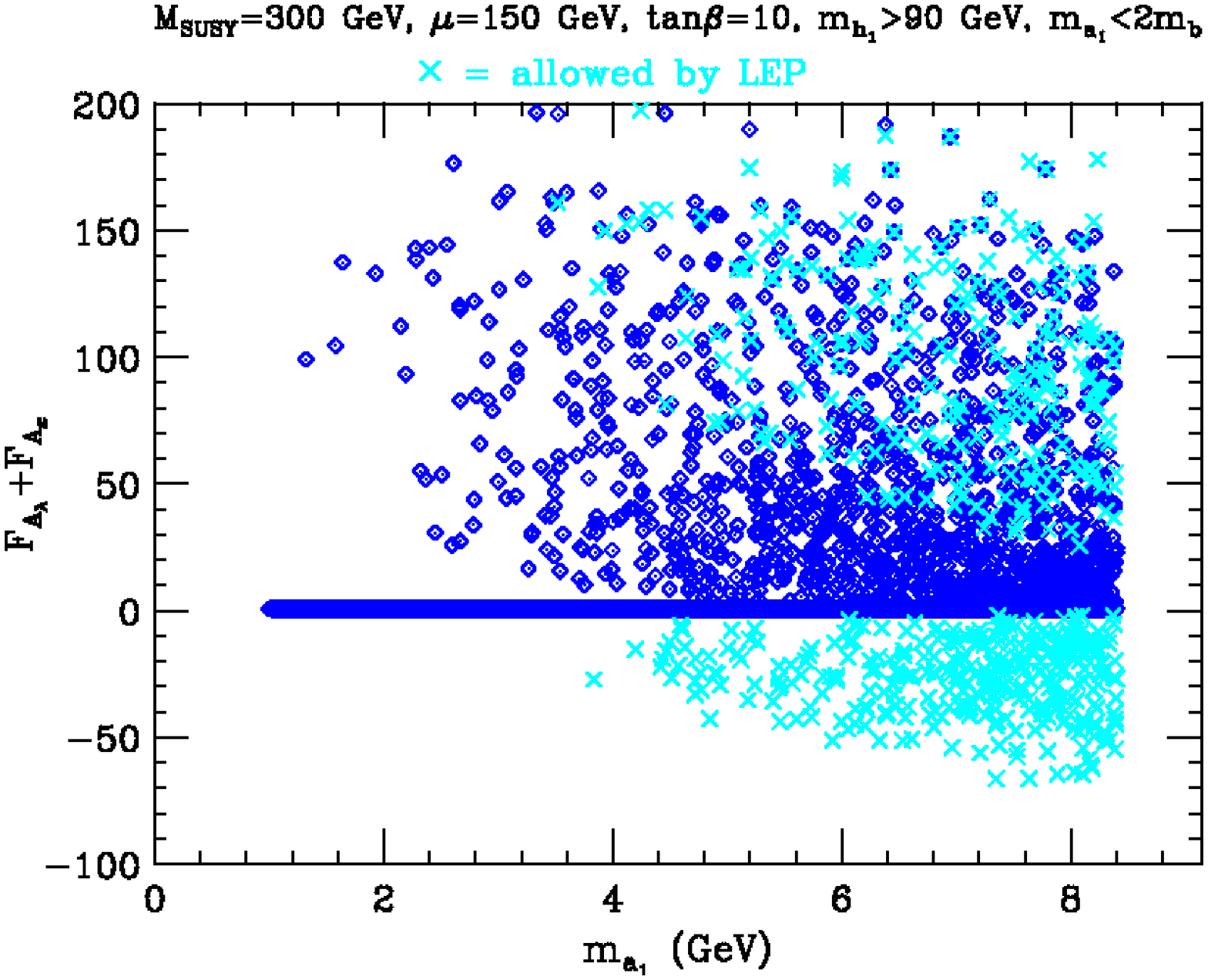}
\caption{In the left-hand frame, we plot $\fmax$ vs. $\mai$
  for the points with $\fmax<100$.  In the right-hand frame, we plot
  $\falam+\fakap$ vs. $\mai$ for these same points.}
\label{fakfalma}
\end{figure}

\begin{figure}
\includegraphics[width=2.4in,angle=90]{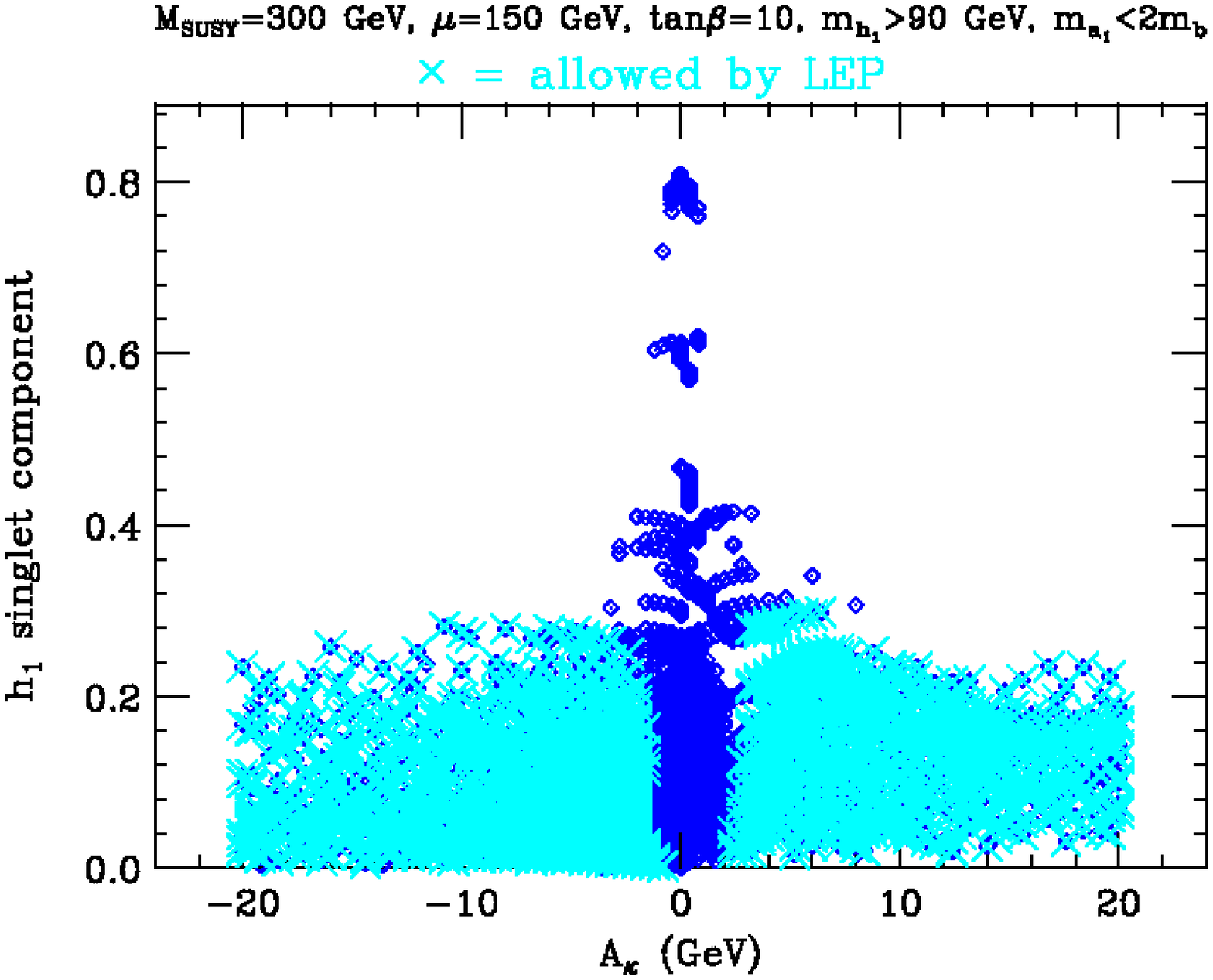}
\hspace{0.2cm}
\includegraphics[width=2.4in,angle=90]{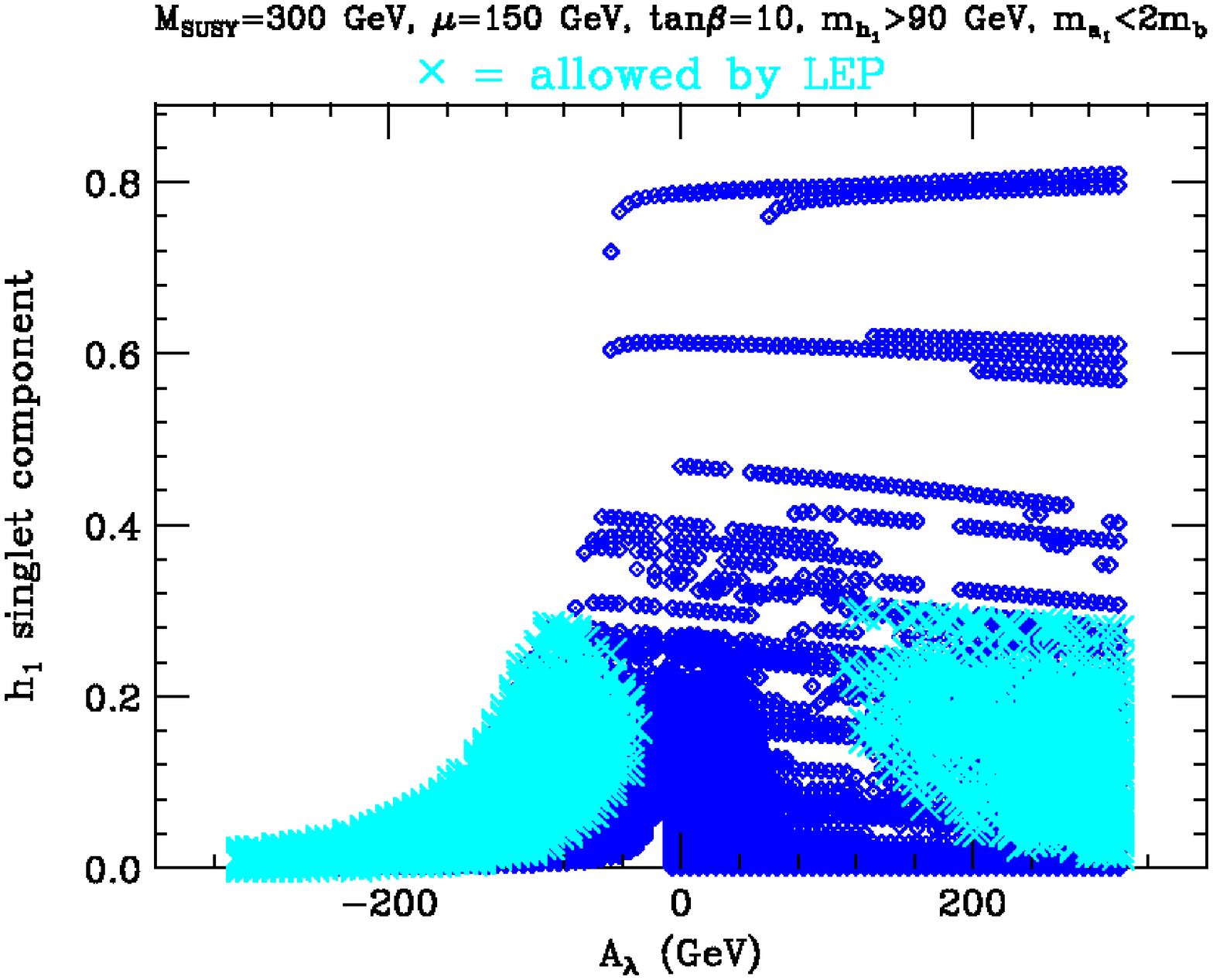}
\caption{Singlet component of $h_1$ vs. $A_\kappa$ and
$A_\lambda$. Point conventions as in Fig.~\ref{fig:AkAl_and_kl}.} \label{fig:h1singlet_vs_Ak_and_Al}
\end{figure}

However, as discussed in the analytic section of the paper,
fine-tuning with respect to GUT-scale parameters, denoted there as $F_{\mai}$, 
need not be as large
as $\fmax$.  In particular, there are many model scenarios in which
$F_{\mai}$ is reduced compared to $\fmax$ by cancellations between
the dependence of $\alam(\mz)$ on the dominant GUT-scale parameter $p$
and the dependence of $\akap(\mz)$ on this same $p$, in the simplest
cases yielding $F_{\mai}\sim \falam+\fakap$.  In the most
naive approximation, $\mai^2$ is linear in $\alam$ and linear in
$\akap$ and therefore $\falam\sim -\fakap$ and $F_{\mai}$ is
then quite small. Thus, it is
important to understand the correlations between $\falam$ and
$\fakap$ in order to assess the extent to which  $\falam+\fakap$
can be small.  In the left-hand
window of Fig.~\ref{fakfal}, we plot $\fakap$ vs. $\falam$ 
for points with $\fmax<100$. We see
that many of the points have $\falam\sim
-\fakap$. The corresponding values of $\falam+\fakap$ appear
in the right-hand plot. We observe that the LEP-allowed points with
small negative $\akap$
(which are those with the smallest magnitudes of $\falam$ and
  $\fakap$ separately)  
yield still smaller $\falam+\fakap$ and therefore
possibly very small $F_{\mai}$ for appropriate GUT-scale models.
For all LEP-allowed $\akap<0$ points, $\falam+\fakap$ is much smaller than 
$\fmax$.

It is also useful to understand the extent to which $\fmax$ and 
$\falam+\fakap$ depend on $\mai$.  This dependence is revealed in
Fig.~\ref{fakfalma}. We see that the smallest $\fmax$ values are
achieved for the larger $\mai$ values up near $2\mb$. 
Small values of $\falam+\fakap$ are distributed over the broader range
of $4\gev\lsim \mai<2\mb$. In either case, the fine-tuning 
associated with $\mai<2\mb$ suggests some preference for
$\mai>2\mtau$.  This is important in that $\hi\to \ai\ai\to4\tau$
decays may prove to be visible at the LHC, whereas $\hi\to\ai\ai$
with $\ai\to c\anti c,gg,\ldots$ will be much harder to detect. 

\begin{figure}
\includegraphics[width=2.4in,angle=90]{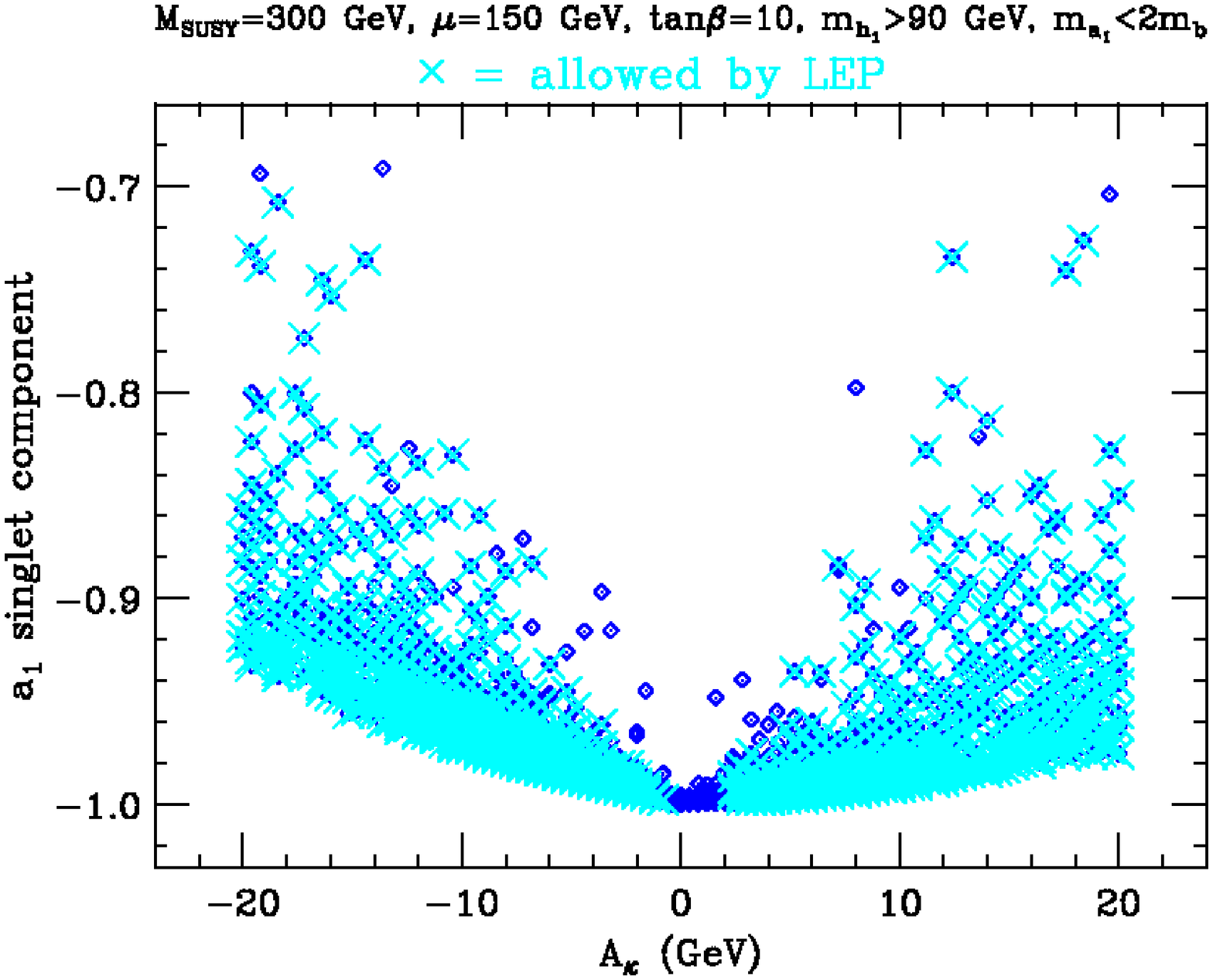}
\hspace{0.2cm}
\includegraphics[width=2.4in,angle=90]{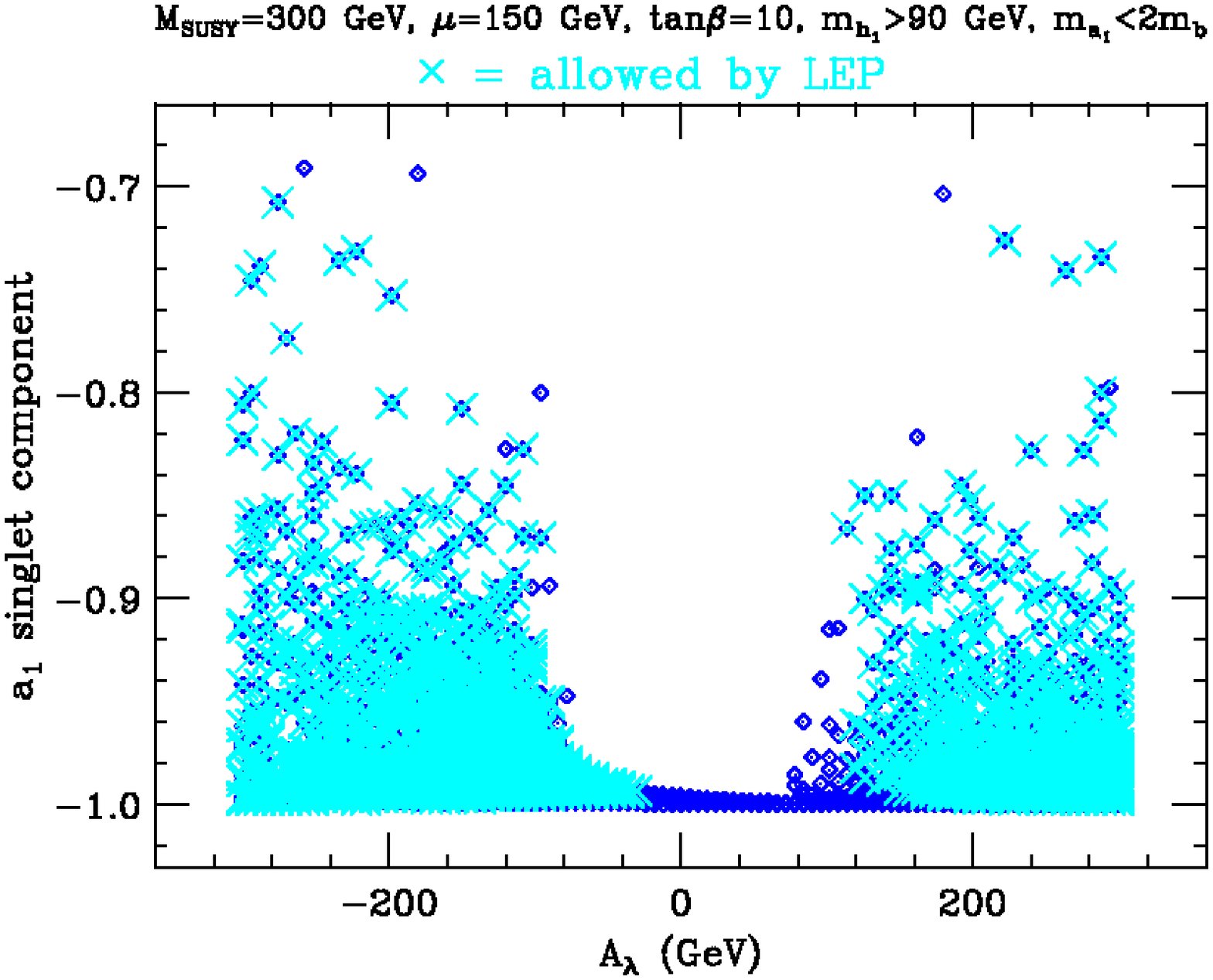}
\caption{Singlet component of $a_1$ vs. $A_\kappa$ and
$A_\lambda$. Point conventions as in Fig.~\ref{fig:AkAl_and_kl}.} \label{fig:a1singlet_vs_Ak_and_Al}
\end{figure}

\begin{figure}
\includegraphics[width=2.4in,angle=90]{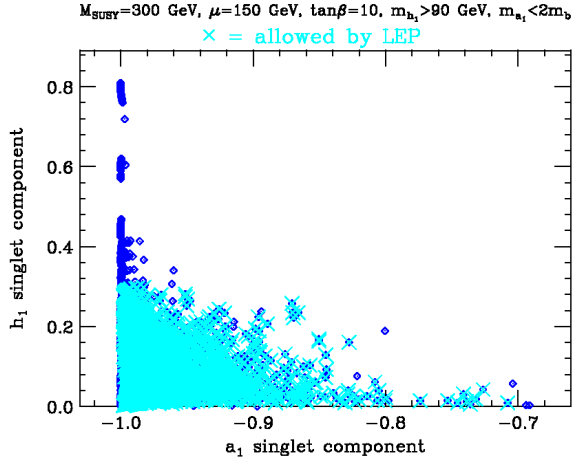}
\caption{Singlet component of $h_1$ vs. singlet component of
$a_1$. Point conventions as in Fig.~\ref{fig:AkAl_and_kl}.} \label{fig:h1singlet_vs_a1singlet}
\end{figure}

\begin{figure}
\includegraphics[width=2.4in,angle=90]{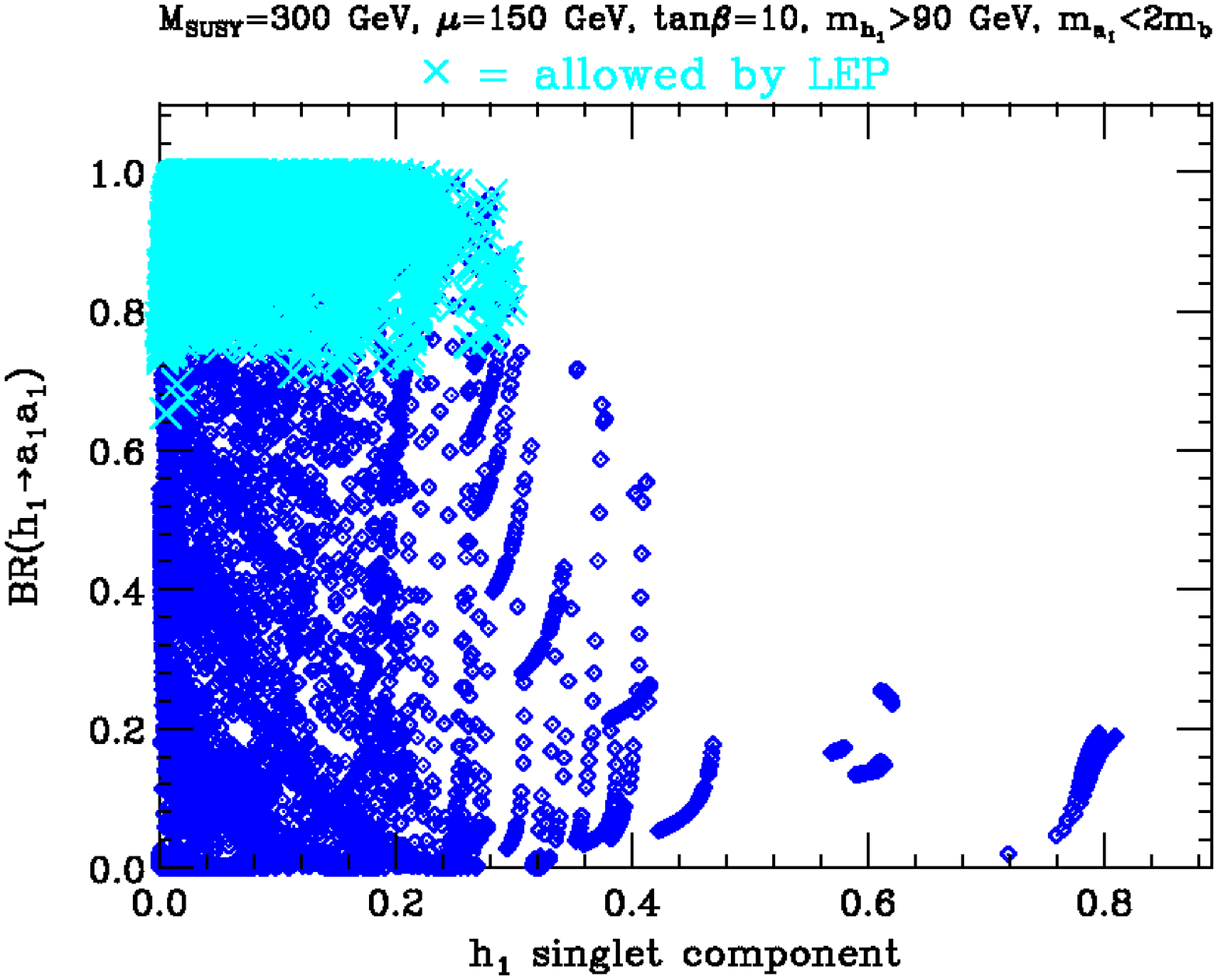}
\hspace{0.2cm}
\includegraphics[width=2.4in,angle=90]{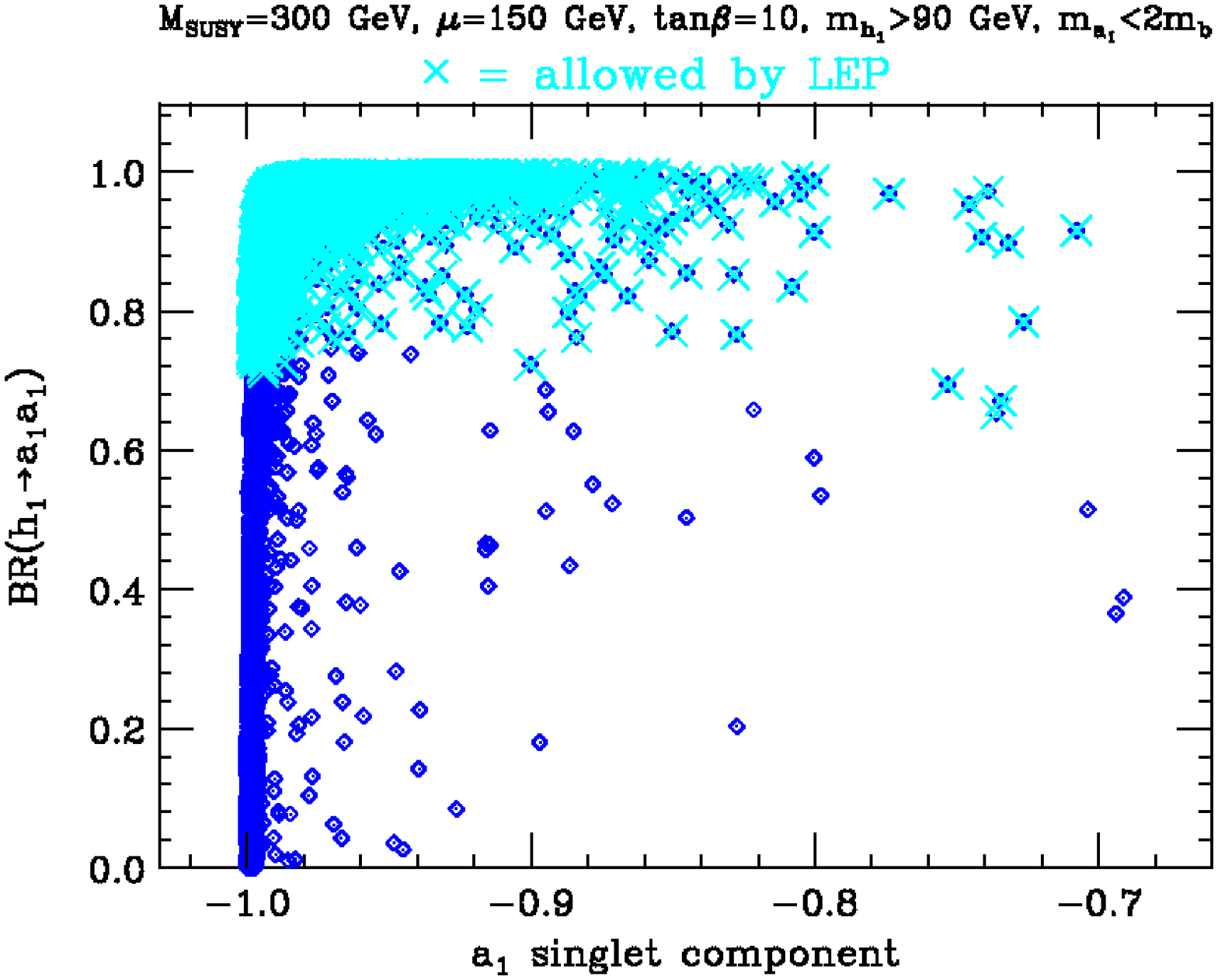}
\caption{$\br(\hi\to\ai\ai)$ vs. singlet component of $h_1$ and $a_1$.
Point conventions as in Fig.~\ref{fig:AkAl_and_kl}.} \label{fig:Br_vs_h1singlet_and_a1singlet}
\end{figure}

Higgs phenomenology is crucially dependent upon the singlet and
doublet compositions of the lightest CP-even and CP-odd Higgs mass
eigenstates.  The lightest CP-even Higgs boson is preferably
SM-like although it can have up to 20 \% singlet component, see
Fig.~\ref{fig:h1singlet_vs_Ak_and_Al}, while the lightest CP-odd
Higgs boson has to be very close to being a singlet, at least 98
\% singlet in the region where the least tuning is necessary, see
Fig.~\ref{fig:a1singlet_vs_Ak_and_Al}. 
(Here, $\sin\theta_A$, see Eq.~(\ref{eq:a1_composition}), is the singlet component
of the $\ai$ at the amplitude level. The {\it probability} for the $\ai$ to
be singlet is $\sin^2\theta_A$.  Similar definitions are used in the
case of the $\hi$.) This correlation between the
$\hi$ and $\ai$ compositions is explicit in the
plot of the $h_1$ singlet component versus the $a_1$ singlet
component given in Fig.~\ref{fig:h1singlet_vs_a1singlet}. 
This figure shows that as the $\ai$ becomes less singlet the $\hi$
must have less singlet component. The left-hand plot of
Fig.~\ref{fig:Br_vs_h1singlet_and_a1singlet} also makes it clear that
the $\hi$ singlet component must be small (or zero) in order to maximize
$\br(\hi\to\ai\ai)$. The right-hand plot of
Fig.~\ref{fig:Br_vs_h1singlet_and_a1singlet} shows that the $\ai$ singlet
component must be at least $70\%$ in amplitude, \ie\ 50\%
in probability, for the range of parameters scanned with most points
corresponding to an $\ai$ that is mainly singlet.

\begin{figure}
\includegraphics[width=2.4in,angle=90]{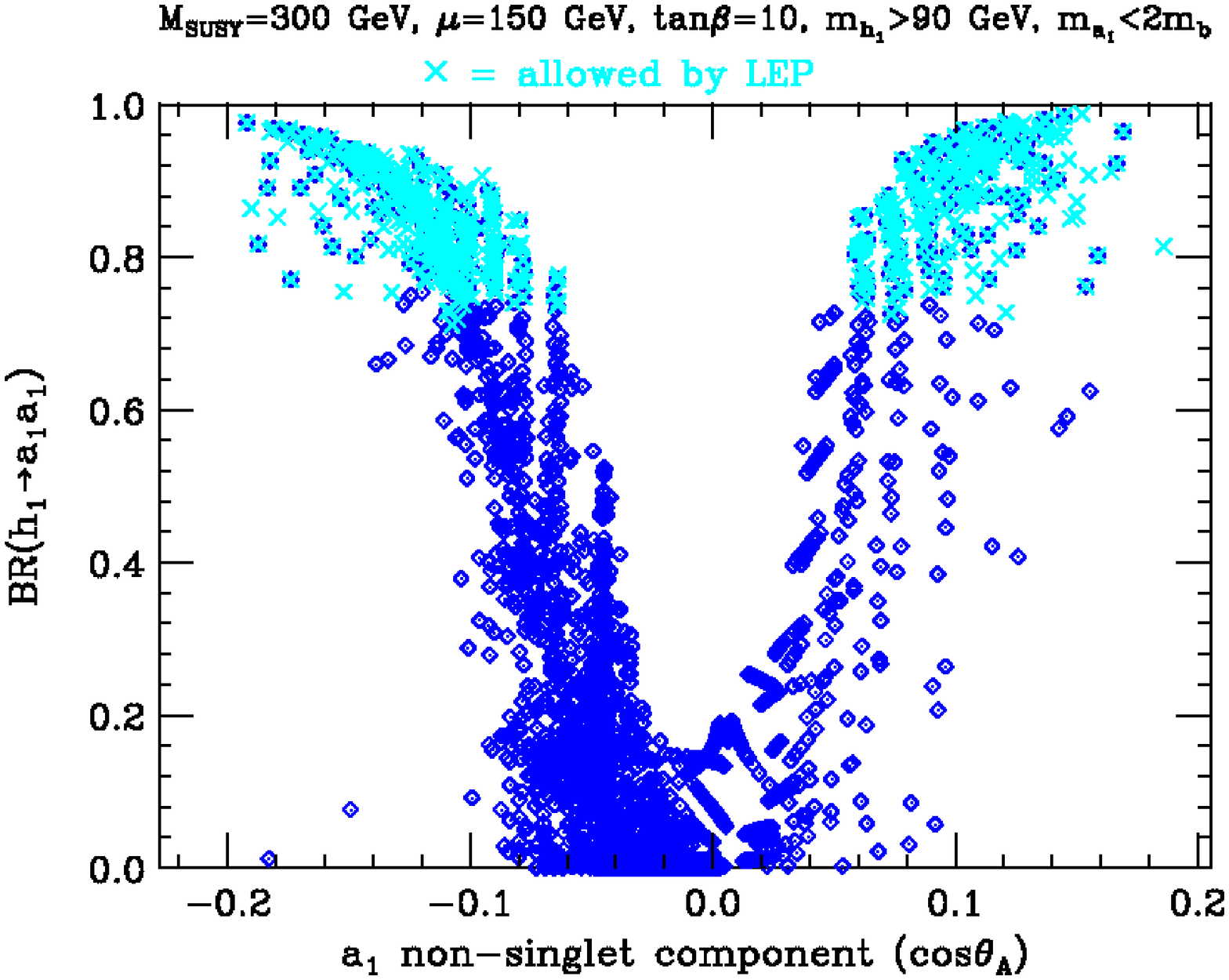}
\hspace{0.2cm}
\includegraphics[width=2.4in,angle=90]{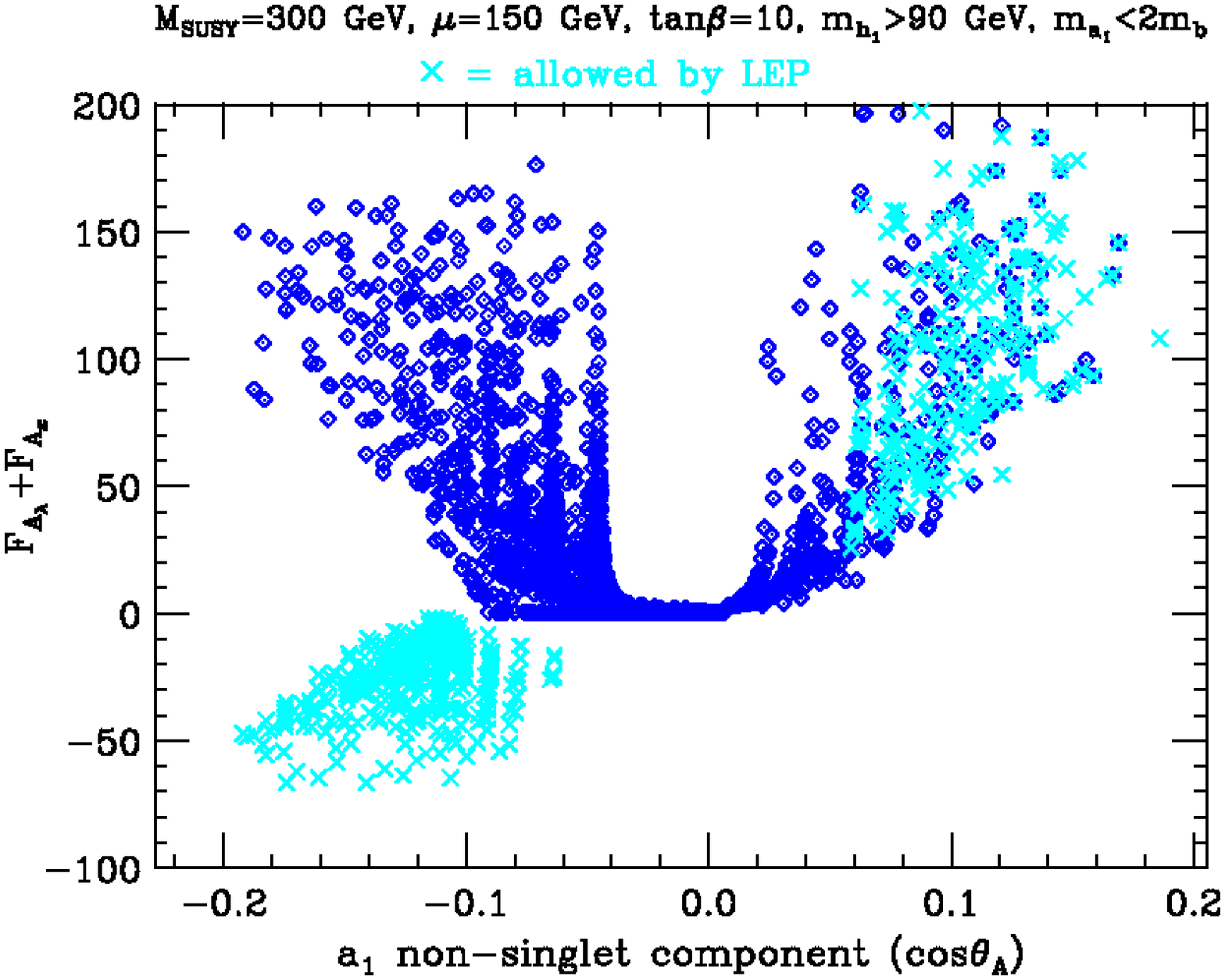}
\caption{$\br(\hi\to\ai\ai)$ and $\fakap+\falam$ vs. the non-singlet
  component of the $a_1$.
Point conventions as in Fig.~\ref{fig:AkAl_and_kl}.} \label{ctaplots}
\end{figure}

A particularly revealing pair of plots are those of
Fig.~\ref{ctaplots} where we show how $\br(\hi\to\ai\ai)$ and
$\fakap+\falam$ depend on $\cos\theta_A$, the non-singlet component of
the $\ai$. We see from the first plot that there is a lower bound on
$|\cos\theta_A|$ of order $0.06$ (this is the bound for $\tanb=10$ ---
the precise number depends on $\tanb$) in order for $\br(\hi\to\ai\ai)$ to
be large enough that LEP limits are evaded.  The existence of the
lower bound follows from the fact, discussed earlier, that $\kappa$
and $A_\lambda$ must have opposite signs in order for the scenario to
be viable. Since, the non-singlet component of the $\ai$ is
proportional to $A_\lambda - 2 \kappa s$, see Eq.~(\ref{ctaform}), it
cannot go to zero for the given correlation of signs.  From the second
plot, we see that the $\fakap+\falam$ measure of fine tuning for
$\mai$ has a very distinct minimum value close to 0 for
$\cos\theta_A\sim-0.1$. Combined with the dependence of
$\fakap+\falam$ on $\mai$ displayed in Fig.~\ref{fakfalma}, we see a
possible preference for $\mai>2\mtau$ and $\cos\theta_A\sim -0.1$ in
order to be reasonably certain that fine-tuning is not required in
order to achieve large $\br(\hi\to\ai\ai)$ and $\mai<2\mb$.  We note
that the coupling of the $\ai$ to down type quarks is proportional to
$\tanb\cos\theta_A$ which is never smaller than about $0.6$ in
magnitude for the points that escape LEP limits. Thus, even though the
$\ai$ is largely singlet, it has substantial down-quark and lepton
couplings.  In a follow-up paper~\cite{Dermisek:2006py}, we show that
this implies an $\mai$-dependent lower limit on the branching ratio
for $\Upsilon\to \gam\ai$ decays.  This lower limit can probably be
probed at future, if not existing, $B$ factories if $\mai$ is not too
close to $M_\Upsilon$, especially if $\mai>2\mtau$ so that
$\ai\to\tauptaum$ decays are dominant.

\begin{figure}
\includegraphics[width=2.4in,angle=90]{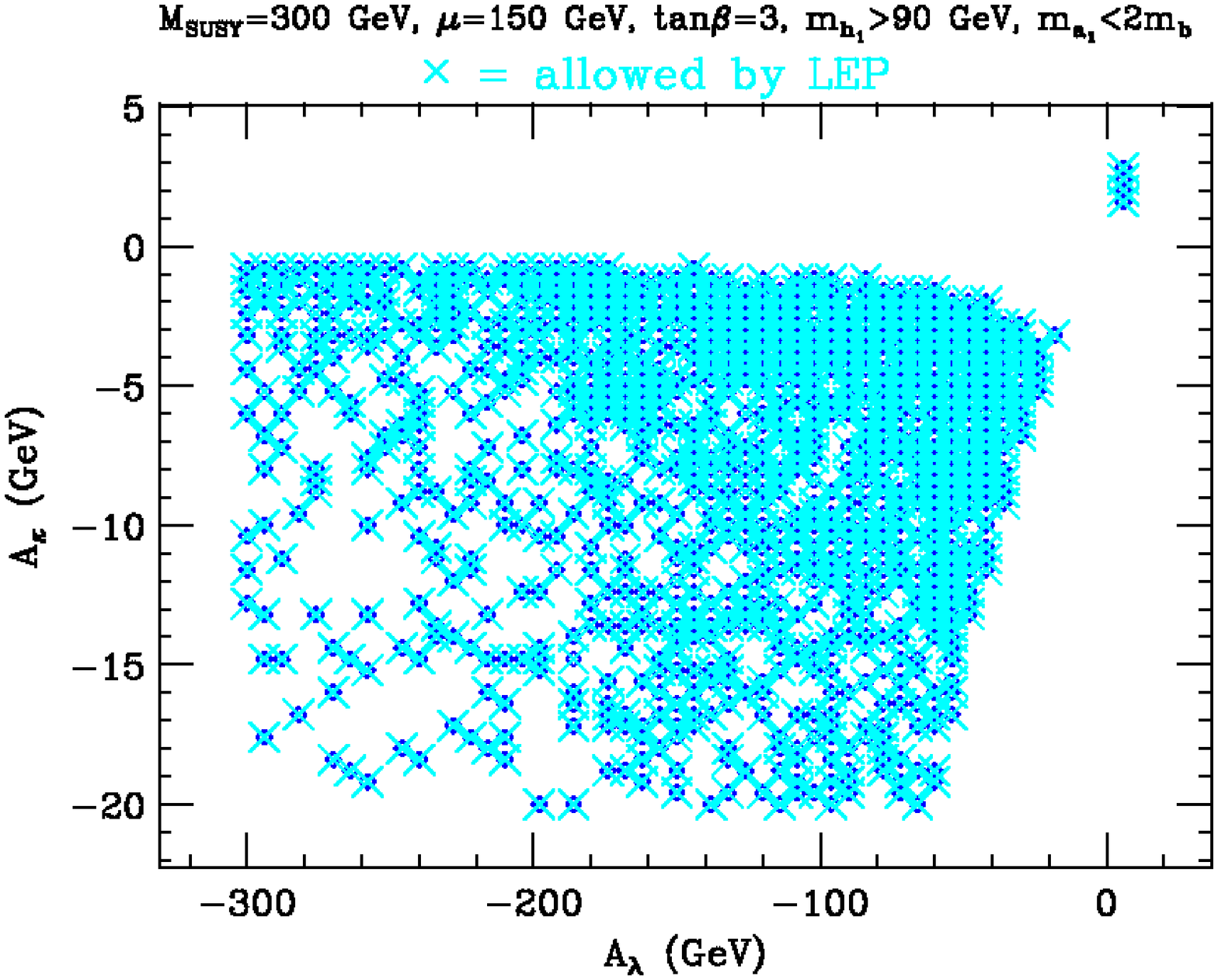}
\hspace{0.2cm}
\includegraphics[width=2.4in,angle=90]{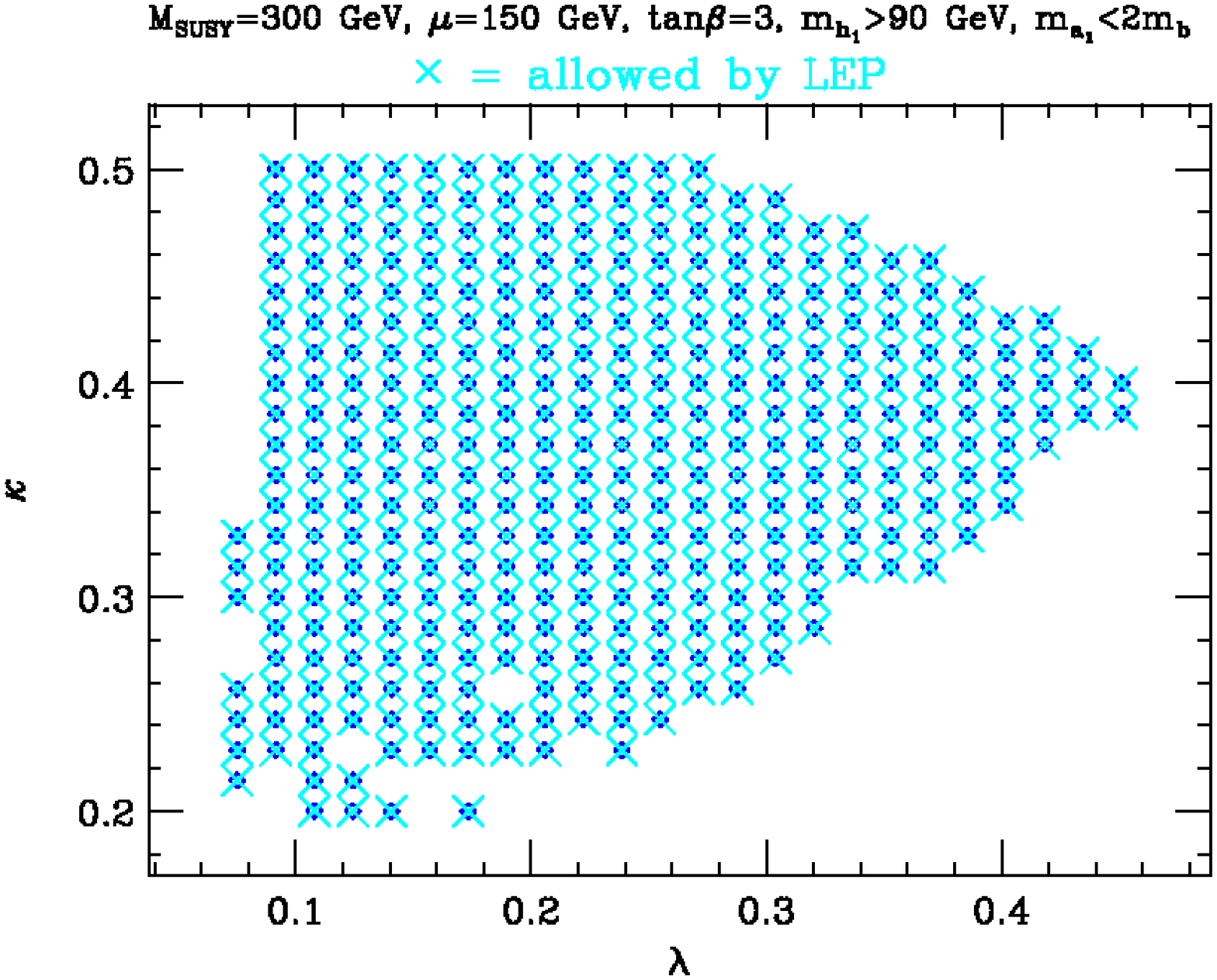}
\caption{Allowed parameter space in the $A_\kappa - A_\lambda$ and $\kappa
  - \lambda$ planes for $\tanb=3$ and $\mu=150\gev$.
Point conventions as in Fig.~\ref{fig:AkAl_and_kl}.}
\label{fig:AkAl_and_kl_tb3}
\end{figure}

\begin{figure}
\includegraphics[width=2.4in,angle=90]{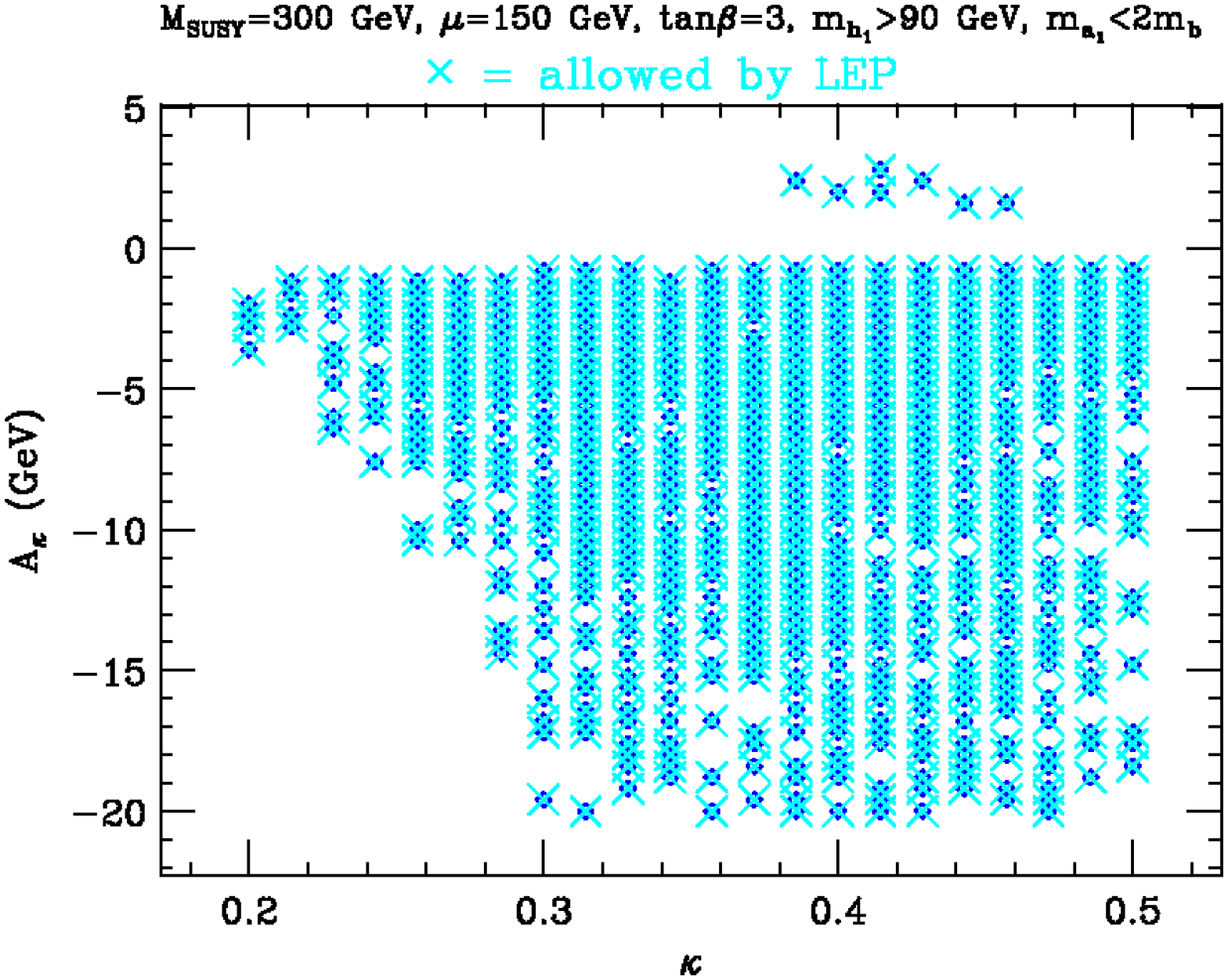}
\hspace{0.2cm}
\includegraphics[width=2.4in,angle=90]{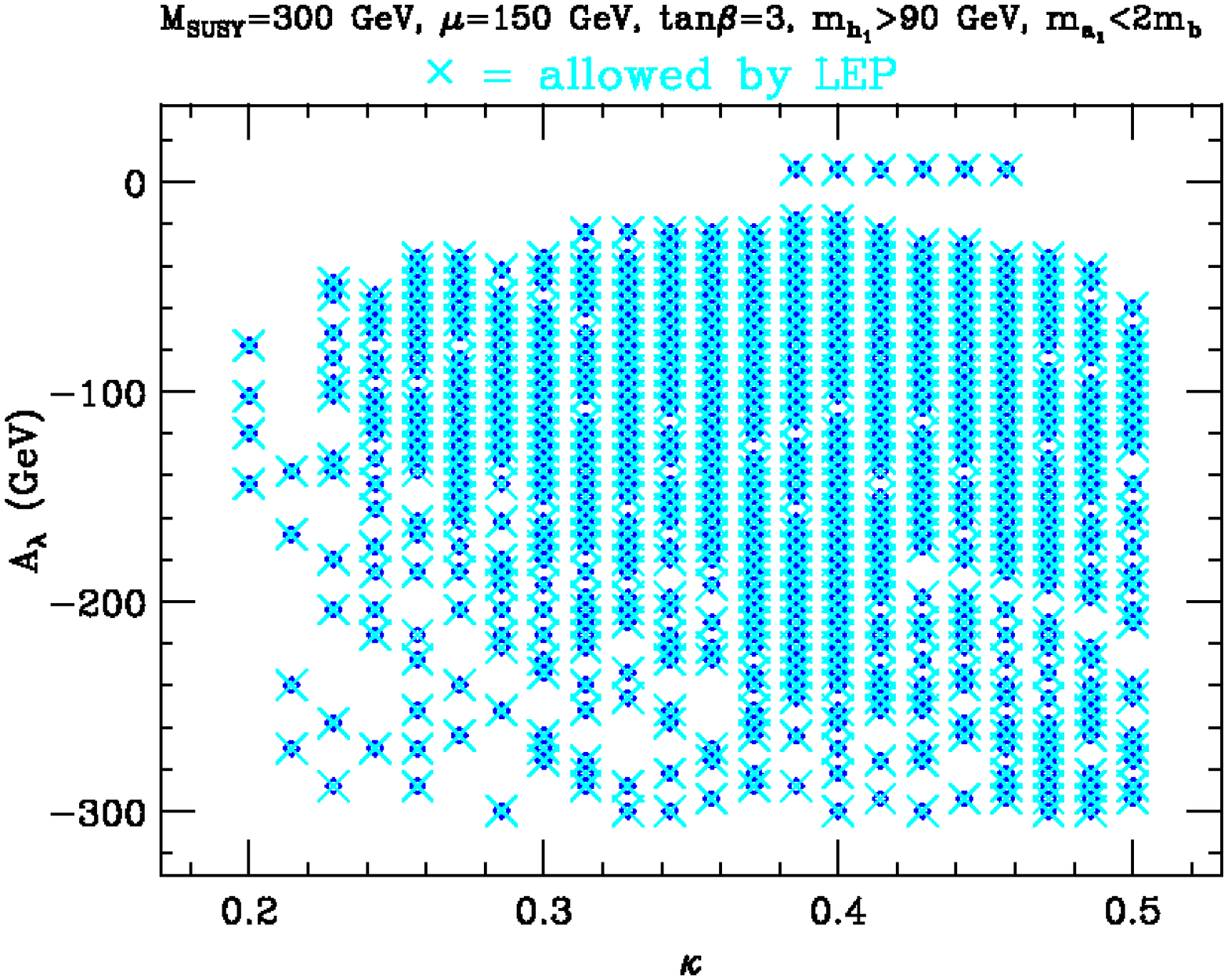}
\caption{Allowed parameter space in the $A_\kappa - \kappa$ and $A_\lambda
  - \kappa$ planes for $\tanb=3$ and $\mu=150\gev$.
Point conventions as in Fig.~\ref{fig:AkAl_and_kl}.}
\label{fig:Akk_and_Alk_tb3}
\end{figure}

\begin{figure}
\includegraphics[width=2.4in,angle=90]{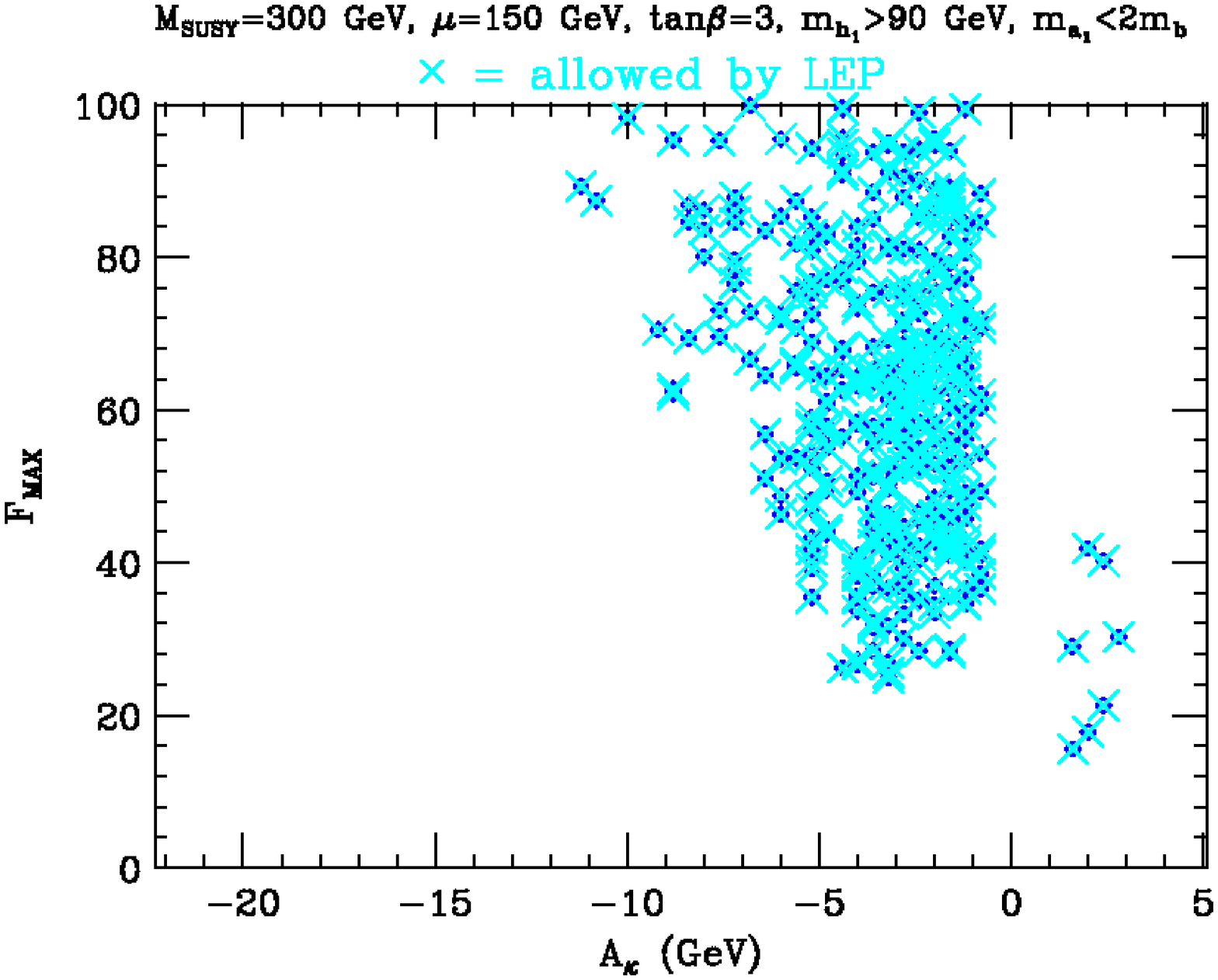}
\hspace{0.2cm}
\includegraphics[width=2.4in,angle=90]{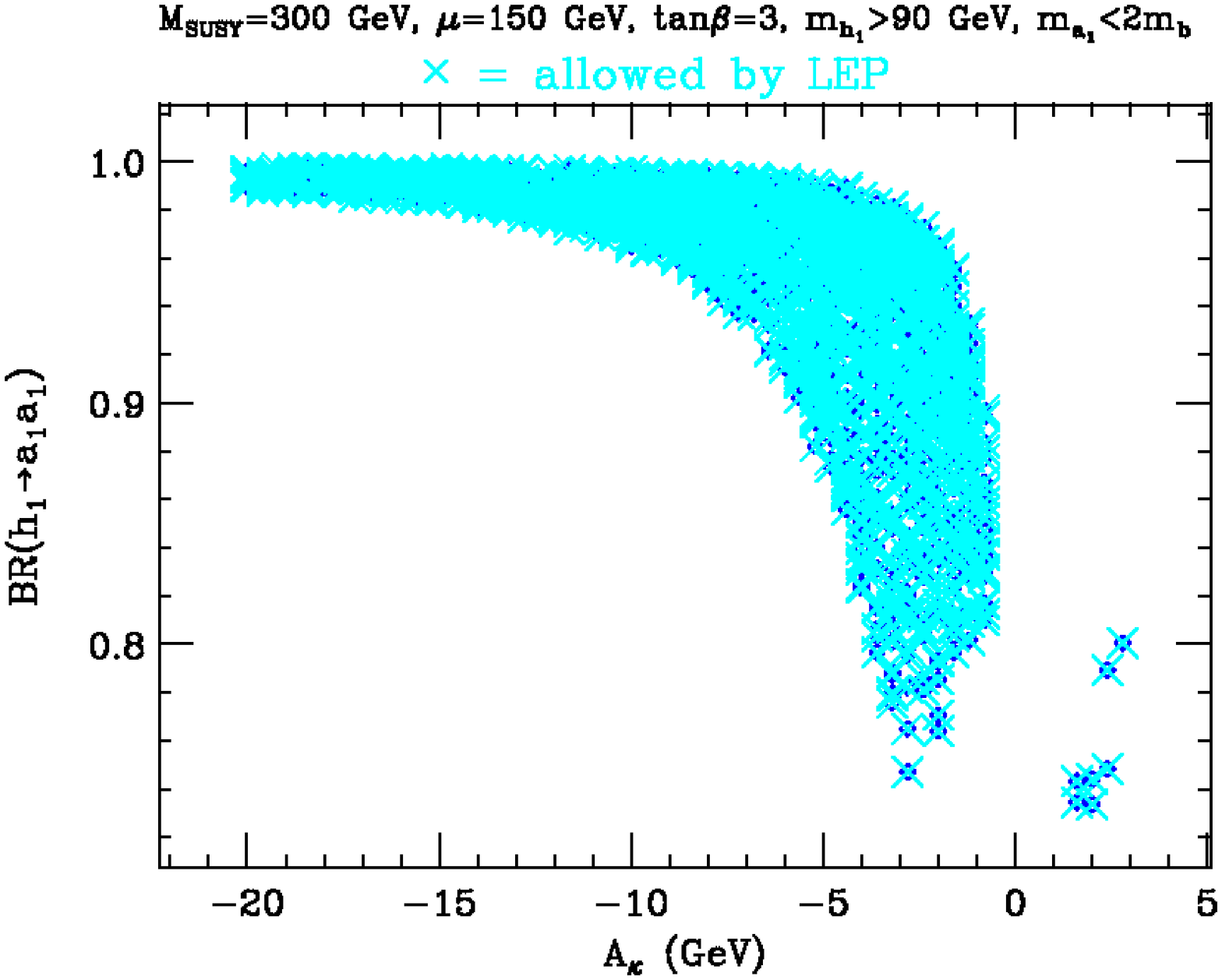}
\caption{$\fmax$ and $\br(\hi\to\ai\ai)$ vs. $A_\kappa$ for $\tan
  \beta = 3$ and $\mu=150\gev$.
Point conventions as in Fig.~\ref{fig:AkAl_and_kl}.}
\label{fig:fmaxbrtb3}
\end{figure}

\begin{figure}
\includegraphics[width=2.4in,angle=90]{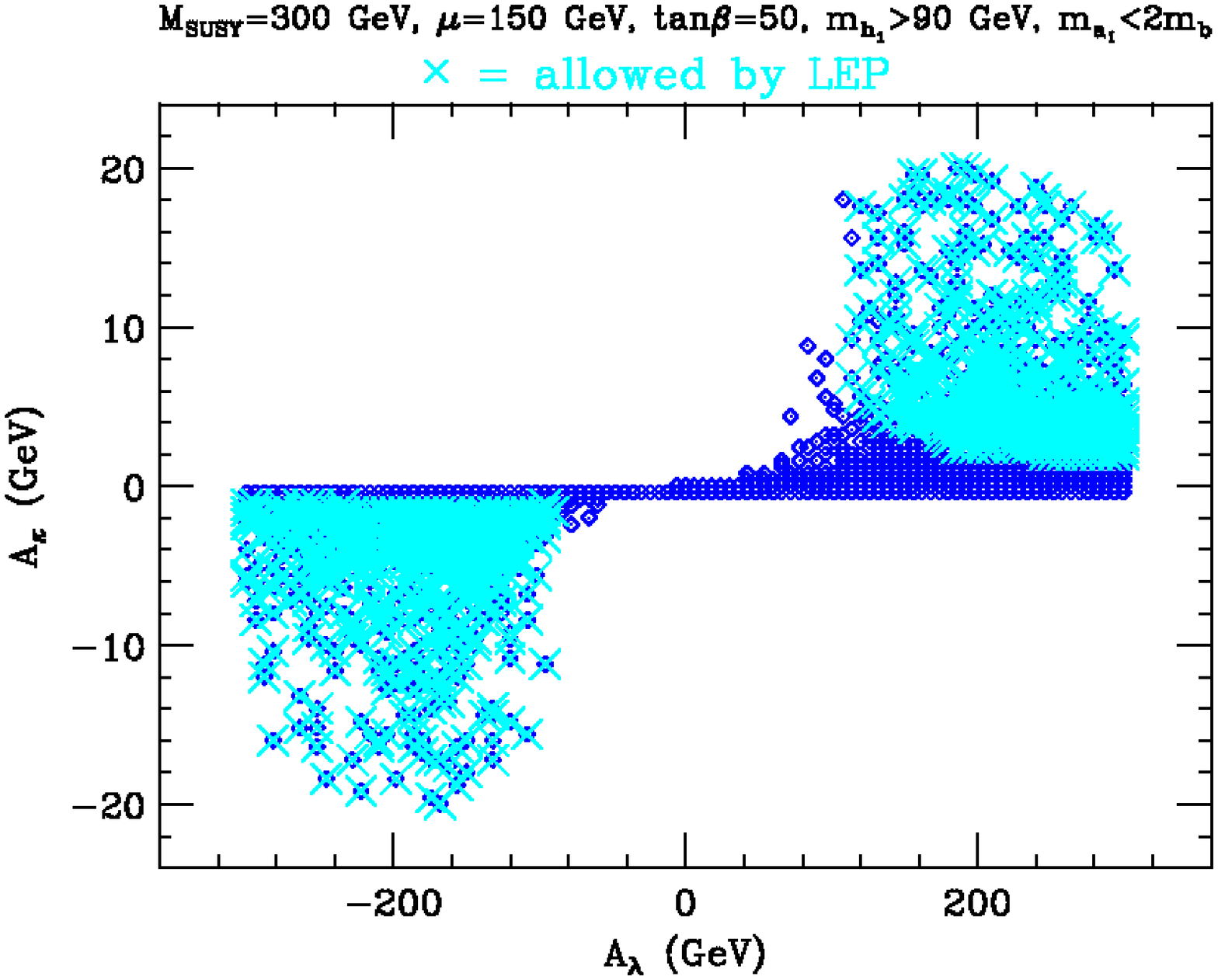}
\hspace{0.2cm}
\includegraphics[width=2.4in,angle=90]{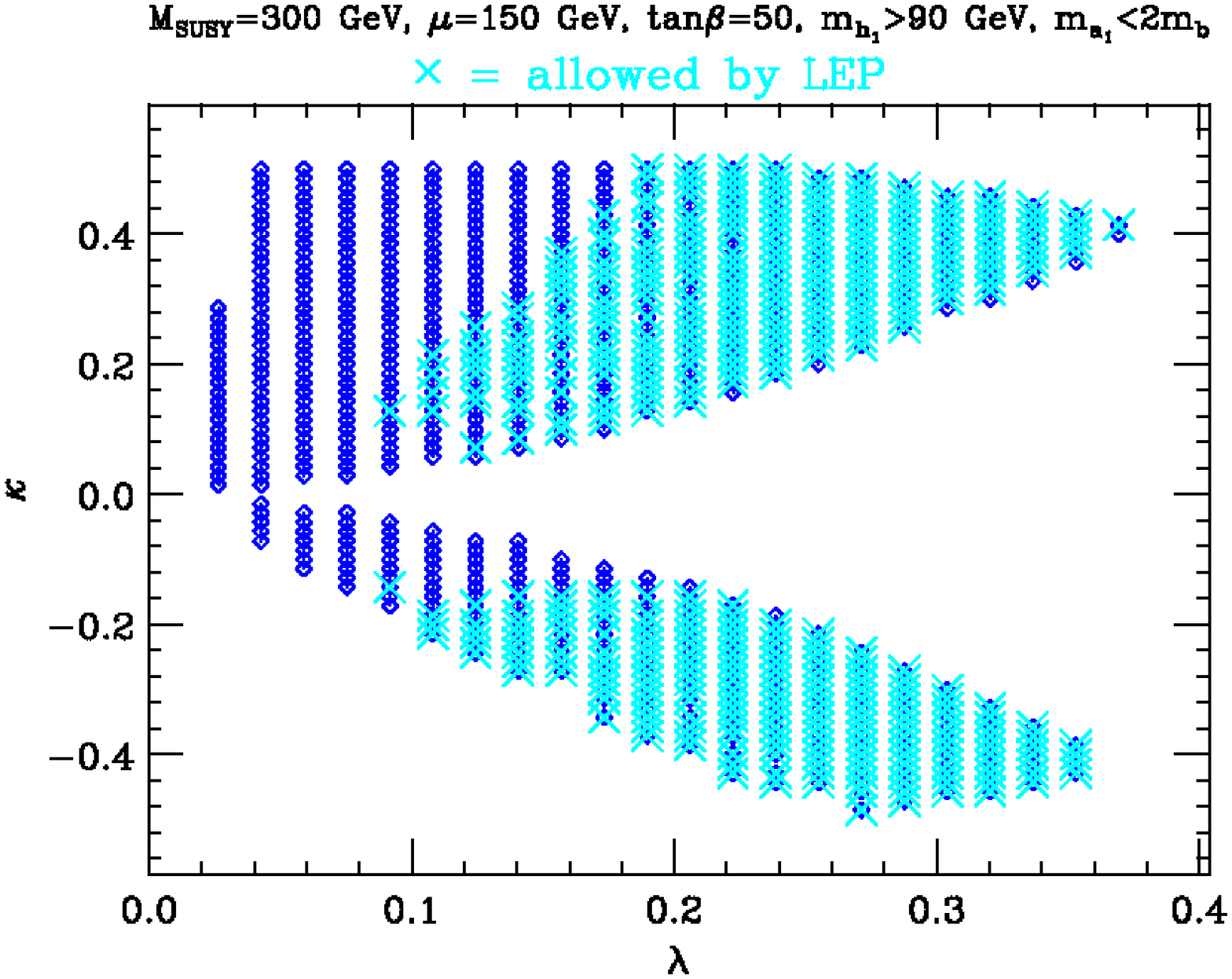}
\caption{Allowed parameter space in the $A_\kappa - A_\lambda$ and $\kappa
  - \lambda$ planes for $\tanb=50$ and $\mu=150\gev$.
Point conventions as in Fig.~\ref{fig:AkAl_and_kl}.}
\label{fig:AkAl_and_kl_tb50}
\end{figure}

\begin{figure}
\includegraphics[width=2.4in,angle=90]{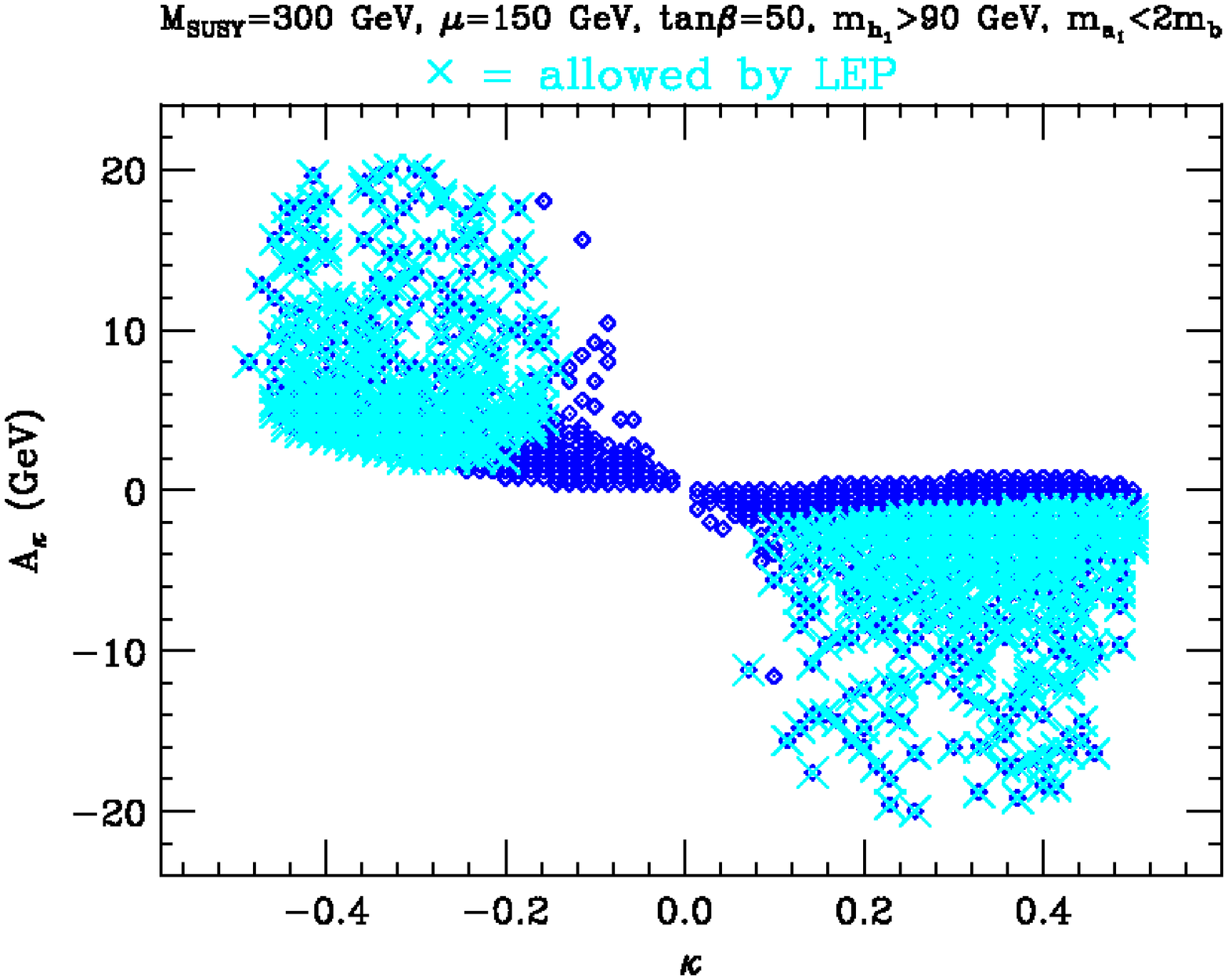}
\hspace{0.2cm}
\includegraphics[width=2.4in,angle=90]{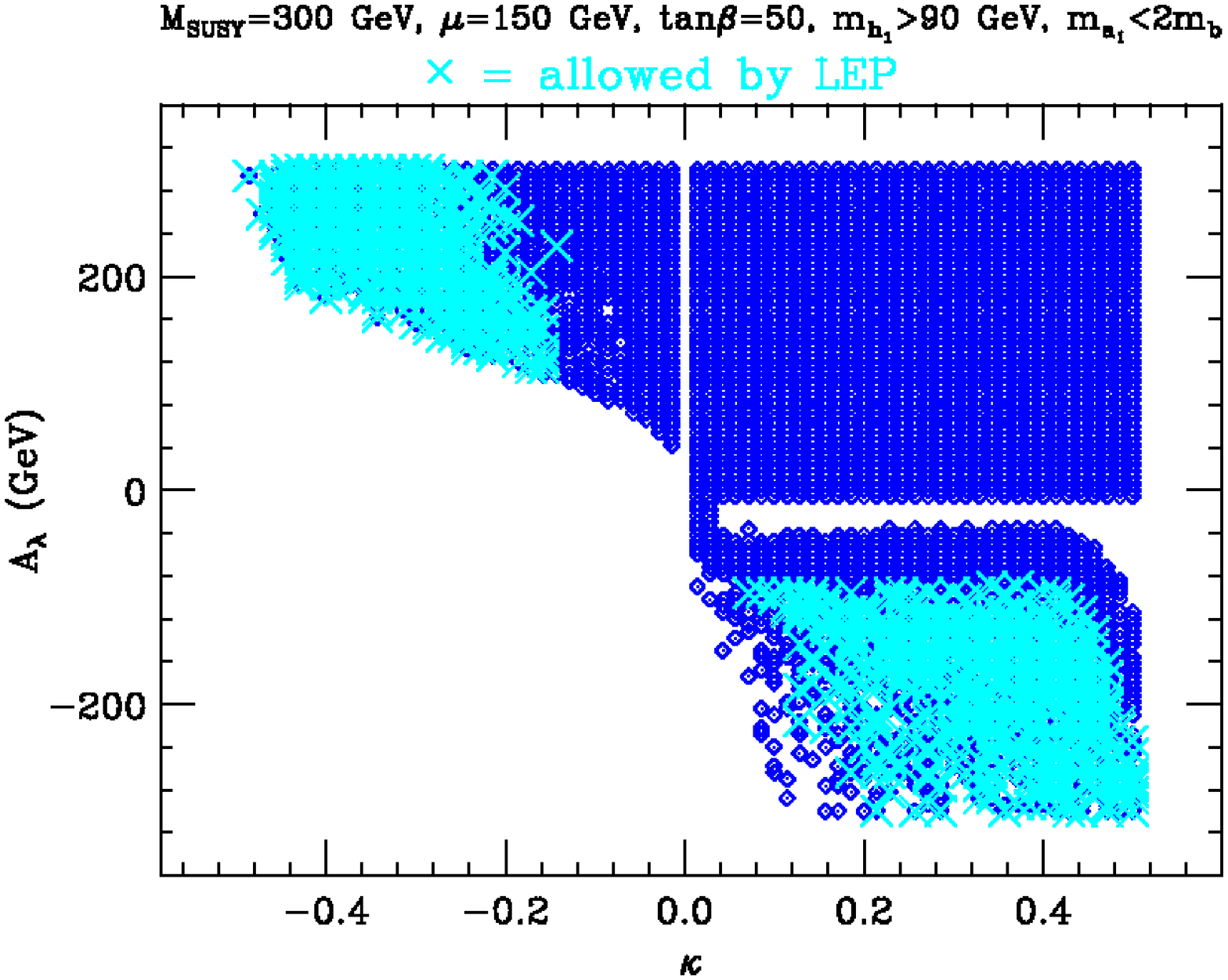}
\caption{Allowed parameter space in the $A_\kappa - \kappa$ and $A_\lambda
  - \kappa$ planes for $\tanb=50$ and $\mu=150\gev$.
Point conventions as in Fig.~\ref{fig:AkAl_and_kl}.}
\label{fig:Akk_and_Alk_tb50}
\end{figure}

\begin{figure}
\includegraphics[width=2.4in,angle=90]{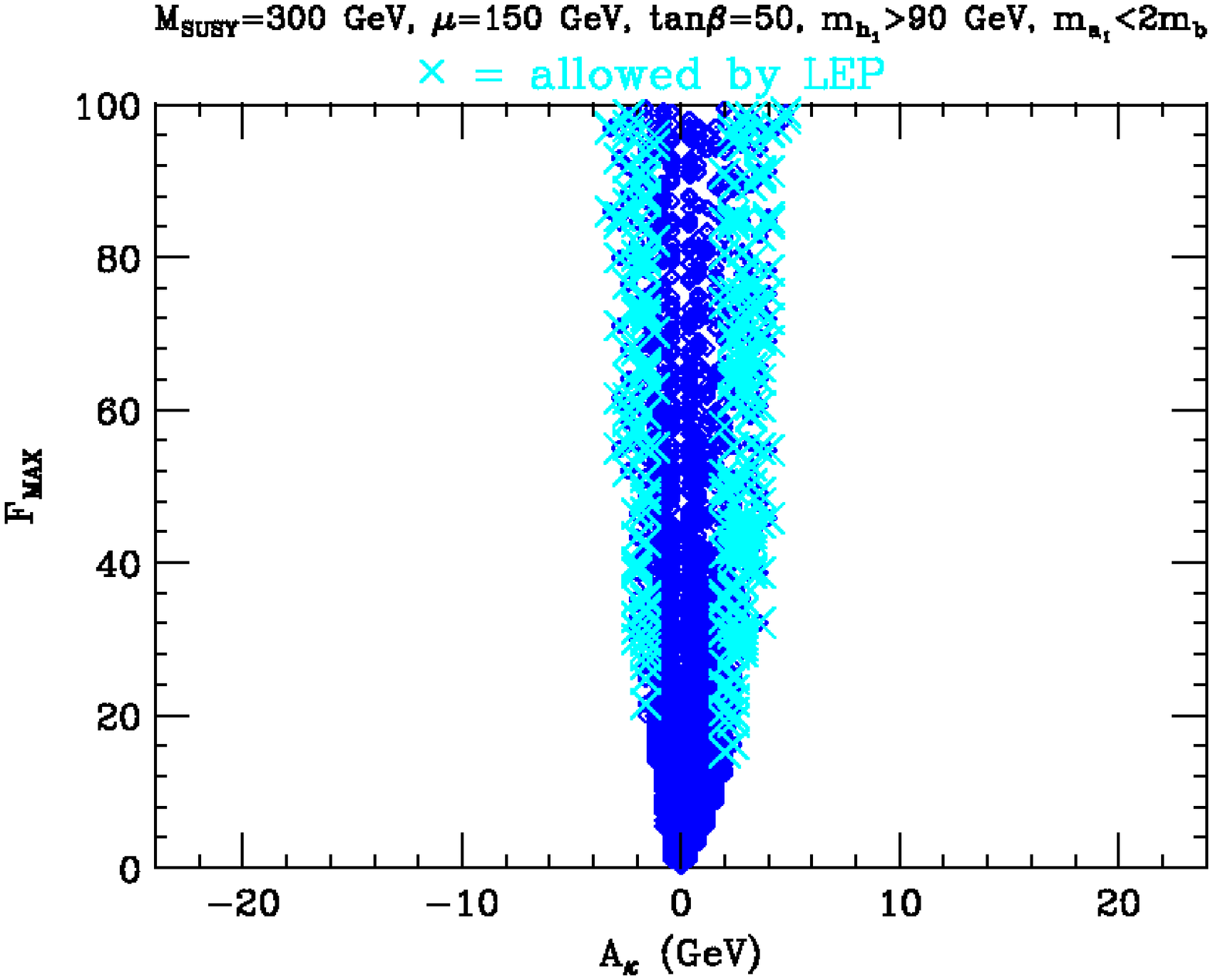}
\hspace{0.2cm}
\includegraphics[width=2.4in,angle=90]{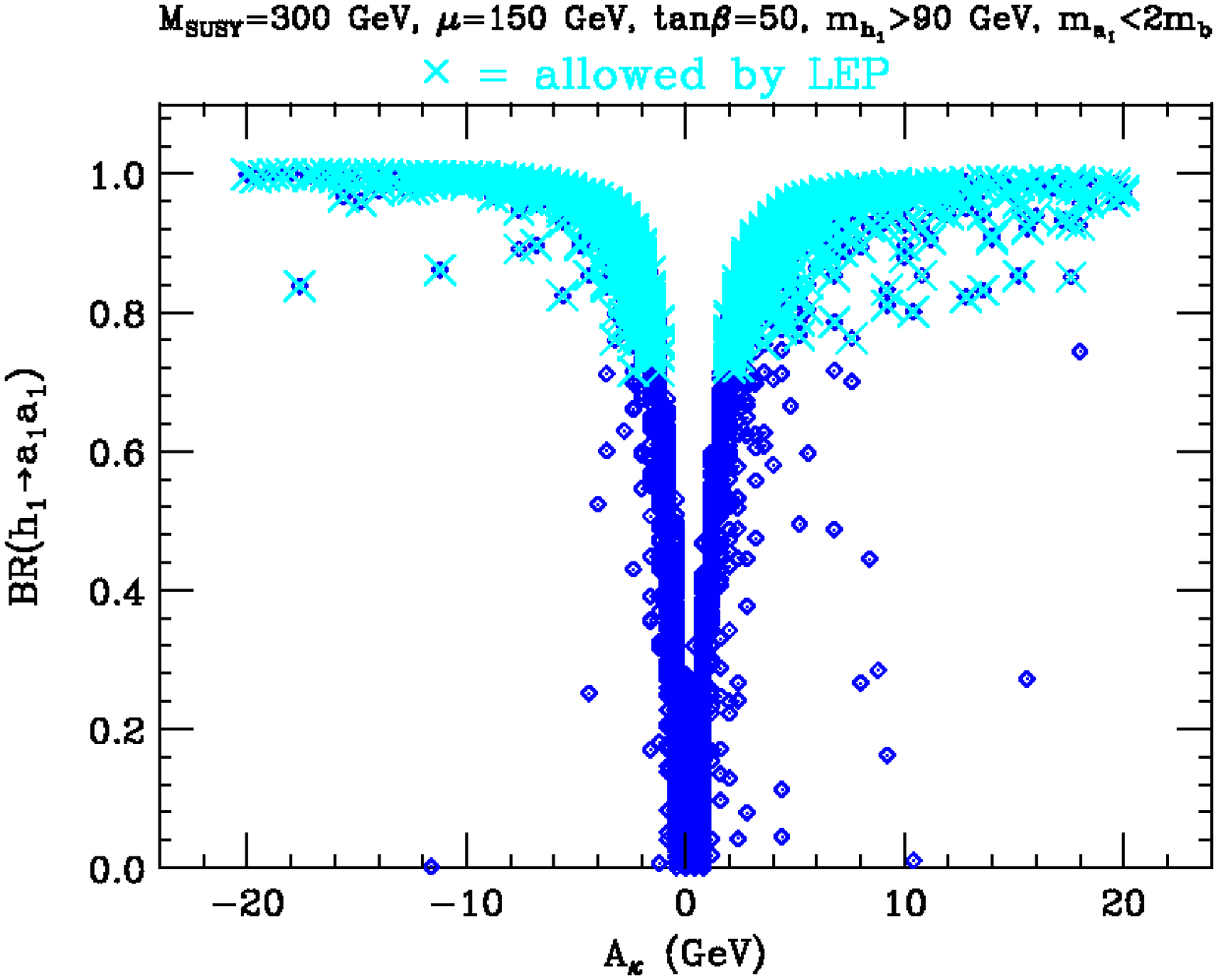}
\caption{$\fmax$ and $\br(\hi\to\ai\ai)$ vs. $A_\kappa$ for $\tan
  \beta = 50$ and $\mu=150\gev$.
Point conventions as in Fig.~\ref{fig:AkAl_and_kl}.}
\label{fig:fmaxbrtb50}
\end{figure}

\begin{figure}
\includegraphics[width=2.4in,angle=90]{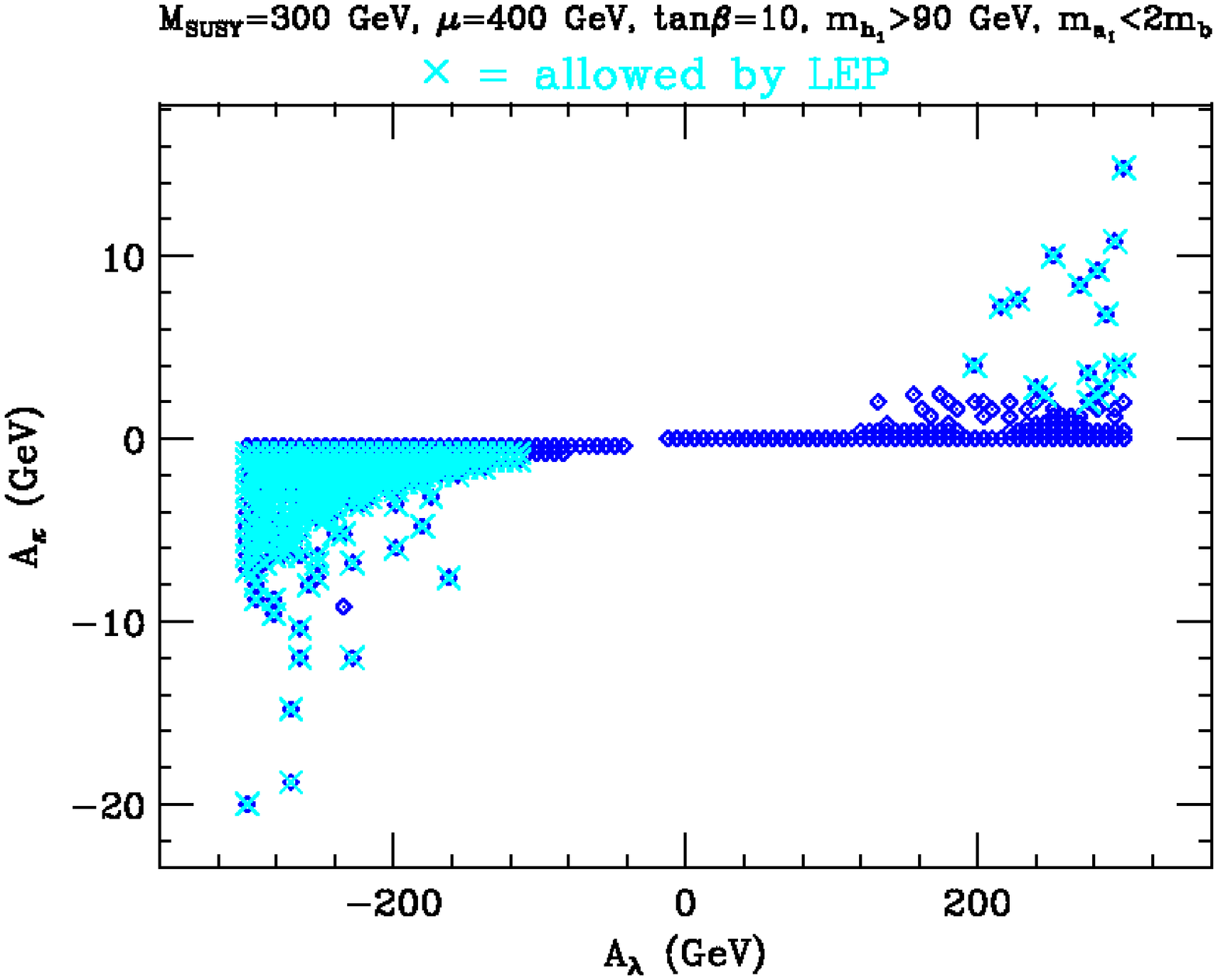}
\hspace{0.2cm}
\includegraphics[width=2.4in,angle=90]{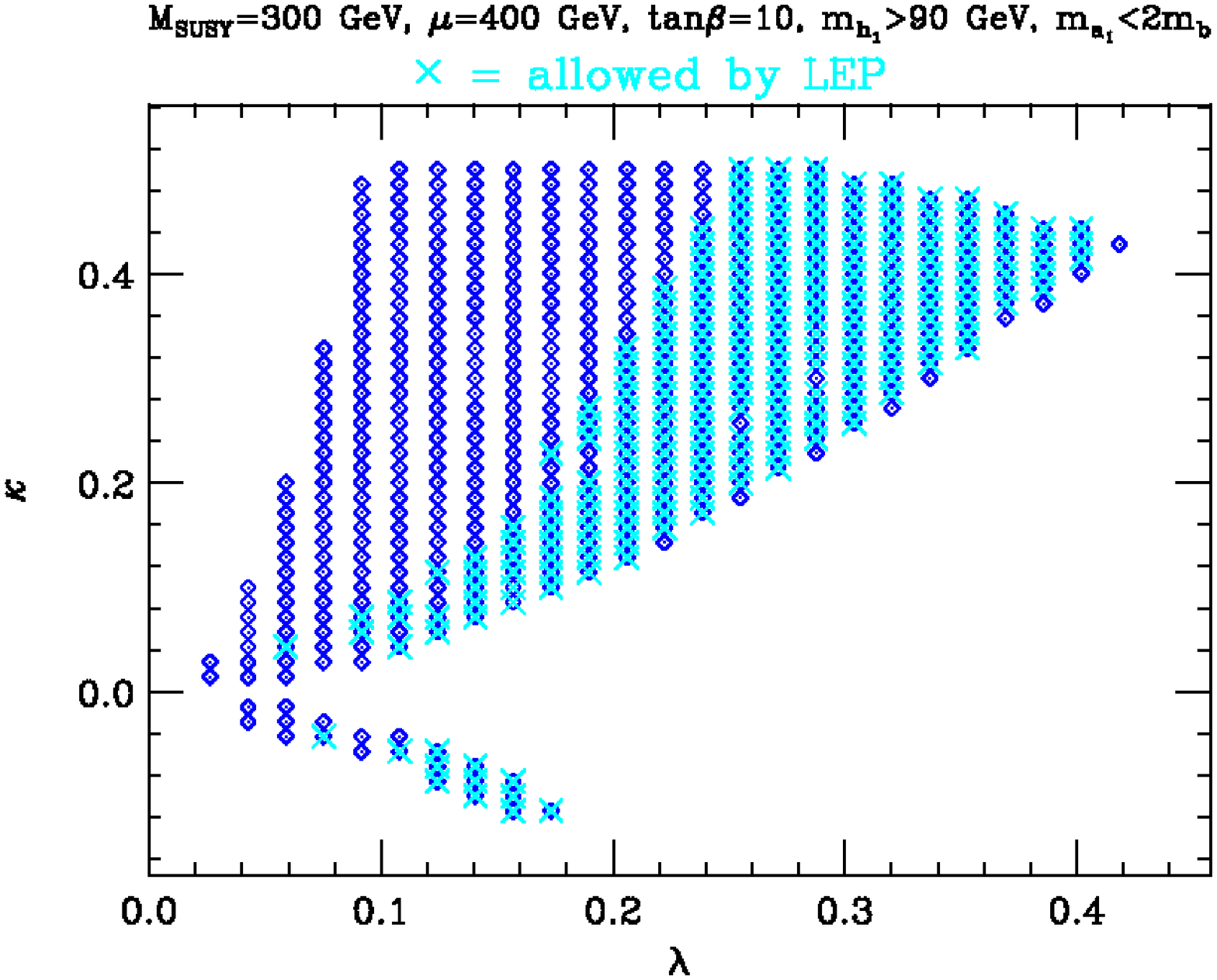}
\caption{Allowed parameter space in the $A_\kappa - A_\lambda$ and
  $\kappa - \lambda$ planes for $\tanb=10$ and $\mu=400\gev$.
Point conventions as in Fig.~\ref{fig:AkAl_and_kl}.}
\label{fig:AkAl_and_klmu400}
\end{figure}

\begin{figure}
\includegraphics[width=2.4in,angle=90]{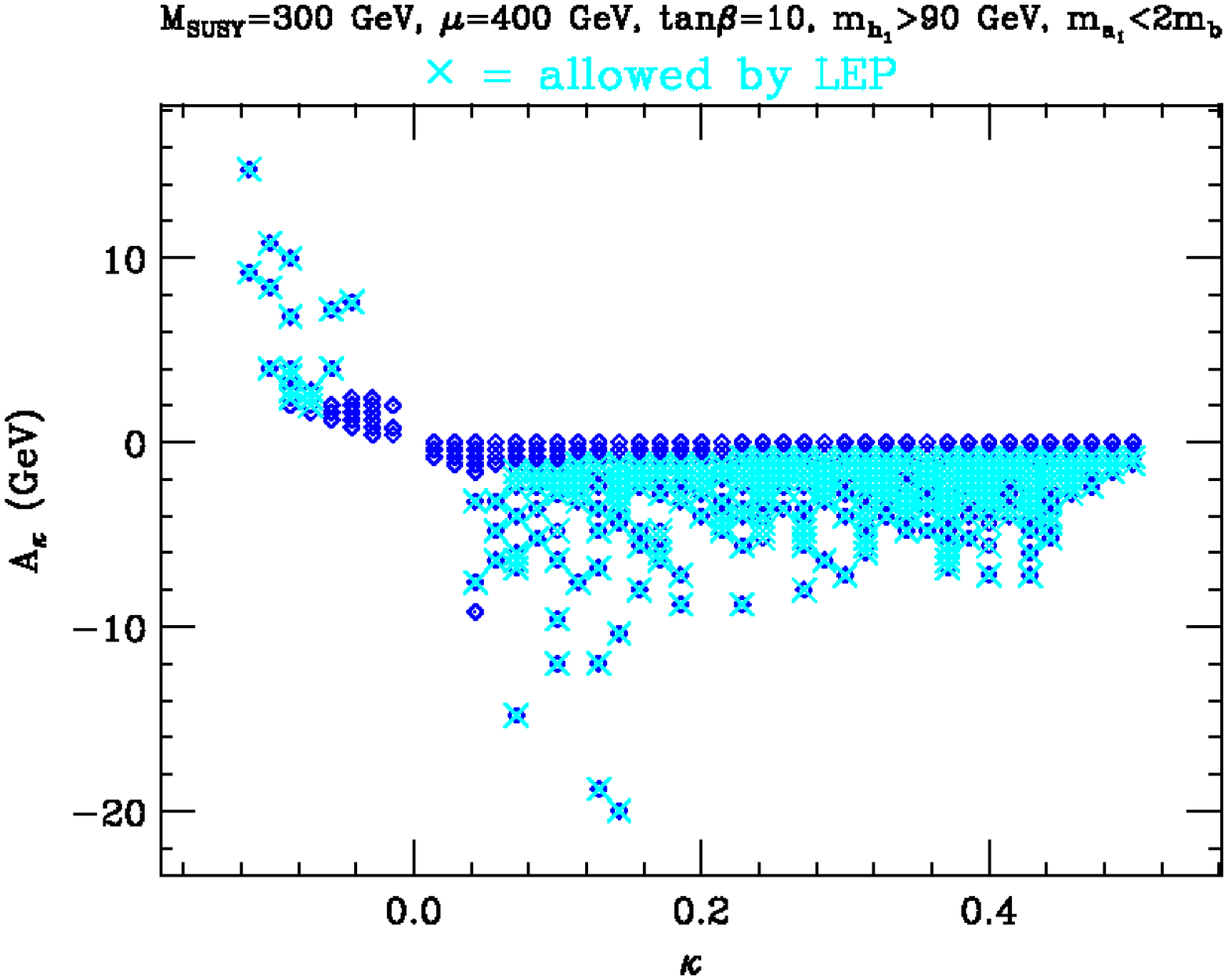}
\hspace{0.2cm}
\includegraphics[width=2.4in,angle=90]{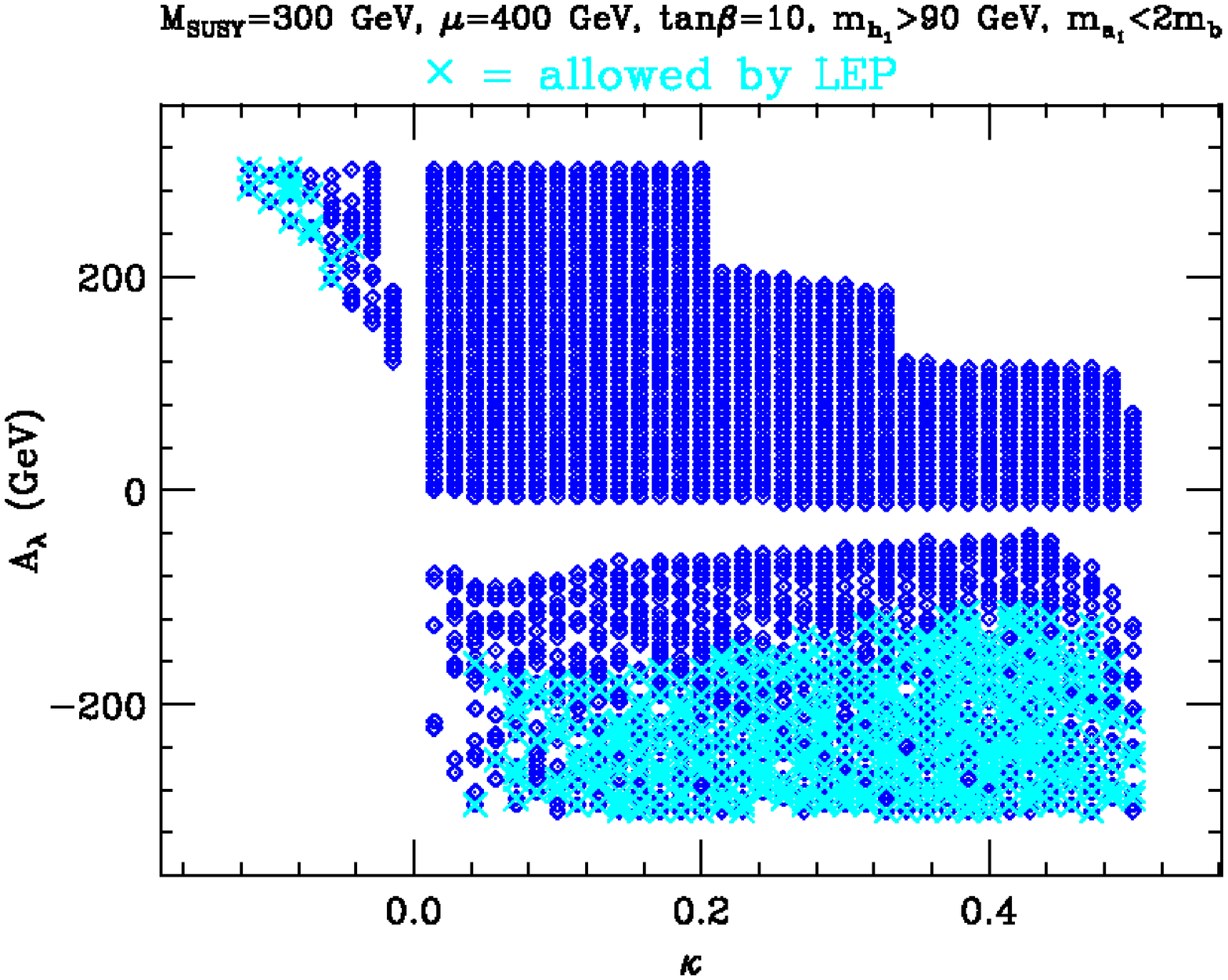}
\caption{Allowed parameter space in the $A_\kappa - \kappa$ and
  $A_\lambda - \kappa$ planes for $\tanb=10$ and $\mu=400\gev$.
Point conventions as in Fig.~\ref{fig:AkAl_and_kl}.}
\label{fig:Akk_and_Alkmu400}
\end{figure}

\begin{figure}
\includegraphics[width=2.4in,angle=90]{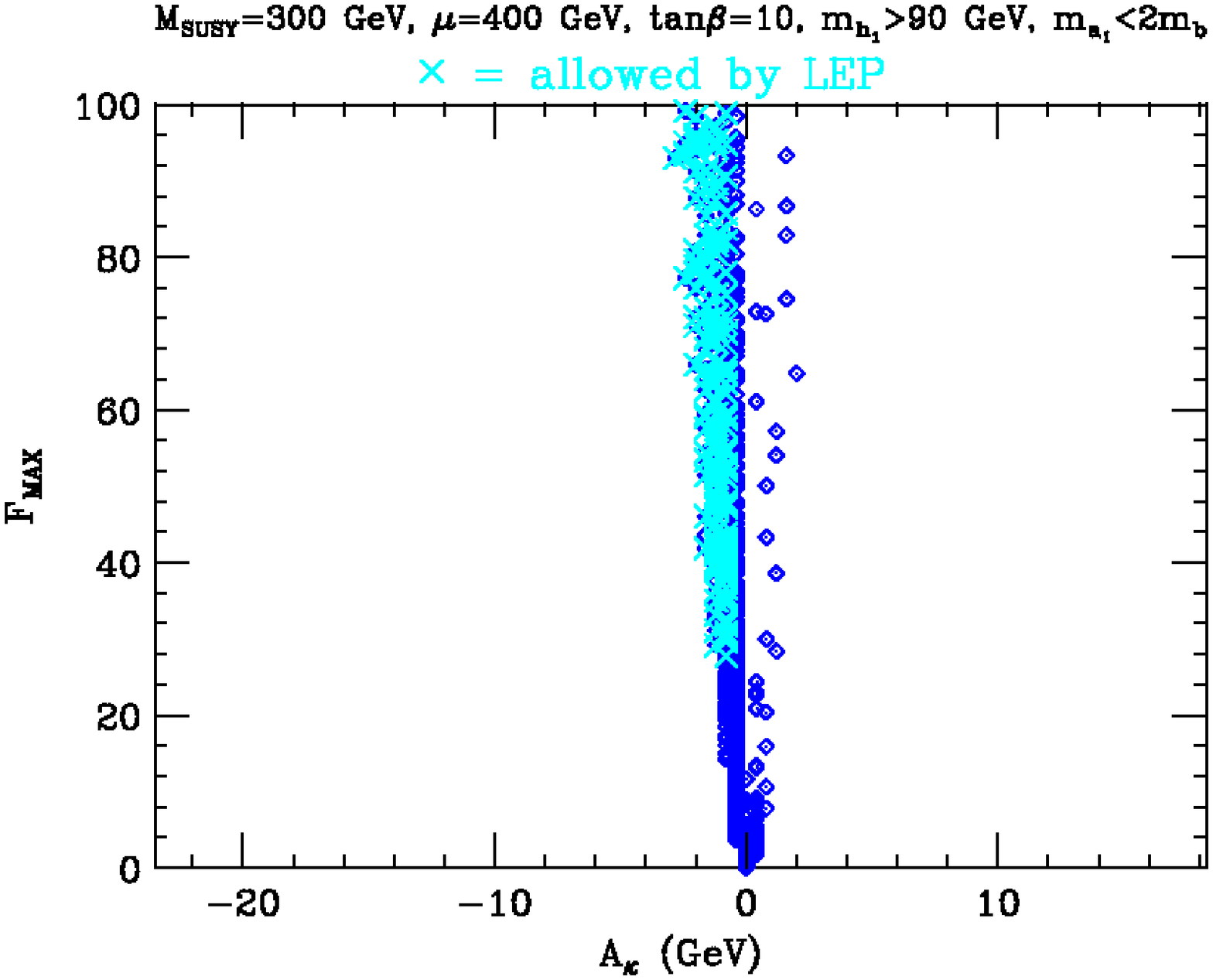}
\hspace{0.2cm}
\includegraphics[width=2.4in,angle=90]{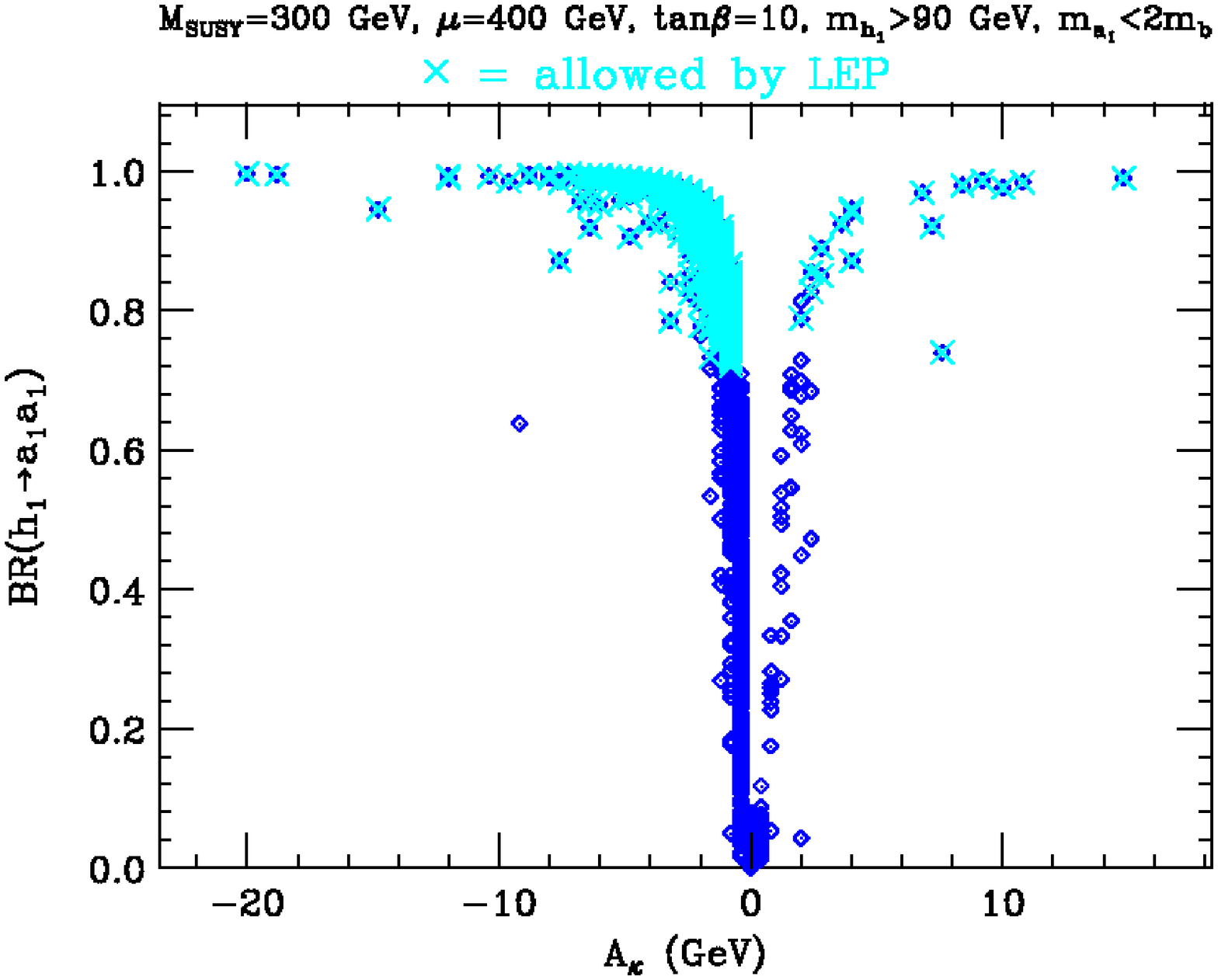}
\caption{$\fmax$ and $\br(\hi\to\ai\ai)$ vs. $A_\kappa$ for $\tan
  \beta = 10$ and $\mu=400\gev$.
Point conventions as in Fig.~\ref{fig:AkAl_and_kl}.}
\label{fig:fmaxbrtb10mu400}
\end{figure}

\begin{figure}
\includegraphics[width=2.4in,angle=90]{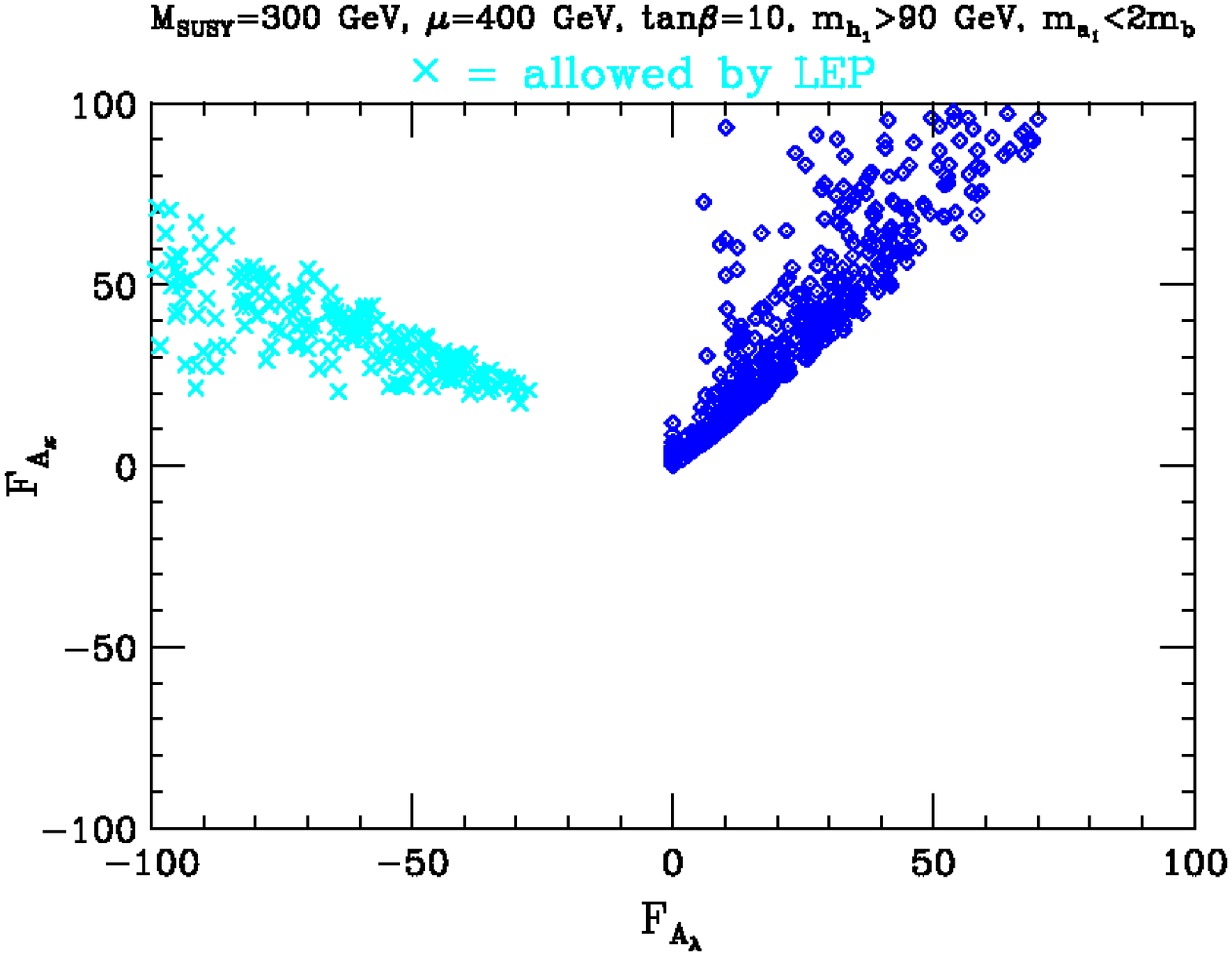}
\hspace{0.2cm}
\includegraphics[width=2.4in,angle=90]{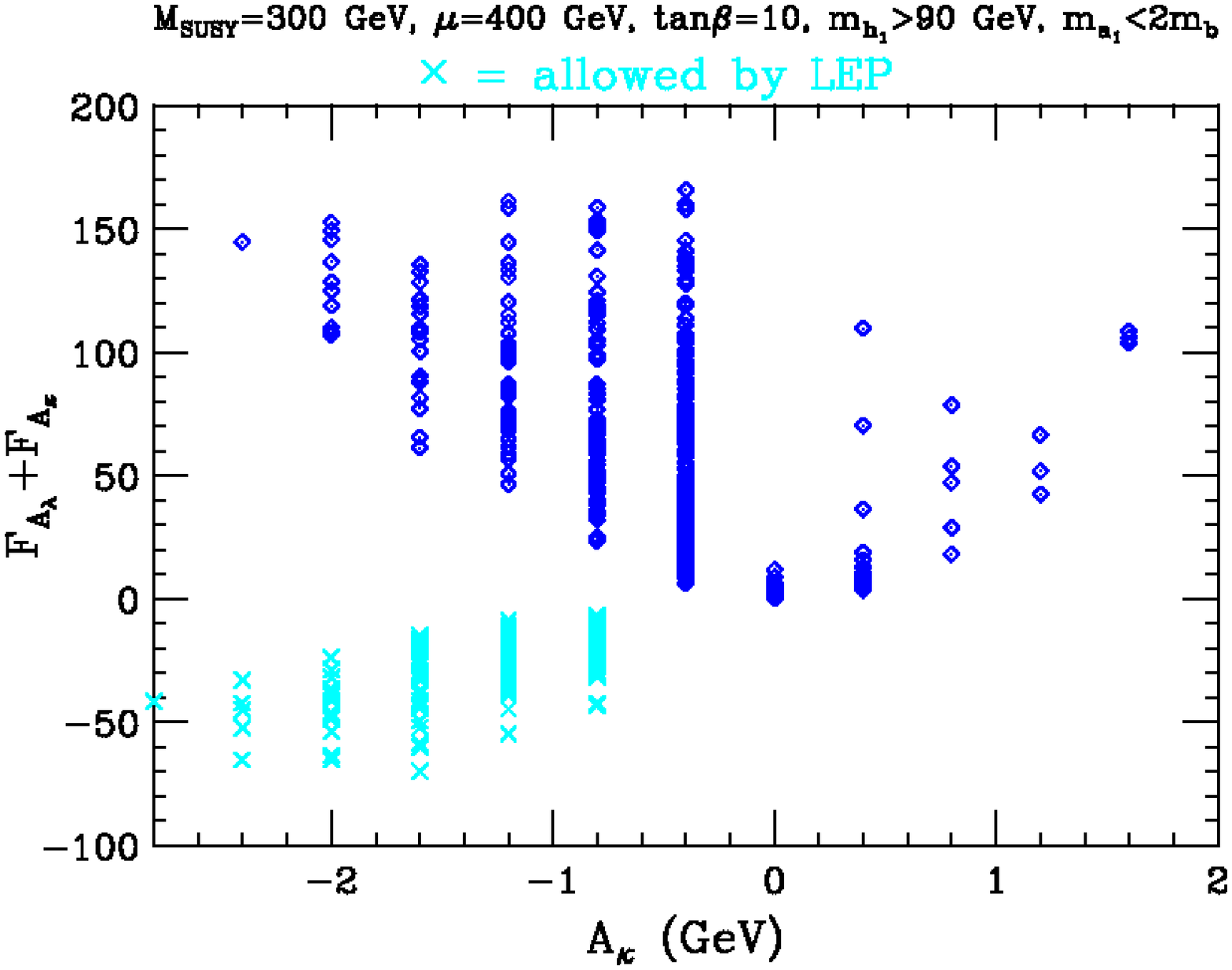}
\caption{We plot $\fakap$ vs. $\falam$ (left window)
and $\fakap+\falam$ vs. $\akap$ (right window) for $\mu=400\gev$, and   $\tanb=10$.
Point conventions as in Fig.~\ref{fig:AkAl_and_kl}.}
\label{fakfal400}
\end{figure}

Finally, in order to see the effect of varying $\tan \beta$ and
$\mu$, we give a selection of some of the same plots for: $\mu=150\gev$
and $\tanb=3$; $\mu=150\gev$ and $\tanb=50$; and $\mu=400\gev$ with
$\tanb=10$. These appear in 
Figs.~\ref{fig:AkAl_and_kl_tb3} --- \ref{fakfal400}.
We note that there is an exact symmetry under $\mu\to -\mu$ (implying
$s\to -s$ in the $\lam>0$ convention we employ), $\akap\to -\akap$
and $\alam\to -\alam$. Thus, only positive $\mu$ values need be
considered when both signs of $\alam$ and $\akap$ are scanned.

Comparing the plots for $\tan \beta =3$,
Figs.~\ref{fig:AkAl_and_kl_tb3} --- \ref{fig:fmaxbrtb3}, and the plots
for $\tan \beta = 50$, Figs.~\ref{fig:AkAl_and_kl_tb50} --
\ref{fig:fmaxbrtb50}, with the previous plots for $\tan \beta = 10$ we
see that the region of allowed parameter space for $\kappa > 0$,
$A_\lambda <0$ and $A_\kappa <0$ does not change much when varying
$\tan \beta$. On the other hand the allowed region of parameter
space with $\kappa < 0$, $A_\lambda >0$ and $A_\kappa >0$ changes
dramatically. It does not even exist for $\tan \beta =
3$~\footnote{There is a tiny allowed region with positive $A_\kappa$ and
positive but very small $A_\lambda$. This region is not a
continuation of the region present in the plots for $\tan \beta = 10$.}. 
The reason is that the
first two terms in Eq.~(\ref{eq:M22sq}) are less suppressed for
smaller $\tan \beta$ and they combine into a term proportional to
$A_\lambda + 4 \kappa s$. The last term in Eq.~(\ref{eq:M22sq})
dominates only for larger values of $A_\kappa$. If the last term
does not dominate, the first two terms are positive only for very
large $A_\lambda$ making the region with $\kappa < 0$, $A_\lambda
>0$ and $A_\kappa
>0$ very constrained. For exactly the same reason, this region is
larger for $\tan \beta = 50$ compared to $\tan \beta = 10$ and the
allowed parameter space expands to lower values of $A_\kappa$
as particularly noticeable in the $\akap$ vs. $\kap$ plots; compare
Figs.~\ref{fig:Akk_and_Alk_tb50},  \ref{fig:Akk_and_Alk_tb3} and \ref{fig:Akk_and_Alk}. In
the region of parameter space with $\kappa > 0$, $A_\lambda <0$
and $A_\kappa <0$ the condition $A_\lambda + 4 \kappa s > 0$ is
always satisfied when the necessary condition for having positive
mass squared eigenvalues, $A_\lambda + \kappa s
> 0$, is satisfied. Therefore, this region of parameter space is
not very sensitive to $\tan \beta$.

Quite interestingly, once $\mai<2\mb$ is achieved for $\tan \beta
= 3$, $\br(\hi\to\ai\ai)$ is automatically large enough ($\gsim
0.75$ for the most part) and, correspondingly, $\br(\hi\to b\bar
b)$ is sufficiently suppressed that $Zh\to Zb\bar b$ constraints
at LEP are satisfied; see Fig.~\ref{fig:fmaxbrtb3}. As in the
$\tanb=10$ case, $\fmax$ is smallest for the smallest possible
$\akap$ and approaches $\sim 15$ for a couple of the points with
$\akap>0$. Overall, it would appear that $\mai<2\mb$ is a bit more
difficult to achieve without tuning when $\tanb=3$ vs. when
$\tanb=10$. For $\tan \beta = 50$, see Fig.~\ref{fig:fmaxbrtb50},
small $\fmax$ ($\lsim 20$) is achieved for a narrower range of
$\akap$ and $\akap>0$ gives the smallest $\fmax$ values (whereas
for $\tanb=10$ there was no particular preference with regard to
the sign of $\akap$).

Finally, we discuss the effect of increasing $\mu$. We consider
$\mu=400\gev$ at $\tanb=10$ and present results in
Figs.~\ref{fig:AkAl_and_klmu400} --- \ref{fakfal400}. These
figures clearly show that the range of parameter space for which
$\mai<2\mb$ shrinks with increasing $\mu$. This is easy to understand
from Eq.~(\ref{eq:ma1}) or Eq.~(\ref{eq:ma1_better}). For fixed
$\lam$, increasing $\mu$ results in an increase of $s$. Consequently
the term proportional to $A_\lambda$ in the formula for $\mai$ is
further suppressed while the term proportional to $A_\kappa$ is
enhanced.  In order to compensate for this so as to keep $\mai$ small,
smaller values of $\kappa$ and $A_\kappa$ and larger values of
$\lambda$ and $A_\lambda$ are required as compared to the $\mu =
150\gev$ case.  Moreover, the $\mai<2\mb$ region of parameter space
with $\kap<0$, $\alam>0$ and $\akap>0$ is even further constrained by
the condition $\alam + \kappa s > 0$ which requires larger $A_\lambda$
for larger $s$.  These effects are clearly visible in
Fig.~\ref{fig:AkAl_and_klmu400} and Fig.~\ref{fig:Akk_and_Alkmu400}.
The increased value of $A_\lambda$ required for this scenario to work
for larger $\mu$ leads to larger tuning necessary to achieve
$\mai<2\mb$. Fig.~\ref{fig:fmaxbrtb10mu400} shows a dramatic narrowing
of the $\akap$ region for which moderate $\fmax$ values are achieved
and that the best $\fmax$ value for points having large enough
$\br(\hi\to\ai\ai)$ that LEP limits are evaded is of order $\sim 30$
in contrast to the $\fmax\sim 15$ values achieved for $\mu=150\gev$.
However, we see from Fig.~\ref{fakfal400} that the values of $\falam$
and $\fakap$ are more strongly correlated so that $\falam+\fakap$ can
again take on small values. As we have seen, the latter implies small
$F_{\mai}$ in models in which the soft-SUSY breaking parameters are
correlated or $\alam(\mz)$ and $\akap(\mz)$ are dominated by a single
soft-SUSY-breaking parameter.



\begin{figure}
\includegraphics[width=2.4in,angle=90]{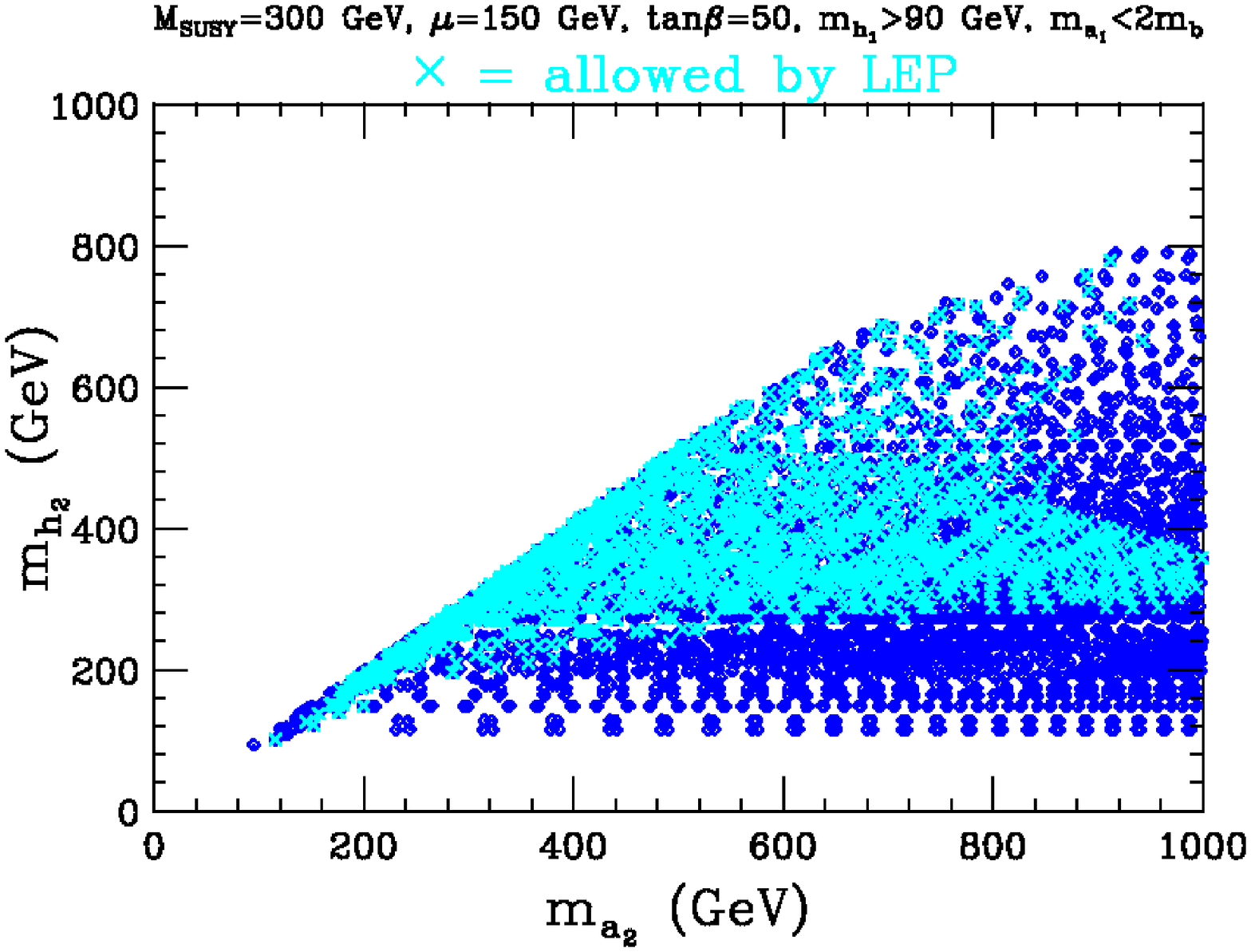}
\includegraphics[width=2.4in,angle=90]{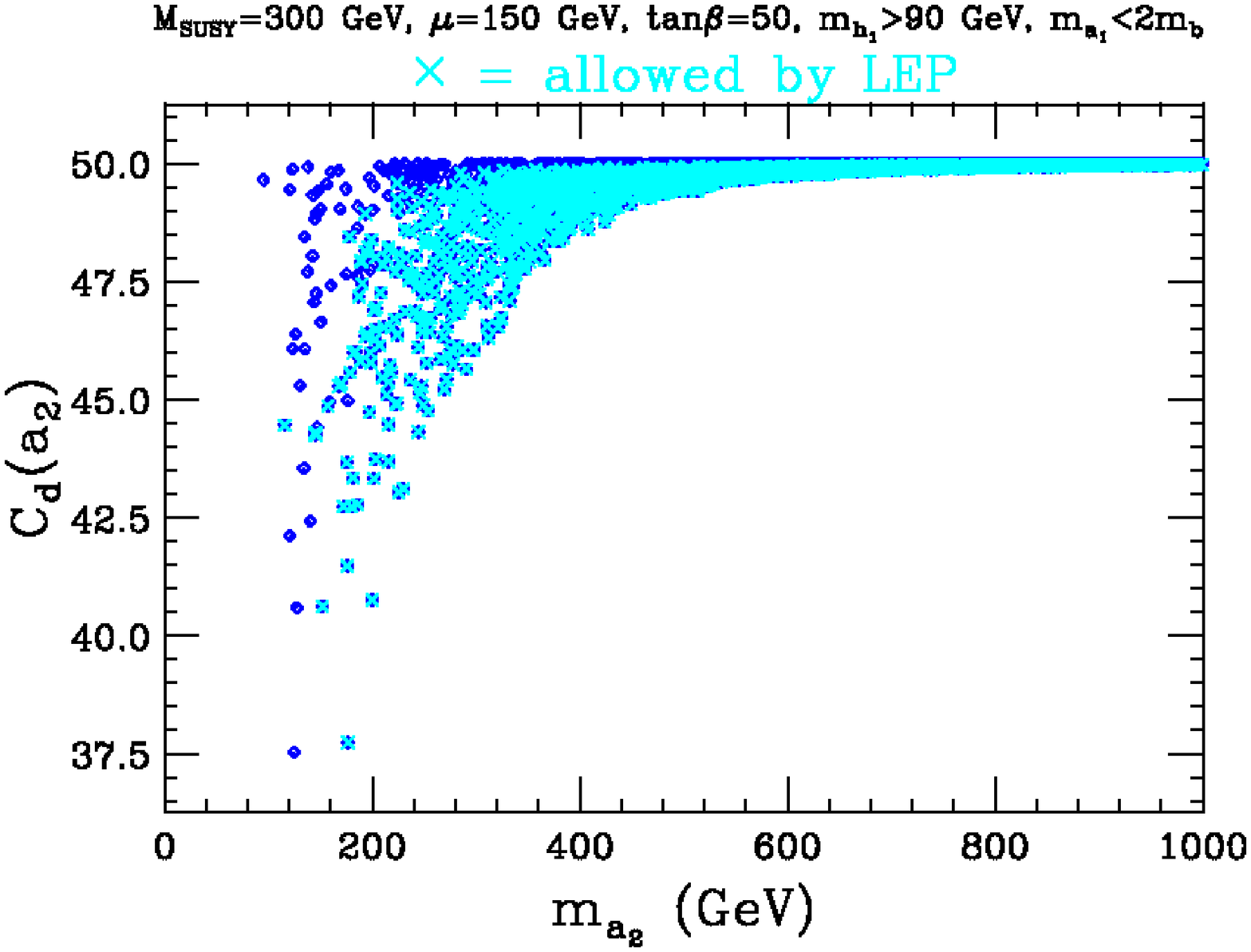}
\includegraphics[width=2.4in,angle=90]{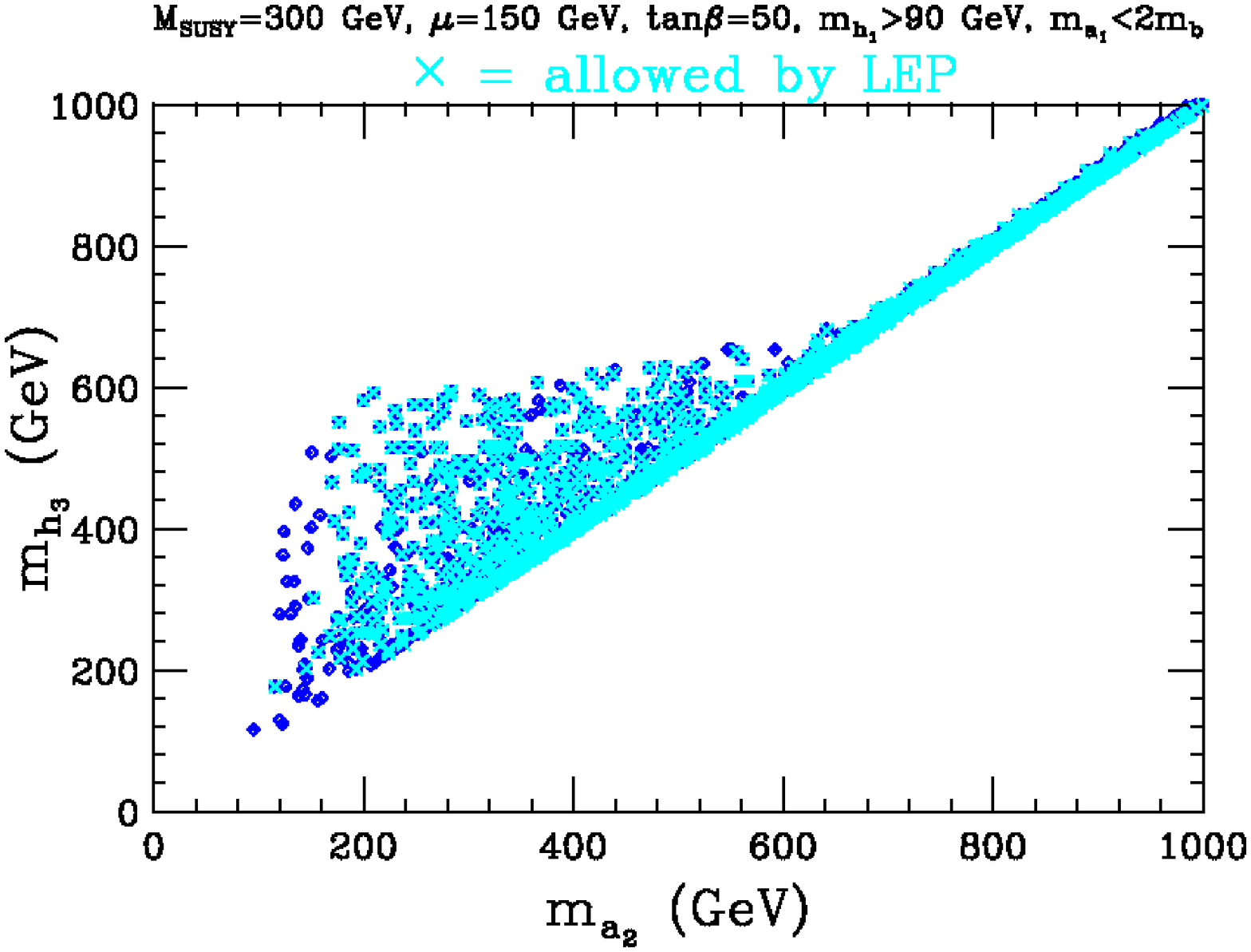}
\includegraphics[width=2.4in,angle=90]{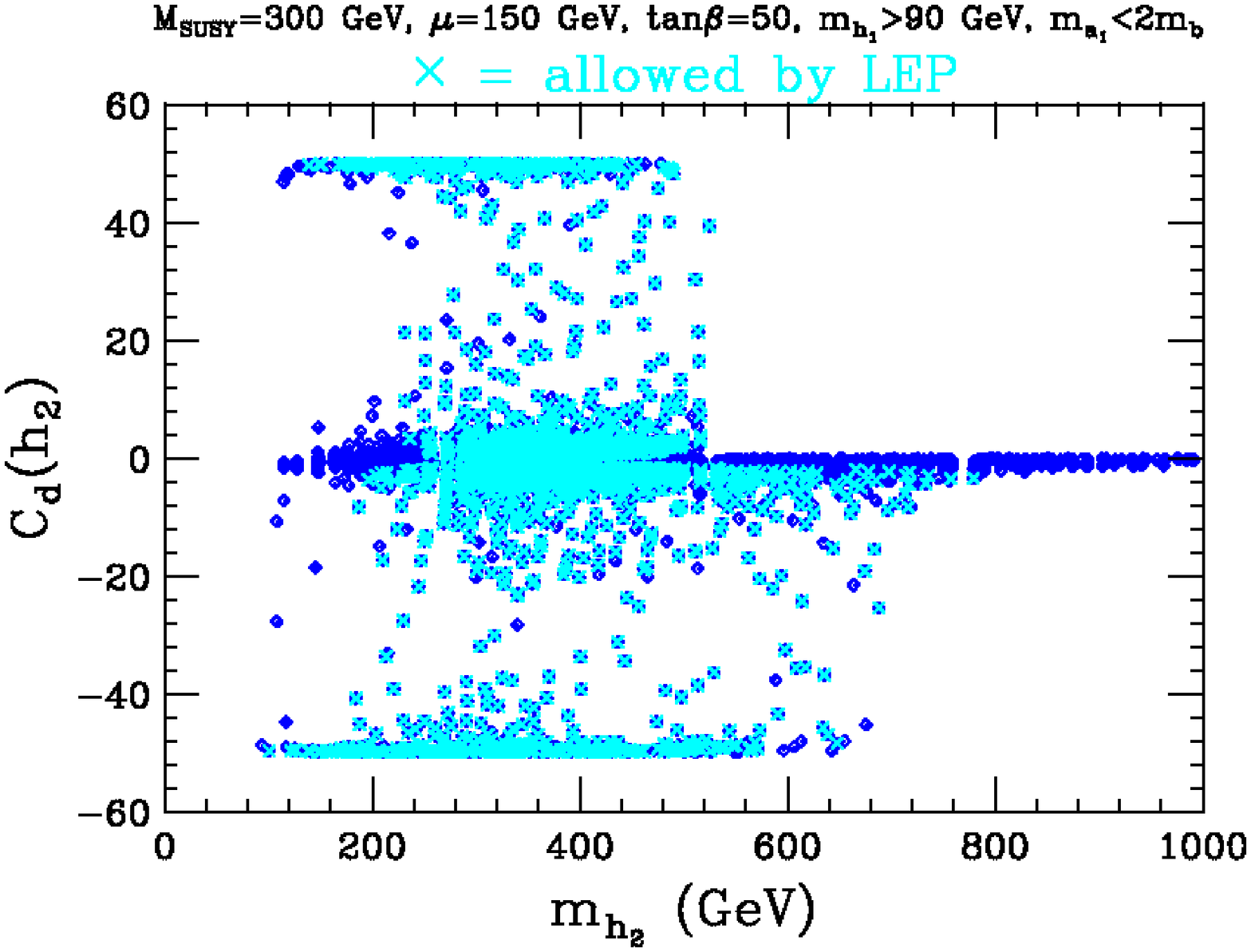}
\includegraphics[width=2.4in,angle=90]{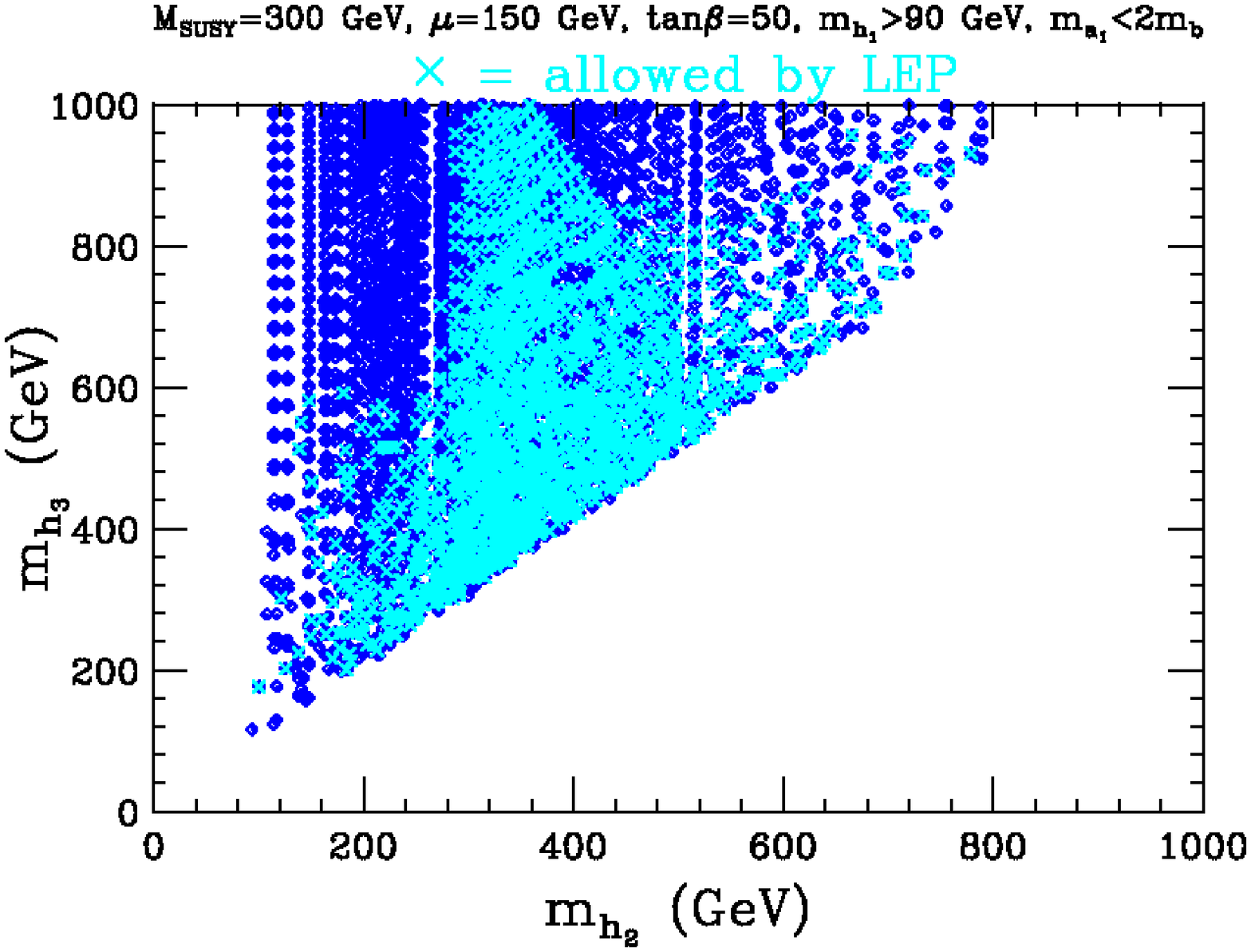}
\includegraphics[width=2.4in,angle=90]{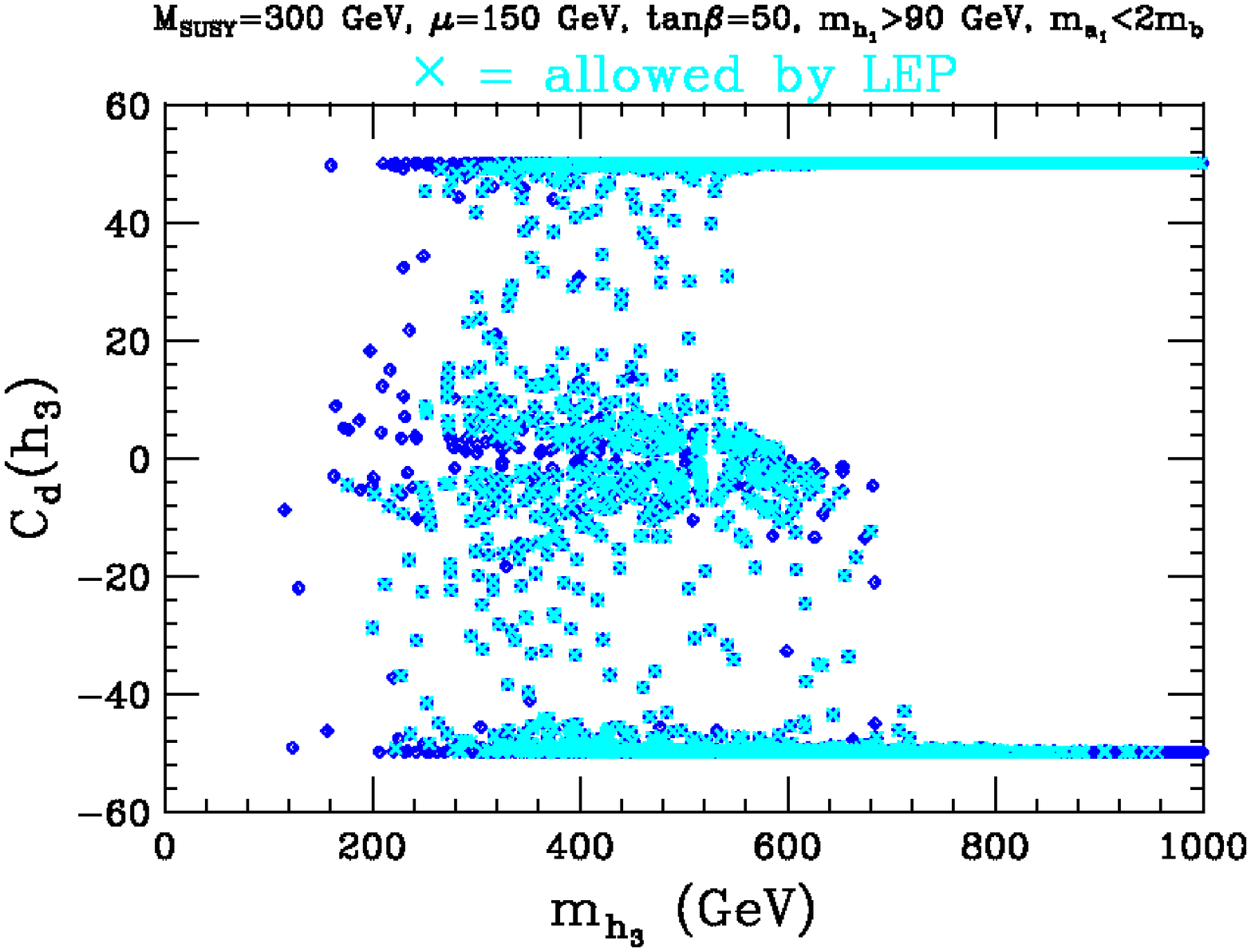}
\caption{In the left-hand set of plots, we present
 $\mhii$ vs. $\maii$ (top), $\mhiii$ vs. $\maii$
  (middle) and $\mhiii$ vs. $\mhii$ (bottom). In the right-hand set of
  plots, we present $C_d(\aii)$ vs. $\maii$ (top), $C_d(\hii)$ vs. $\mhii$
  (middle) and $C_d(\hiii)$ vs. $\mhiii$ (bottom). All plots are for  $\tan
  \beta = 50$ and $\mu=150\gev$.
Point conventions as in Fig.~\ref{fig:AkAl_and_kl}.}
\label{tb50mhvsmh}
\end{figure}

Given that detection of the $\hi$ at hadron colliders will be quite
challenging for the large-$\br(\hi\to\ai\ai)$ scenarios that we focus
on, an interesting question is then whether or not the other Higgs
bosons $h_{2,3}$ and $a_2$ might be detectable.  This will depend on
their masses and on their couplings.  We will see that within the
scenarios considered, their masses can range from somewhat above
$100\gev$ to quite large values.  As regards their couplings, since
the $\hi$ has quite SM-like $WW,ZZ$ couplings for the scenarios being
considered (as illustrated earlier), the $h_{2,3}$ will have quite
weak $WW,ZZ$ couplings, and of course the $a_2$ has no tree-level
couplings to $WW,ZZ$. Thus, the question is whether production
mechanisms relying on $q\anti q $ couplings of the $\hii,\hiii,\aii$
could lead to an observable signal.  This same question of course
applies to the $\hh$ and $\ha$ of the MSSM.  The answer there is that
high $\tanb$ is required, in which case $gg\to b\anti b \hh,b\anti b
\ha$ production is highly enhanced since the $\hh,\ha$ have $b\anti b$
coupling strength of order $\tanb$ times the SM-like strength. We wish
to explore the extent to which this also applies in the NMSSM for the
$\hii,\hiii,\aii$ in scenarios with large $\br(\hi\to\ai\ai)$.  For
this purpose, we present some additional plots in the case of
$\tanb=50$. We will denote the strength of the $b\anti b$ coupling
(generally any down-type quark or lepton) of any given Higgs boson
relative to the SM-like strength by $C_d(h)$, where
$h=\aii,\hii,\hiii$ will be considered.  The relevant plots appear in
Fig.~\ref{tb50mhvsmh}.

\begin{figure}
  \centerline{\includegraphics[width=2.4in,angle=90]{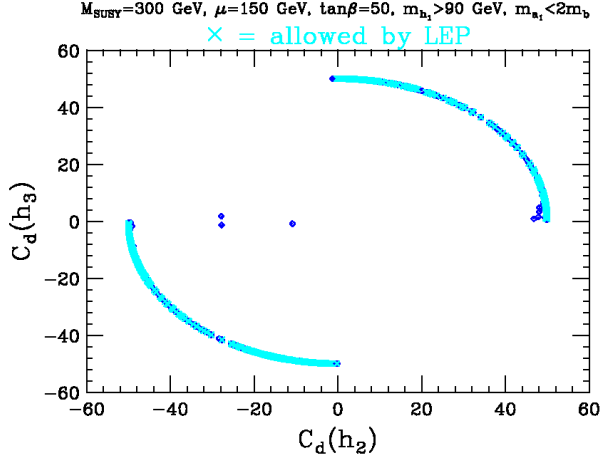}}
\caption{We plot $C_d(\hiii)$ vs. $C_d(\hii)$  for $\tan
  \beta = 50$ and $\mu=150\gev$.
Point conventions as in Fig.~\ref{fig:AkAl_and_kl}.}
\label{tb50cd3vscd2}
\end{figure}

In the left-hand plots of Fig.~\ref{tb50mhvsmh}, we give $\mhii$ vs. $\maii$,
$\mhiii$ vs. $\maii$ and $\mhiii$ vs. $\mhii$.  The light (cyan)
points are those that have both $\mai<2\mb$ and $\br(\hi\to \ai\ai)$ large
enough to escape the LEP limits on the $Z+2b$ channel. 
For such points, the smallest allowed values of $\maii$, $\mhii$ 
and $\mhiii$ are about $120\gev$, $105\gev$ and $180\gev$. We see that
at low $\maii<200\gev$, $\mhii\sim\maii$, while at large $\maii>600\gev$ one finds
$\maii\sim \mhiii$. At intermediate $\maii$, one can have either
degeneracy. On the right-hand side of 
Fig.~\ref{tb50mhvsmh}, we plot the  relative
coupling strength $C_d$ of each of the heavier Higgs bosons 
as a function of its mass.  Correlating with the mass plots, we observe
that for $\maii\lsim 200\gev$ it is always the $\hii$ that is
degenerate with the $\aii$ and that both have $C_d\sim \tanb$. For
$\maii\gsim 600\gev$, we always have $\mhiii\sim \maii$ and
$C_d(\hiii)$ and $C_d(\aii)$ are both $\sim \tanb$.  For $\maii$ in
the intermediate mass range, the situation is more complicated and we
can get cases where $\maii\sim\mhii\sim\mhiii$ and while
$C_d(\aii)\sim \tanb$ one finds that the $\hii$ and $\hiii$ can share
the $\tanb$ enhancement factor. This is illustrated in
Fig.~\ref{tb50cd3vscd2}.

Overall, it is clear that if $\tanb$ is large
a search at the LHC in channels such as $gg\to b\anti b
+Higgs$ could reveal a signal so long as $\maii$ is not too large.
For $\maii\lsim 250\gev$, the Tevatron might
also be able to detect this kind of signal when $\tanb$ is large enough.

\section{Conclusions \label{conclusions}}

Eliminating EWSB fine tuning in supersymmetric models has become an
important issue. In the NMSSM, this is most easily achieved
by allowing the lightest Higgs boson, $\hi$, to have mass of order 
$100\gev$, as naturally predicted after radiative corrections for stop
masses in the range of a few hundred GeV.  The modest stop masses imply no
significant fine-tuning. However, such an $\hi$ is typically Standard
Model like in its $ZZ$ coupling and can escape LEP limits only if 
the dominant decay is $\hi\to \ai\ai$ (so that the $\hi\to b\anti b$
decay is sufficiently suppressed to escape LEP limits)
and if $\mai<2\mb$ (so that the
$\ai\ai$ final state does not feed into the $Z+b's$ final state that
is strongly constrained by LEP data). 
In this paper, we have considered the degree to which the
GUT-scale soft-SUSY-breaking parameters must be tuned
in order to have $\mai<2\mb$ and $\br(\hi\to\ai\ai)>0.7$ (the rough
requirement for suppressing the $\hi\to b\anti b$ mode sufficiently).
We have found that such a scenario need not have
significant tuning. We began by assessing the tuning required of
the $\mz$-scale parameters $\alam(\mz)$ and $\akap(\mz)$ that
primarily control both $\mai$ and $\br(\hi\to\ai\ai)$. This tuning was
quantified via  
\beq
\fmax\equiv \max\left\{|\falam|,|\fakap|\right\}\,,\quad F_{A}\equiv {\ptl
  \log\mai^2\over \ptl \log A}\,,
\eeq
evaluated at scale $\mz$.
We found that so long as
$\mu$ is not too large ($|\mu|\lsim 200\gev$) then the values of
$\akap(\mz)$ and $\alam(\mz)$ need only be tuned to a level of order
$\fmax\sim 10 - 20$, corresponding to tuning in the range of
5\% to 10\%, for the magnitudes of $\akap(\mz)$ and $\alam(\mz)$ that are
of order those automatically generated by radiative evolution from
small GUT scale values. Further, these same RGE-generated values automatically give the required
large values of $\br(\hi\to\ai\ai)$. We also discussed how
these tuning estimates based on  $\fmax$ will generically greatly
overestimate the tuning with respect to GUT-scale soft-SUSY-breaking
parameters. In any SUSY scenario in which $\alam(\mz)$ and
$\akap(\mz)$ deriving from RG evolution 
are dominated by a GUT-scale parameter or
model-correlated set of GUT-scale parameters, generically denoted by
$p$, the tuning with respect to $p$ is given by
$f_p\sim \falam+\fakap$ and for the $\alam(\mz)$ and $\akap(\mz)$ 
regions with $\mai<2\mb$ and $\br(\hi\to\ai\ai)>0.7$ one finds
$\falam\sim -\fakap$. As a result $f_p$ can easily be very modest in
size, even for quite large $\mu$ --- the opposite-sign correlation
becomes increasingly effective as $\mu$ increases.

Thus, we have demonstrated that GUT-scale parameters can be chosen so
that fine-tuning can essentially be eliminated for
the NMSSM scenario of a light Higgs with $\mhi\sim 100\gev$,
decaying primarily via $\hi\to\ai\ai$ with $\mai<2\mb$ and
$\br(\hi\to\ai\ai)\gsim 0.7$ (both being required to escape LEP
limits). We regard this scenario (or something similar)
as a highly attractive possibility for the Higgs sector. It solves
both the $\mu$ problem and the fine-tuning problem.  It does, however,
introduce the need for proving viability of LHC discovery modes
involving $\hi\to \ai\ai\to 4\tau$ or $4~jets$, where the jets could
be $c, g,s ,\ldots$. The $4~jets$ mode is certain to be
quite difficult to probe at the LHC -- only diffractive $pp\to pp X$
might provide a signal in the form of a bump in the $\mx$ distribution
at $\mhi$. In this regard, it is important to note that our work shows
that minimizing the GUT-scale fine-tuning measure $f_p$ appears to
mildly prefer $\mai>2\mtau$.  Thus, a strong effort should be made to
develop $\hi\to \ai\ai\to 4\tau$ discovery modes at the LHC and to see
if LEP data can constrain this possibility. Of course, discovery of
the $\hi$ at the ILC through a bump at $\mx=\mhi$ in
the  $\epem\to Z^*\to Z X$ channel will be extremely easy.

\acknowledgments

RD thanks K.~Agashe, P.~Schuster, M.~Strassler and N.~Toro for discussions. RD
is supported by the U.S. Department of Energy, grant
DE-FG02-90ER40542.  Thanks go to the Galileo Galilei Institute (JFG) and the Aspen
Center for Physics (JFG and RD) for hospitality and support during the course of
this research. JFG also thanks M. Schmaltz for helpful conversations.
JFG is supported by DOE grant \#DE-FG02-91ER40674 and by the
U.C. Davis HEFTI program.

\vspace{0.2cm}



\end{document}